\documentclass{aastex701}
\usepackage{CJK}
\usepackage{graphicx}
\usepackage{enumerate}
\usepackage{amssymb, amsmath}
\usepackage{natbib}
\usepackage{color}
\usepackage{ulem}
\usepackage{appendix}
\usepackage{hyperref}
\usepackage[stable]{footmisc}
\usepackage{comment}
\usepackage{footnote}
\usepackage{lineno}

\begin{document}

\newcommand{\bjdtdb}{\ensuremath{\rm {BJD_{TDB}}}}
\newcommand{\feh}{\ensuremath{\left[{\rm Fe}/{\rm H}\right]}}
\newcommand{\teff}{\ensuremath{T_{\rm eff}}}
\newcommand{\rsun}{\ensuremath{\,R_\Sun}}
\newcommand{\lsun}{\ensuremath{\,L_\Sun}}
\newcommand{\mj}{\ensuremath{\,M_{\rm J}}}
\newcommand{\rj}{\ensuremath{\,R_{\rm J}}}
\newcommand{\fave}{\langle F \rangle}
\newcommand{\fluxcgs}{10$^9$ erg s$^{-1}$ cm$^{-2}$}
\newcommand{\acen}{$\alpha$ Cen}
\newcommand{\mearth}{M$_\oplus$}
\newcommand{\rearth}{R$_\oplus$}
\newcommand{\mum}{$\mu$m}
\newcommand{\epseri}{$\epsilon$ Eridani}
\newcommand{\deleri}{$\delta$ Eridani}
\newcommand{\tentos}{\ensuremath{10^{-7}}}
\newcommand{\tentoe}{\ensuremath{10^{-8}}}
\newcommand{\tenton}{\ensuremath{10^{-9}}}
\newcommand{\tentot}{\ensuremath{10^{-10}}}
\newcommand{\octofitter}{\texttt{Octofitter}}
\newcommand{\logg}{\ensuremath{\log g}}
\newcommand{\fsed}{\ensuremath{f_{\rm sed}}}
\newcommand{\mh}{[M/H]}
\newcommand{\sensitivitySanghi}{\ensuremath{\approx\!3\times10^{-7}}}

\title{An Updated Model for $\epsilon$ Eridani b and Prospects for Imaging with the Roman Coronagraph}

\author[0000-0002-3414-784X]{Jorge Llop-Sayson}
\affiliation{Jet Propulsion Laboratory, California Institute of Technology, Pasadena, CA 91109, USA}
\email{jorge.llop.sayson@jpl.nasa.gov}

\author[0009-0007-6766-2040]{Andre Fogal}
\affiliation{University of Victoria, 3800 Finnerty Road, Victoria, BC, V8P 5C2, Canada}
\affiliation{NRC Herzberg Astronomy and Astrophysics, 5071 West Saanich Road, Victoria, BC, V9E 2E7, Canada}
\email{afogal@uvic.ca}

\author[0000-0001-5864-9599]{James Mang}
\altaffiliation{NSF Graduate Research Fellow}
\affiliation{Department of Astronomy, University of Texas at Austin, Austin, TX 78712, USA}
\email{j_mang@utexas.edu}

\author[0000-0001-5684-4593]{William Thompson}
\affiliation{NRC Herzberg Astronomy and Astrophysics, 5071 West Saanich Road, Victoria, BC, V9E 2E7, Canada}
\email{william.thompson@nrc-cnrc.gc.ca}

\author[0000-0001-5173-2947]{Clarissa R. Do \'O}
\affiliation{Cahill Center for Astronomy and Astrophysics, California Institute of Technology, Pasadena, CA 91125, USA}
\email{cdoo@caltech.edu}

\author[0000-0002-1838-4757]{Aniket Sanghi}
\altaffiliation{NSF Graduate Research Fellow}
\affiliation{Cahill Center for Astronomy and Astrophysics, California Institute of Technology, Pasadena, CA 91125, USA}
\email{asanghi@caltech.edu}

\author[0000-0002-4031-6400]{Isabela G. Huckabee}
\affiliation{Department of Astronomy and Carl Sagan Institute, Cornell University, 122 Sciences Drive, Ithaca, NY 14853, USA}
\email{igh7@cornell.edu}

\author[]{Laurent Pueyo}
\affiliation{Space Telescope Science Institute, Baltimore, Maryland, United States}
\email{pueyo@stsci.edu}

\author[0000-0002-5627-5471]{Charles Beichman}
\affiliation{NASA Exoplanet Science Institute, IPAC, Pasadena, CA 91125, USA}
\affiliation{Jet Propulsion Laboratory, California Institute of Technology, Pasadena, CA 91109, USA}
\email{chas@ipac.caltech.edu}

\author[0000-0001-5966-837X]{Geoffrey Bryden}
\affiliation{Jet Propulsion Laboratory, California Institute of Technology, Pasadena, CA 91109, USA}
\email{geoffrey.bryden@jpl.nasa.gov}

\author[0000-0001-7591-2731]{Marie Ygouf}
\affiliation{Jet Propulsion Laboratory, California Institute of Technology, Pasadena, CA 91109, USA}
\email{Marie.Ygouf@jpl.nasa.gov}

\author[0000-0001-8612-3236]{Andr\'as G\'asp\'ar}
\affiliation{Steward Observatory, University of Arizona, Tucson, AZ, 85721, USA}
\email{agaspar@arizona.edu}

\author[0000-0002-9977-8255]{Schuyler Wolff}
\affiliation{Steward Observatory, University of Arizona, Tucson, AZ, 85721, USA}
\email{sgwolff@arizona.edu}

\author[0000-0002-0834-6140]{Jarron Leisenring}
\affiliation{Steward Observatory, University of Arizona, Tucson, AZ, 85721, USA}
\email{jarronl@arizona.edu}

\author[0000-0002-8895-4735]{Dimitri Mawet}
\affiliation{Cahill Center for Astronomy and Astrophysics, California Institute of Technology, Pasadena, CA 91125, USA}
\affiliation{Jet Propulsion Laboratory, California Institute of Technology, Pasadena, CA 91109, USA}
\email{dmawet@astro.caltech.edu}

\author{Tiffany Meshkat}
\affiliation{IPAC, California Institute of Technology, 1200 E. California Blvd., Pasadena, CA 91125, USA}
\email{meshkat@ipac.caltech.edu}

\author[0000-0002-4404-0456]{Caroline V. Morley}
\affiliation{Department of Astronomy, University of Texas at Austin, Austin, TX 78712, USA}
\email{cmorley@utexas.edu}

\author[0000-0003-2233-4821]{Jean-Baptiste Ruffio}
\affiliation{Department of Astronomy \& Astrophysics, University of California, San Diego, La Jolla, CA 92093, USA}
\email{jruffio@ucsd.edu}

\begin{abstract}
\epseri~b, the nearest known Jupiter analog, has its orbit and mass constrained from radial velocity (RV) and absolute astrometry, and its atmosphere from JWST/NIRCam imaging upper limits in \citet{Sanghi2026}. Here we follow up that work with a self-consistent model of \epseri~b that treats all available data within a single Bayesian framework.
We extend the RV data with new measurements, update the treatment of the astrometry data with a new model, and include the NIRCam observations with a grid of evolutionary and atmospheric models. 
We find a mass of $0.91\pm0.04\,M_{\rm Jup}$ and an orbit consistent with previous work. 
The imaging data helps constrain planet effective temperature, atmospheric metallicity and surface gravity. The constraints are dependent on model assumptions: for an atmosphere in chemical equilibrium, an otherwise low statistically significant feature in one of the epochs is recovered as the planet at high confidence. When assuming chemical disequilibrium, the posteriors exhibit a bimodal distribution, with one mode consistent with zero flux and the other coinciding with the flux of the tentative feature.
Consistent with previous work, the clear atmosphere models are found to be viable at very enhanced metallicity, whereas the cloudy models yield more moderate values. 
Applying these models to the Roman Coronagraph, we find that the predicted reflected-light fluxes place every cloudy atmosphere our fit allows within the instrument's expected sensitivity. A non-detection would be very constraining: a final contrast sensitivity of 2$\times$\tenton\ would rule out all cloudy atmospheres under our model assumptions.

\end{abstract}

\section{Introduction}
\setcounter{footnote}{0}

\epseri~is a K2V star \citep{Gray2003} at a distance of 3.2~pc, host to a multi-belt debris disk and a Jupiter-mass companion, \epseri~b, on a 7.4~yr orbit \citep{Mawet2019}. In age, spectral type, and overall
architecture it resembles the early Solar System, and owing to its proximity and apparent brightness ($V=3.73$) it has been a recurring target for studies of how planetary systems comparable to our own form
and evolve. As the nearest known Jupiter analog, \epseri~b is in addition a high-priority target for direct imaging \citep{Mamajek2024}.

\epseri~hosts one of the first debris disks to be detected. The star is one of the four infrared-excess sources identified by IRAS \citep[the ``Fab Four''; ][]{Gillett1986, Backman1993}, and subsequent observations from the infrared to the millimeter have established a multi-component structure, with warm dust interior to $\sim$25~AU and a cold belt extending to $\sim$70~AU \citep{Backman2009, MacGregor2015,Su2017}. ALMA
imaging resolves the outer belt as a narrow ring centered near 69~AU, with a width of order 12~AU and an inclination of $\sim$34\arcdeg\ \citep{Booth2017, Booth2023}. The confinement of this ring has been
attributed to an unseen planet orbiting at 40--50~AU, although recent JWST mid-infrared imaging finds no signature of a massive perturber beyond 5~AU and a dust distribution consistent with inward transport by
stellar-wind drag \citep{Wolff2025}.

The Jovian planet was identified through the star's radial velocity (RV). \citet{Hatzes2000} reported a long-period RV variation consistent with a giant planet of $\simeq0.86\,\mj$ on an eccentric ($e\simeq0.6$),
$\sim$6.9~yr orbit; the high eccentricity was difficult to reconcile with the disk geometry, however, and the star's pronounced magnetic activity raised the prospect of contamination by stellar jitter \citep{Anglada-Escude2012,Zechmeister2013}. \citet{Mawet2019} subsequently combined three decades of RV data with deep imaging upper limits in a joint Bayesian analysis that simultaneously modeled the planetary signal and the correlated stellar noise, deriving a mass of $0.78^{+0.38}_{-0.12}\,\mj$, a semi-major axis
of $3.48\pm0.02$~AU, a period of $7.37\pm0.07$~yr, and a low eccentricity
of $0.07^{+0.06}_{-0.05}$ compatible with the disk; \citet{Llop-Sayson2021}
refined this solution by adding absolute astrometry data from Hipparcos, and Gaia. These analyses nonetheless left the orbital orientation, in particular the inclination, and ascending node, poorly determined. 
\citet[][hereafter T25]{Thompson2025} performed a full model fit to the RV record together with absolute astrometry from Hipparcos, the Hubble FGS, and Gaia DR2/DR3, within the \octofitter~\citep{Thompson2023} framework. The new derived mass was of $1.00\pm0.10\,\mj$, a near-circular orbit, and an orbital plane closely aligned with the cold outer belt.

Notwithstanding these dynamical constraints, \epseri~b has not been detected directly. Even at its modest angular separation ($\sim$1\arcsec), it has proven beyond the sensitivity of current ground-based high-contrast imagers \citep{Mawet2019,Llop-Sayson2021,Tschudi2024}, leaving JWST as the most capable facility for a thermal-infrared search. Two NIRCam coronagraphic imaging programs have targeted the planet: a Guaranteed Time Observation (GTO) program \citep{Llop-Sayson2025} and a follow-up Director's Discretionary Time (DDT) program
\citep[][hereafter S26]{Sanghi2026} that reached a $5\sigma$ contrast of $\sim3\times10^{-7}$ at 1\arcsec\ in the F444W band, the most stringent 4--5\,\mum\ limit reported for the system; neither program yielded a detection. In S26, the interpretation of the NIRCam upper limit resulted in the first constraints on the planet's atmosphere. Using a revised stellar age of $1.1\pm0.1$~Gyr from gyrochronology, S26 find that the evolutionary models predict an effective temperature of 150--200~K for a Jupiter-mass planet, and that the absence of a 4--5\,\mum\ source is most easily explained by an atmosphere enriched in heavy elements, the presence of water clouds, or a combination of the two. A NIRSpec program (PI: Ruffio) is currently pursuing a spectroscopic detection of the planet. To date, no confirmed direct detection of \epseri~b has been obtained.

The orbital and atmospheric constraints summarized above have been derived independently: the mass and orbit from RV and astrometry \citep{Thompson2025}, and the atmospheric properties from imaging upper limits \citep[][incorporating the T25 dynamical mass]{Sanghi2026}. This work is a direct follow-up to the aforementioned two articles: we present a self-consistent model of \epseri~b that combines its orbit and mass with a grid of atmospheric models, using the JWST/NIRCam images to constrain the planet's effective temperature, carbon-to-oxygen ratio, surface gravity, metallicity, and radius. 

Such a model is of immediate relevance in advance of the Nancy Grace Roman Space Telescope, whose Coronagraph Instrument \cite[hereafter the Roman Coronagraph;][]{Bailey2023} is expected to demonstrate reflected-light imaging at planet-to-star flux ratios of $10^{-8}$--$10^{-9}$; a nearby, bright star hosting a giant planet near 1\arcsec\ separation is a well-suited target for such a demonstration. Indeed, with their updated atmospheric constraints, S26 showed that \epseri~b has a Band 1 contrast $\sim 10^{-9}$--$10^{-8}$ (both cloudy and clear scenarios, at optimal phase) and strongly motivated Roman Coronagraph observations of the planet in the early ``observation phase". Here, we apply the results of our joint model fits to the prospective observations with the Roman Coronagraph to give a prediction of the sensitivity of the upcoming instrument for our different model assumptions.

This paper is organized as follows. In Section~\ref{sec:methods} we describe the observations and \octofitter~setup. In Sections~\ref{sec:results} and~\ref{sec:discussion} we present and discuss the results. In Section~\ref{sec:roman} we apply the resulting model to the detectability of the planet with the Roman
Coronagraph. We summarize our conclusions in Section~\ref{sec:conclusions}.
 
\section{Observations and Methods}
\label{sec:methods}
\subsection{Radial Velocities}
\label{sec:rvs}

\begin{figure}\begin{center}
   \begin{tabular}{c} \includegraphics[height=7.0cm,trim={0cm 0cm 0cm 0cm}]{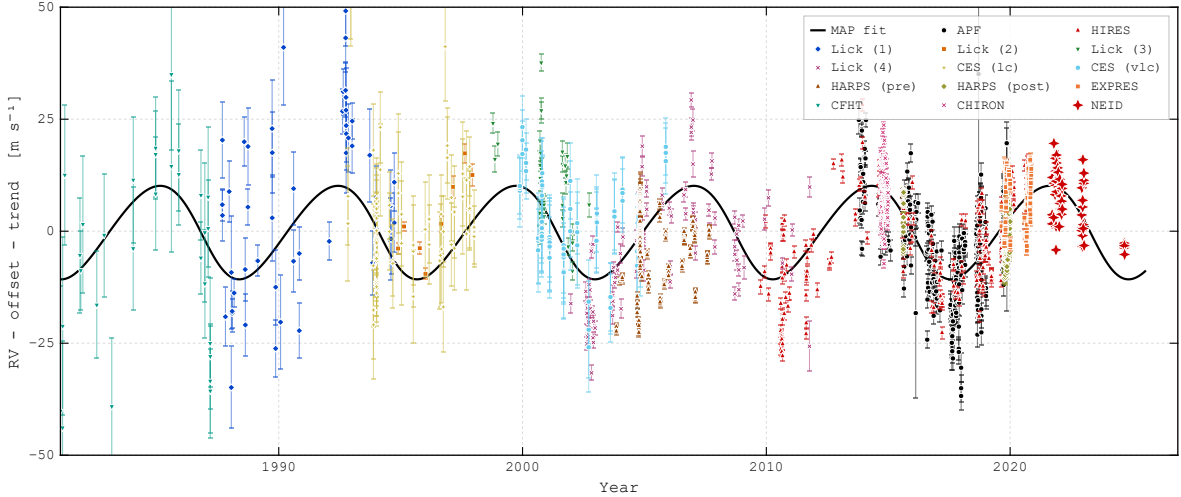}
   \end{tabular}
   \end{center}
   \caption{\epseri~radial velocities used in the \octofitter~model fit. The NEID data (red crosses) is the only new data with respect to previous model fits for this system. 
   \label{fig:rv}
   }  
\end{figure}

The radial velocity (RV) data used in this work is identical to that compiled by T25, with the exception of the new NEID measurements presented in Sec.~\ref{sec:neid}. The dataset combines decades of precision Doppler monitoring across thirteen distinct instrumental groupings: CFHT \citep{Campbell1988}; four Lick subsets split by epoch and dewar configuration as described in \citet{Fischer2014}; CES (LC and VLC, treated separately) from \citet{Zechmeister2013}; HARPS pre- and post-2015 upgrade from \citet{Trifonov2020}; HIRES and APF from \citet{Mawet2019} and \citet{Llop-Sayson2021}; CHIRON from \citet{Giguere2016}; and EXPRES from \citet{Roettenbacher2022}. As shown in Fig.~\ref{fig:rv}, these datasets, together with the NEID data, provide nearly continuous coverage of \epseri's orbit from 1981 to 2024, with each new RV measurement adding leverage on the planet's reflex motion against the prominent stellar activity signal.

We adopt the same treatment of the RV data as T25, which we briefly summarize here. Each of the thirteen instrumental groupings is fit with its own offset ($\gamma$), jitter ($\sigma$), and linear trend ($m$) parameters, adding 39 nuisance variables to the joint orbit fit. The stellar activity signal, dominated by rotational modulation at $\approx 11$\,d with a month-long decay timescale, is modeled with a quasi-periodic Gaussian process using the Celerite framework \citep{Foreman-Mackey2017}, with hyper-parameters $B$, $C$, $L$, and $P_{\rm rot}$ shared across all instruments. Over the 40-year baseline considered here, perspective acceleration from \epseri's high proper motion ($\approx 1''$/yr) and large barycentric radial velocity ($\approx 16$\,km\,s$^{-1}$) becomes non-negligible; T25 handle this term explicitly within the model rather than relying on the pre-existing static corrections applied to each archival dataset, and we follow the same approach. We refer the reader to T25 for a complete description of the priors and the implementation in the Octofitter framework \citep{Thompson2023}.

\subsubsection{NEID Radial Velocities}
\label{sec:neid}

We extend the RV baseline of T25 with the measurements from the NEID spectrograph \citep{Schwab2016}, a high-resolution ($R \approx 100{,}000$), environmentally-stabilized optical (380--930\,nm) Doppler spectrometer mounted on the WIYN 3.5\,m telescope at Kitt Peak National Observatory. NEID was specifically designed to deliver the sub-m\,s$^{-1}$ precision required for the detection of Earth-mass planets around nearby Sun-like stars, and its early performance on \epseri\ was characterized by \citet{Jiang2024} in the context of stellar activity diagnostics. The new data span multiple epochs from 2021 September to 2024 September, extending the RV baseline by approximately three years past the last observations included in T25; the individual measurements are listed in Table~\ref{tab:rvs}. Some of these RVs were used in the \citet{Jiang2024} work. We treat the NEID dataset as a fourteenth independent grouping in the orbit fit, with its own offset, jitter, and slope parameters, and include it under the same shared activity GP as the rest of the RV data.

\subsection{Astrometric Data}
\label{subsec:astrometric_data}
We utilize the same astrometric datasets as T25: Hipparcos \citep{vanLeeuwen2007}, Hubble/FGS \citep{Benedict2006}, and Gaia DR2 and DR3. However, we adopt the composite catalog and orbit-fitting framework of \citet{Thompson2026} (referred as G23H), summarized in Sec.~\ref{subsubsec:g23h}.

From Hipparcos we use the intermediate astrometric data (IAD) sourced from the Java Tool following \citet{Nielsen2020}, together with the Hipparcos proper motions and long-baseline Hipparcos--Gaia scaled position difference from the DR3 Hipparcos--Gaia Catalog of Accelerations \citep[HGCA;][]{Brandt2021}. As in T25, the HGCA values are pre-calibrated against the Gaia DR3 reference frame, which keeps our long-term proper motion anchor consistent with Gaia without introducing reference-frame alignment as free parameters. For the Gaia epochs, we use DR2 and DR3 proper motions, the DR3--DR2 scaled position difference, and the Gaia DR3 radial velocity variability; all entering the fit through G23H.

The Hubble/FGS data are handled exactly as in T25. We use the residuals to the five-parameter parallax and proper-motion solution of \citet{Benedict2006}, and fit only the acceleration within the FGS baseline while marginalizing over a linear trend in each axis. This sidesteps absolute calibration of the FGS data against the Gaia DR3 frame, analogous to including a linear trend in RV fits.

\subsubsection{The G23H Composite Catalog and Likelihood}
\label{subsubsec:g23h}

G23H extends the treatment of Gaia and Hipparcos data in T25 along three axes that we exploit here: (1) calibration of the Gaia DR2 proper motions and the DR3--DR2 scaled position difference into the Gaia DR3 reference frame, (2) inclusion of the Gaia DR3 RV variability from \citet{Chance2025} as an orthogonal companion diagnostic, and (3) a joint forward model that properly handles the correlation between DR2 and DR3 measurements. G23H also offers a likelihood term based on the Gaia astrometric excess noise (the UEVA, following \citealt{Kiefer2025}); we do not use it for \epseri, for reasons we explain below.

The calibration is the most consequential change for our purposes. Where T25 relied on the global rotation of \citet{Lindegren2020} with uncertainties matching the inflations applied in the HGCA, G23H performs a spatially resolved fit: a global rotation in 84 magnitude bins motivated by the magnitude-dependent reference-frame drift of \citet{Cantat-Gaudin2021}, followed by a local Gaussian mixture model in each HEALPix level-6 bin, yielding both a local shift and a per-bin uncertainty inflation factor. The same procedure is applied to the DR3--DR2 scaled position difference, which is sensitive to accelerations on sub-baseline timescales entirely within the Gaia mission. As shown in \citet{Thompson2026}, the calibrated DR2 and DR3--DR2 quantities follow Gaussian distributions against DR3 across magnitudes, which is the behavior our likelihood assumes. For \epseri, with its known long-term acceleration and a planet orbital period near the Gaia DR3 baseline, the DR3--DR2 scaled position difference is where most of the new acceleration information lives.

We deliberately exclude the UEVA term from our fit. \citet{Thompson2026} itself caution that the catalog calibration becomes unreliable for stars brighter than $\sim$4th magnitude, where bright-star processing changed substantially between DR2 and DR3 and the calibration sample is small. \epseri~($V \approx 3.7$) sits squarely in that regime, and the bulk of its astrometric excess noise is expected to be instrumental rather than astrophysical. Including UEVA in the likelihood would force the sampler to attribute this instrumental scatter to a companion, which can produce false positives, inflate inferred masses, and bias the rest of the orbital posterior. We therefore drop $\ln \mathcal{L}_{\rm UEVA}$ for this target. We note that this also means we do not benefit from G23H's orbital-model-driven de-inflation of the DR3 covariances, but we judge that conservatism preferable to a known systematic.

The DR3 RV variability from \citet{Chance2025} is inexpensive to include and provides an orthogonal check against close-in stellar-mass impostors. The joint DR2--DR3 covariance properly accounts for the fact that the two catalogs share the majority of their along-scan measurements (see \citet{Thompson2026} for the per-source estimator of $\rho_{\rm DR2,DR3}$).

The full astrometric log-likelihood is therefore the sum
\begin{equation}
    \ln \mathcal{L}_{\rm astro} = \ln \mathcal{L}_{\rm PM} + \ln \mathcal{L}_{\rm IAD} + \ln \mathcal{L}_{\rm RV} + \ln \mathcal{L}_{\rm FGS},
\end{equation}
with $\ln \mathcal{L}_{\rm PM}$ itself decomposed into Hipparcos, Hipparcos--Gaia, joint DR2--DR3, and DR3--DR2 position-difference terms as in \citet{Thompson2026}. 

\subsection{Imaging Data}
\label{sec:imaging_data}
We use the two epochs of JWST/NIRCam coronagraphic data of \epseri\ in this work: the GTO program (program \#1193) reported in \citet{Llop-Sayson2025}, and the DDT program (program \#9431) presented in S26 \citep{10.17909/g6k6-bx64}. Both epochs observed the system simultaneously at F210M and F444W with the MASK335R coronagraph, and used $\delta$~Eridani as the PSF reference star. For the GTO data we solely use the subarray data, executed on February 2024; a three-roll sequence optimized for the inner region, and a 5-point dither
pattern on $\delta$~Eri, which reaches $\sim\!1\times10^{-6}$ contrast at $1''$ in F444W \citep{Llop-Sayson2025}. The DDT data, executed on 2025~August~29, were specifically designed to avoid the bright NIRCam coronagraph PSF lobes, or hexpeckles, at the predicted position of \epseri~b from the orbit fit of T25; with telescope V3 angles of $268.\!^{\circ}09$ and $258.\!^{\circ}09$, and a 9-point dither pattern on $\delta$~Eri, this dataset reaches
$\approx\!3\times10^{-7}$ at $1''$ in F444W \citep{Sanghi2026}.

Other available high-contrast imaging datasets of \epseri\ do not add constraining power on top of these two NIRCam epochs. The Keck/NIRC2 vortex $Ms'$ search of \citet{Llop-Sayson2021} was the deepest pre-JWST 4--5~$\mu$m imaging of the system; however, its $5\sigma$ upper limit at the planet separation is now surpassed by more than an order of magnitude by the NIRCam F444W contrast curve. SPHERE/ZIMPOL polarimetric imaging from \citet[][]{Tschudi2024} reaches an impressive contrast in the optical; the atmospheric models used in this work, however, do not provide reliable predictions of the emitted
or scattered flux at those wavelengths for a $\sim\!150$--$200$~K planet, and hence we cannot translate the ZIMPOL upper limit into a meaningful constraint on the parameters of interest. Similarly, the F210M contrast reach in the two NIRCam epochs, albeit very deep levels of contrast, it does not provide much constraining power given the bandpass. The JWST/MIRI data taken as part of the same GTO program \citep[]{Wolff2025} didn't use the coronagraph and the contrast at the planet expected separation is well above the flux predicted by any of our F444W-consistent atmospheric models. We thus base the analysis in this work on the F444W data from the GTO and DDT NIRCam epochs.

The reduction of both epochs follows the procedure described in
\citet{Greenbaum2023} and \citet{Ygouf2024}, which we briefly
summarize here. For the GTO epoch we use the reduction presented in \citet{Llop-Sayson2025}: stage-0 to stage-2 processing through the \texttt{SpaceKLIP}-modified \texttt{jwst} pipeline \citep{Bushouse2025, Kammerer2022}, image registration via cross-correlation against an \texttt{STPSF} model, and PSF subtraction with \texttt{pyKLIP} \citep{Wang2015} in a combined ADI+RDI configuration using both the science rolls and the 5-point-dithered $\delta$~Eri reference frames. 
For the DDT epoch we start from the same set of \texttt{calints} stage-2 frames as S26, processed with \texttt{spaceKLIP}~v2.1 following the practices of \citet[][]{Carter2023} and \citet[][]{Gagliuffi2025} (including the ``Likely'' up-the-ramp fit of
\citealt[][]{Brandt2024a, Brandt2024b}, $1/f$ destriping, and Gaussian blurring above the Nyquist criterion); however, we rerun the PSF subtraction with the same \texttt{pyKLIP} ADI+RDI workflow used for the GTO data, for consistency between the two epochs. The KLIP reference library for the DDT subtraction is built from the two science rolls and the 9-POINT-CIRCLE $\delta$~Eri dither sequence. 

The contrast sensitivities are calibrated via injection-recovery
tests, and following \citet{Mawet2014} for the small-sample statistics correction. For each KLIP reduction geometry we inject synthetic point sources at a grid of separations, position angles, and contrast ratios, excluding the position angle range covered by the predicted location of \epseri~b. Algorithmic throughput is absorbed into the injection-recovery loop; the coronagraph mask transmission and off-axis PSF normalization are computed with \texttt{STPSF} \citep{Perrin2014STPSF} and \texttt{webbpsf\_ext} \citep{Leisenring2025}.

To compute the flux and contrast images needed by \octofitter, we employ the Forward Model Matched Filter (FMMF) as implemented in \texttt{pyKLIP} \citep{Ruffio2017}. Speckle subtraction algorithms such as KLIP distort the planet PSF; FMMF cross-correlates the residual data with the forward-modeled planet PSF \citep{Pueyo2016}, which accounts for the distortion introduced by the reduction. The reduction yields two maps of interest: the FMMF map, proportional to the signal-to-noise ratio (SNR) of a putative planet at each location, and the FMCont map, the maximum-likelihood estimate of the planet-to-star flux ratio. We adopt these matched-filter and contrast maps, calibrated for self- and over-subtraction through injection and recovery of simulated planets, as the inputs to the \octofitter~orbit and photometry fit.

\subsection{Evolutionary and Atmospheric Models}
\label{sec:evol_atm_models}
Our atmospheric and evolutionary modeling is a variation of the analysis in S26, providing an alternate way of treating the data and performing atmospheric inference. The main difference is methodological: rather than evaluating the consistency of individual atmospheric model gridpoints with the JWST/NIRCam F444W upper limit, as in S26, we perform a full joint fit of the planet's orbital, dynamical, and photometric properties, treating the NIRCam images as a likelihood contribution alongside the RV and absolute astrometry data. We refer the reader to S26 for the full description of the underlying model framework; here we summarize the elements that matter for the present work and detail the modifications.

We adopt the Sonora Flame Skimmer evolutionary tracks and cloud-free atmosphere grid \citep[][]{Mang2026flame-skimmer}. The evolutionary tracks provide a self-consistent set of cooling histories that span planet masses from $\sim 15~M_{\oplus}$ to $\sim 80~M_{\rm Jup}$, ages from a few Myr to several Gyr, and bulk metallicities $[\mathrm{M/H}] \in \{-1.0, -0.5, 0.0, +0.5, +1.0, +1.5, +2.0\}$~dex. The cloud-free atmosphere grid extends the Sonora Elf Owl framework \citep{Mukherjee2023} to the colder effective temperatures, lower surface gravities, and broader range of metallicities relevant for \epseri\ b. It covers $T_{\rm eff} =
50$--$2400$~K, $\log g = 2.0$--$5.5$~dex (cgs), the same $[\mathrm{M/H}]$ range as the evolutionary tracks, $\mathrm{C/O}$ from $0.5$ to $2.5\times$ solar, and vertical mixing $\log K_{zz} \in
\{2, 4, 7, 8, 9\}$ in cgs units, with separate equilibrium (eq) and disequilibrium (deq) chemistry tables. As in S26, the cloud-free models include rainout chemistry for H$_2$O, CH$_4$, and NH$_3$ even though no condensate clouds are explicitly modeled in this case. 

\begin{figure}\begin{center}
   \begin{tabular}{c} \includegraphics[height=6.0cm,trim={0cm 0cm 0cm 0cm}]{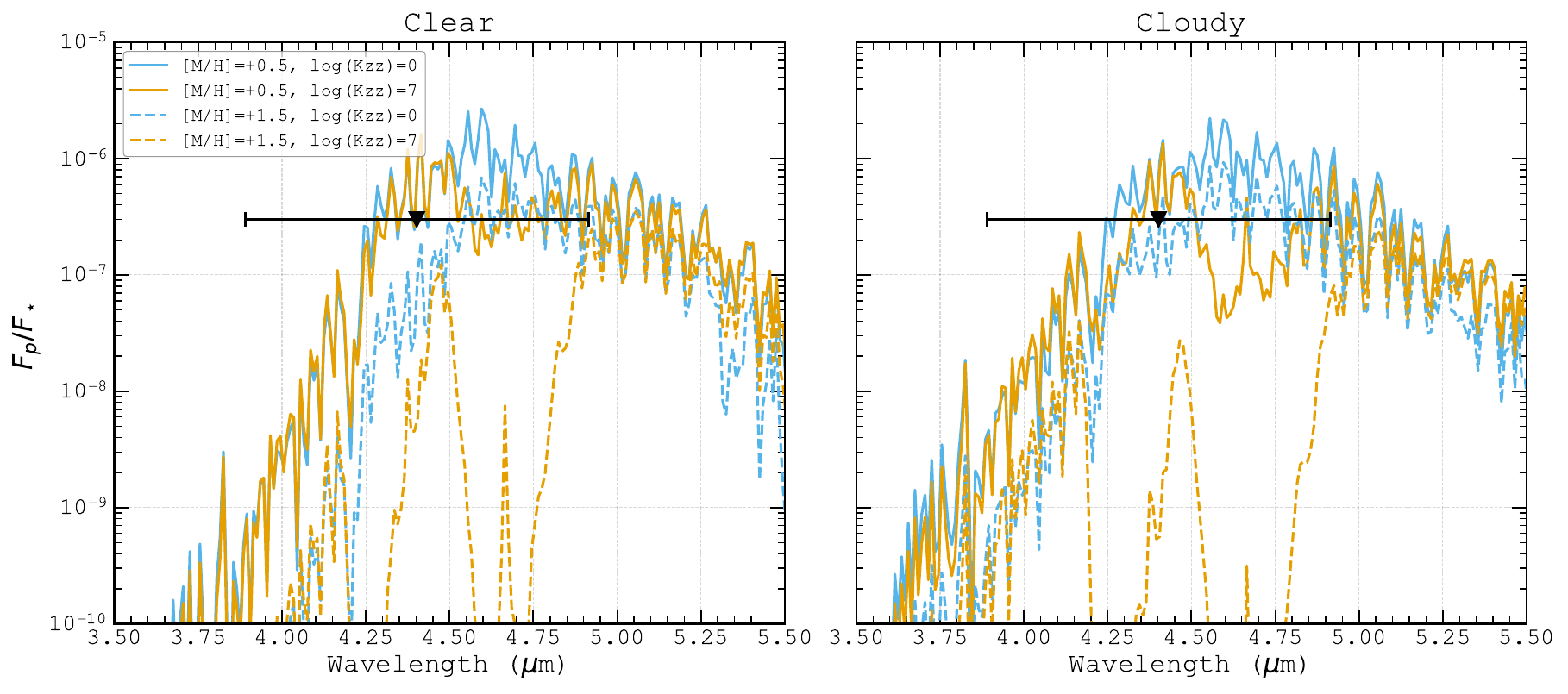}
   \end{tabular}
   \end{center}
   \caption{Clear (\textit{left}) and cloudy (\textit{right}) spectra in the 4 $\mu$m peak region for some representative \epseri~b models. Curves show four combinations of atmospheric metallicity and vertical mixing: [M/H]$=+0.5$ (solid) and $+1.5$ (dashed), each at $\log(K_{zz})=0$ (blue) and $\log(K_{zz})=7$ (orange). The black error bar and downward triangle mark the F444W bandpass and measured NIRCam upper limit. At [M/H]$=+0.5$ the flux is relatively more insensitive to mixing, with the two $K_{zz}$ cases nearly overlapping, whereas at [M/H]$=+1.5$ vigorous mixing suppresses the flux by a further $\sim$1--2~dex within the band, driven by enhanced CO and CO$_2$ absorption across 4--5~$\mu$m. Clouds modestly reduce the spread between models but preserve the overall behavior. 
   \label{fig:spectra_picaso}
   }  
\end{figure} 

At the cold effective temperatures relevant for \epseri\ b, water condenses, and the resulting clouds are the single largest source of uncertainty in the predicted F444W flux {, together with the chemistry, equilibrium versus disequilibrium and mixing (K$_{zz}$) assumptions.}
To quantify their impact on our inference, we therefore complement the cloud-free grid with a custom set of patchy water-cloud atmospheres computed with \texttt{PICASO} 4.0 \citep{Batalha2019,Mukherjee2023,Mang2026picaso} and \texttt{Virga} \citep{Batalha2025,Moran2025}. These are the same models introduced in S26, and we adopt a two-column \citep{Marley2010,Morley2014} prescription with cloud fraction $1 - h = 0.75$. We extend the sedimentation efficiency values to $f_{\rm sed} = 1, 3, 8$, values that are relevant to the sensitivity enabled by the Roman Coronagraph. In Fig.~\ref{fig:spectra_picaso} we show some representative examples of the spectrum of planet b in the vicinity of the 4 $\mu$m emission peak for the clear and cloudy assumption. In Fig.~\ref{fig:albedo_vs_wvl} we show the geometric albedo in the visible for a set of representative models of \epseri~b.

\subsubsection{Custom cloudy atmosphere computation}
\label{sec:cloudy_picaso}
Each cloudy atmosphere is solved self-consistently with the \texttt{PICASO} 4.0 radiative-convective climate solver. For a given $(T_{\rm eff}, \log g)$ we recover the internal temperature by removing the irradiation contribution, $T_{\rm int}^4 = T_{\rm eff}^4 - T_{\rm eq}^4$, with $T_{\rm eq} = T_\star \sqrt{R_\star / 2a}$ the bolometric equilibrium temperature for zero Bond albedo and full redistribution; the day-night redistribution is then carried inside the solver through an irradiation factor appropriate for the global average of an irradiated planet. Mass and radius are taken from the Flame Skimmer tracks at the matching gridpoint, so the surface gravity entering the climate solve is consistent with the cooling history used elsewhere in the fit. For the disequilibrium atmospheres the chemistry is recomputed self-consistently as the $T(P)$ profile evolves, with quenching set by the run-level $K_{zz}$ and with cold-trap and volatile rainout included.
 
\begin{figure}\begin{center}
   \begin{tabular}{c} \includegraphics[height=6.0cm,trim={0cm 0cm 0cm 0cm}]{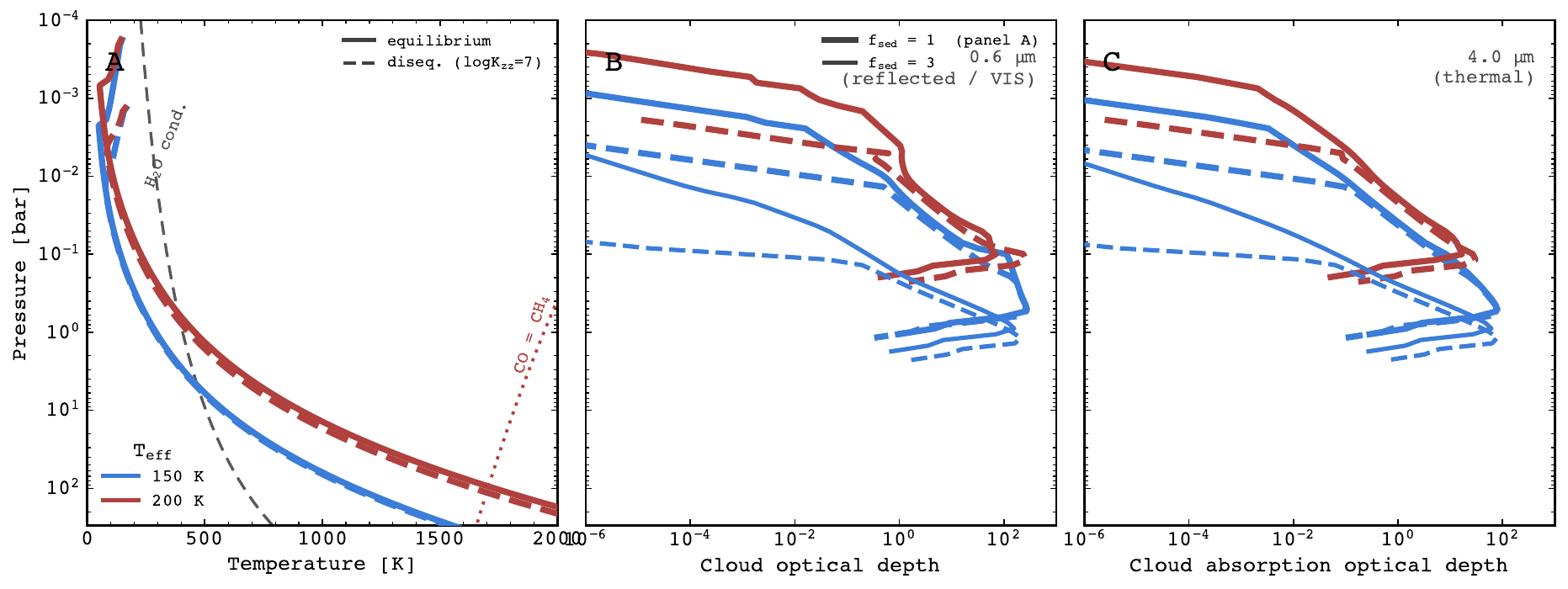}
   \end{tabular}
   \end{center}
   \caption{ To perform model fits with a cloudy atmosphere assumption we compute cloudy atmosphere models with PICASO following the method described in Sec.~\ref{sec:cloudy_picaso}. \textit{Left:} P/T curve for cloudy (\fsed=1) atmospheres, for two effective temperatures in chemical equilibrium and disequilibrium. The effective temperature drives at what levels the clouds form. \textit{Middle:} Cloud optical depth in reflected light for the \fsed=1 and 3. \textit{Right:} Cloud absorption optical depth at 4 $\mu$m, $\tau \times (1-w_0)$. These plots illustrate where the cloud layer sits depending on the temperature and log Kzz. Mixing quenches the condensible vapor abundance at the cloud base, lofting material higher and changing where the optical depth builds up.
   \label{fig:PT}
   }  
\end{figure}

\begin{figure}\begin{center}
   \begin{tabular}{c} \includegraphics[height=7.5cm,trim={0cm 0cm 0cm 0cm}]{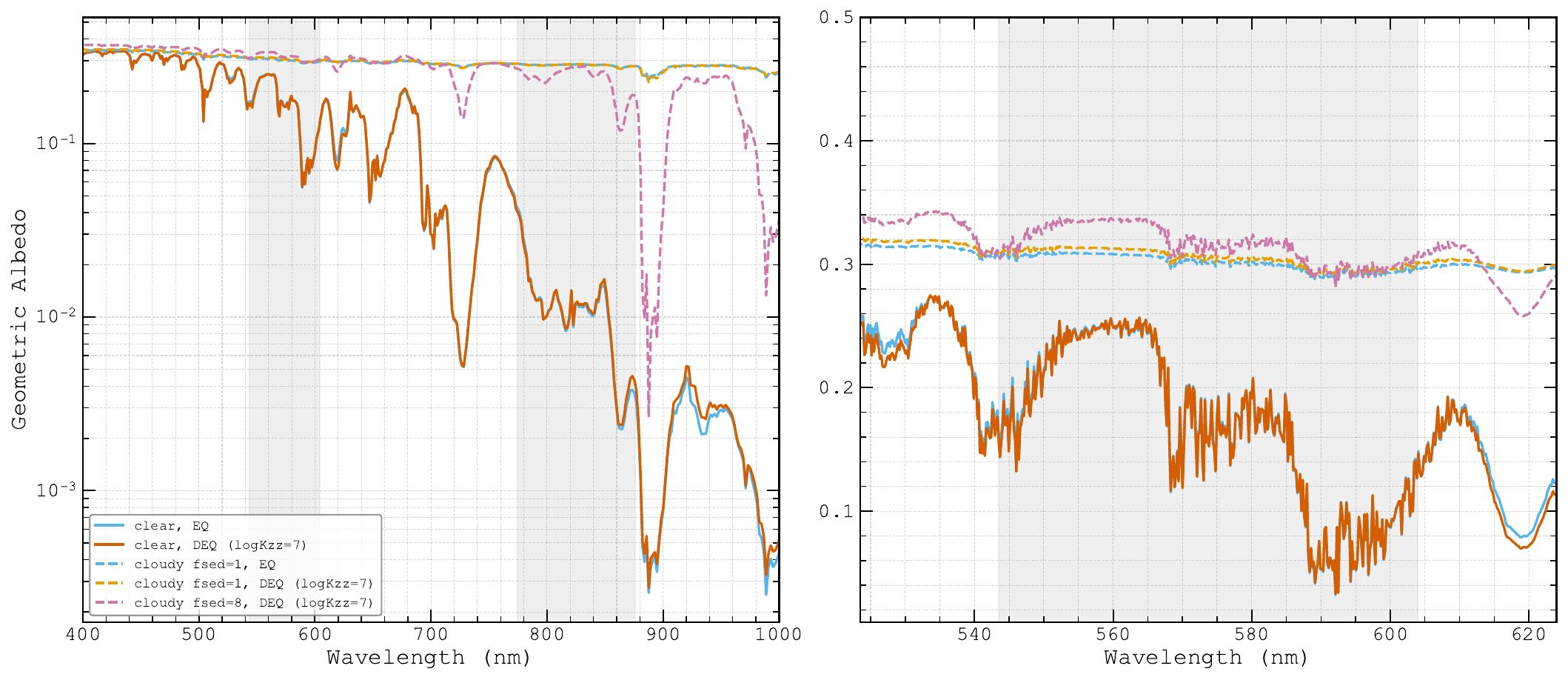}
   \end{tabular}
   \end{center}
   \caption{Geometric albedo for some representative models of \epseri~b. Shaded area represent the Roman Coronagraph's Band 1 and 4 bandpasses. The parameters used to compute thse albedos with PICASO are: \teff=175$^\circ$C, M/H=0.5, C/O=1.5, and \logg=3.5. \textit{Right:} Zoomed-in spectra in the vicinity of the Roman Coronagraph Band 1 (10\% bandwidth at 575 nm).\label{fig:albedo_vs_wvl}
   }  
\end{figure}

The clouds are formed inside the climate solve itself, so that the converged $T(P)$ profile and the condensate structure are mutually consistent. We use three different H$_2$O sedimentation efficiency parameters, $f_{\rm sed} = $1, 3, and 8, where higher values represent less cloudy models with 10 being essentially cloud-free. We treat the partial cloud cover with the two-column patchy prescription, combining a clear fraction $h = 0.25$ with a cloudy fraction $1 - h = 0.75$. Our cloudy model solvers converge well, with $T(P)$ profiles that are consistent with clear models in chemical equilibrium and disequilibrium. In Fig.~\ref{fig:PT} we show some examples of our cloudy atmosphere models. The $T(P)$ curves show an inversion in the top layers of the atmosphere due to the irradiation from the star. The cloud sedimentation level between \fsed~values 1 and 3 is evidently different; however, as discussed in Sec.~\ref{sec:albedo}, the albedo values for both cases will be not so dissimilar for the resulting planet models from our model fits.

To obtain the flux grids for the relevant parameter ranges that are used in the \octofitter~runs, the fluxes are computed at the NIRCam and Roman Coronagraph bandpasses. We integrate flux through the F444W bandpass for the fits that include the NIRCam images. Both thermal emission and reflected light are included but the reflected light contribution is negligible. From the same atmospheres we also compute the fluxes over a grid of phase angles and integrate it through the four Roman Coronagraph bandpasses for the fits presented in Sec.~\ref{sec:roman}. In all Roman Coronagraph bandpasses, the reflected light contribution dominates over the thermal emission. We only use Band 1 (10\% band at 575 nm) for the Roman Coronagraph fits since it was found in the White Paper (see Sec.~\ref{sec:roman} and S26 that this bandpass is the most favorable for a detection.
 
\subsection{Planet Model Fit Implementation}
\label{sec:planet_model_fit}

We fit our planet model to the data using \octofitter~\citep{Thompson2023}, an open-source Julia framework for joint modeling of multi-instrument exoplanet datasets. Octofitter parameterizes a planetary system through Keplerian orbital elements together with stellar and per-planet physical parameters, and ingests heterogeneous observables in a single likelihood: radial velocities, absolute astrometry, proper motion anomalies, and direct imaging detections or images in general. The framework supports both Campbell and Thiele--Innes orbital parameterizations \citep{Thompson2023}; we adopt the Campbell parameterization throughout this work, with a uniform prior on the orbital period $P$ centered on the T25 best-fit value, a uniform prior on eccentricity $e$, a sine prior on inclination $i$ together with a uniform prior on the longitude of the ascending node $\Omega$, a truncated Gaussian prior on the host stellar mass $M_\star$ \citep{Baines2012}, a uniform prior on the companion mass $M_\mathrm{b}$, and uniform priors on the remaining angles. The semimajor axis is derived from $P$ and $M_\star$ via Kepler's third law rather than sampled directly. Posteriors are sampled with stabilized variational non-reversible parallel tempering as implemented in \texttt{Pigeons.jl} \citep{Surjanovic2023}, which is well suited to the multimodal, high-dimensional posteriors that joint RV plus astrometry fits of \epseri\ b are known to produce.

Our orbit fit follows the setup of T25 closely. The RV and absolute-astrometry ingredients are detailed in Sec.~\ref{sec:rvs} and Sec.~\ref{subsec:astrometric_data} respectively; we refer the reader to T25 for the full justification of the priors and the propagation of second-order effects (perspective acceleration, barycentric motion, and the changing parallax and light-travel time over the 40-year baseline), which we retain here unchanged. The substantive differences with respect to T25 are concentrated in four places: (1) the RV baseline is extended by the NEID grouping of Sec.~\ref{sec:neid}, bringing the total to fourteen instrumental groupings; (2) the Gaia--Hipparcos sector is handled by the G23H composite catalog and likelihood (Sec.~\ref{subsubsec:g23h}), which absorbs T25's bespoke DR2--DR3 correlation treatment and adds a calibrated DR3--DR2 scaled position difference; (3) we add the F444W image likelihoods of Sec.~\ref{sec:imaging_data} with the predicted planet flux supplied by the joint evolutionary plus atmospheric grid of Sec.~\ref{sec:evol_atm_models}, as detailed in Sec.~\ref{sec:imaging_likelihood} below; and (4) we add a Gaussian prior on the planet--disk mutual inclination motivated by the MIRI imaging of the debris disk (Sec.~\ref{sec:mutual_inclination_prior}). All other modeling choices, including the literature priors on stellar mass, distance, and systemic velocity, are inherited from T25 unchanged. The full model has of order $\sim\!60$ free parameters, comparable to the 66-parameter most complete model of T25.

We sample the joint posterior with 12 rounds of \texttt{Pigeons} across 32 parallel chains and check convergence with the standard \texttt{Pigeons} diagnostics together with $\hat{R}$ and effective-sample-size statistics on the final chain. 

\subsubsection{Imaging Likelihood}
\label{sec:imaging_likelihood}

The most significant addition compared to T25 is the inclusion of direct imaging constraints from the two JWST/NIRCam epochs (Sec.~\ref{sec:imaging_data}). We incorporate the F444W data as image likelihoods in Octofitter, following the formalism developed in \citet{Ruffio2017}, applied in \citet{Mawet2019} and \citet{Llop-Sayson2021}, and described in Sec.~2.2 of \citet{Thompson2023}. After forward-model PSF subtraction and matched filtering, each pixel of the post-processed image carries two numbers: a maximum-likelihood flux estimate at that location, and a per-pixel noise level set by the residual speckle field; in our case these are the FMCont and FMMF contrast maps of Sec.~\ref{sec:imaging_data} sampled on the 63~mas/pixel NIRCam grid. At every posterior draw, Octofitter evaluates the planet's predicted on-sky position from the orbit, reads off the matched-filter flux at that pixel, and compares it to the model-predicted flux at the same epoch under a Gaussian likelihood weighted by the local noise. The construction produces a meaningful likelihood whether or not the planet is actually detected: a non-detection at the predicted position contributes a Gaussian penalty centered on zero with width set by the local contrast, which disfavors model fluxes well above the contrast floor while leaving fluxes consistent with noise effectively unconstrained. The non-detection is therefore not a hard cut on flux; it is a soft, pixel-dependent constraint whose tightness varies across the field with the achieved sensitivity.

\octofitter~allows two ways to specify the model-predicted flux entering this likelihood, and we make use of both in this work. (1) In the \textit{free-floating flux} mode, the per-band flux $F_b$ is itself a free parameter with a uniform prior, agnostic to any atmospheric or evolutionary model. In other words, there is no connection between the flux and the mass. The sampler propagates the orbit and the flux jointly, and the marginal posterior on $F_b$ behaves as a generalized SNR. If a real source sits at the orbit's predicted position the chain converges on its brightness; otherwise the marginal is consistent with zero out to the local contrast floor, thus treating the data flux as an upper limit. (2) In the \textit{atmospheric-model} mode the flux is replaced by a derived quantity computed from the joint evolutionary plus cloud-free, or cloudy atmospheric grid of Sec.~\ref{sec:evol_atm_models}. 
For each fit we fix the chemical equilibrium or disequilibrium assumption (fixed $\log K_{zz}$) and the clear or cloudy assumption (fixed \fsed). 
{We do not marginalize over these assumptions; the atmospheric posteriors presented in Sec.~\ref{sec:results} are therefore conditional on the chemistry and cloud prescription of each run. We discuss the implications of this choice in Sec.~\ref{sec:lowsnr_feature}.} 
The age is treated as a prior based on S26's updated analysis: 1.1$\pm$0.1 Gyr. 
At each posterior draw of $(M, t, [\mathrm{M/H}], \mathrm{C/O})$, the two-stage interpolator returns $(T_{\rm eff}, \log g, R)$ from the Sonora Flame Skimmer evolutionary tracks and the corresponding F444W flux from the atmosphere grid at the run-level $\log K_{zz}$ slice; this flux is then evaluated against the flux and contrast maps at the orbit's instantaneous projected separation and PA on each of the two NIRCam epochs. The non-detection therefore does not simply truncate the orbit posterior at bright orbital configurations; it propagates back into the atmospheric parameters, disfavoring combinations of atmospheric parameters that would have produced a flux above the contrast floor at the GTO or DDT epochs. This is the same coupling exploited by S26 in their grid-based consistency analysis, but here it is realized inside the joint posterior rather than as a post-hoc check, and it allows us to leave $M_\mathrm{b}$ free instead of fixing it.

We use the free-floating-flux mode primarily as a model-agnostic diagnostic: it returns a flux posterior at the orbit's instantaneous position in each epoch that we compare against the per-epoch contrast curves directly, and that we use to confirm that the imaging likelihood is behaving as expected on these data, and to assess the changes in the orbit model when the images are added. The atmospheric-model mode carries the more meaningful inference. The constraints on $T_{\rm eff}$, $\mathrm{C/O}$, and $[\mathrm{M/H}]$ presented in Sec.~\ref{sec:results}, and the resulting Roman Coronagraph predictions of Sec.~\ref{sec:roman}, are all derived from posteriors sampled in this mode.

\subsubsection{Adding a Constraint to the Disk-Planet Mutual Inclination, $\Phi$}
\label{sec:mutual_inclination_prior}

The orbital fit benefits from an additional prior on the mutual inclination $\Phi$ between the planet's orbit and the debris disk. The MIRI imaging of the \epseri~disk presented in \citet{Wolff2025} shows a remarkably flat, axisymmetric structure that extends inward to near the saturated region of the MIRI image, around 3~AU, with no evidence for a broad gap or warp at the separations probed by the planet's orbit. For a giant planet of roughly Jupiter mass, a substantial mutual inclination would be expected to leave a dynamical imprint on the disk, either as a warp, a vertically puffed component, or a carved gap near the planet's apocenter; none of these features are seen. Hence, while the disk geometry alone does not pin $\Phi$ to zero, it disfavors strongly misaligned configurations.

We therefore adopt a Gaussian prior on the mutual inclination, centered at $\mu = 0^\circ$ with $\sigma = 10^\circ$. This choice is intentionally conservative: it allows the posterior to drift to tens of degrees of misalignment if the astrometry demands it, while penalizing the highly inclined solutions that are inconsistent with the disk morphology. A similar approach was taken by \citet{Wang2020} for the PDS~70 system, where the mutual inclination between the planets and the circumstellar disk was constrained with a comparable Gaussian width. The $10^\circ$ scale is well above the level of coplanarity expected from secular planet-disk interaction for a $\sim 1\,M_\mathrm{Jup}$ perturber, so the prior is not driving the fit toward an artificially flat configuration; it is removing the tail of solutions that the disk image already rules out.

\section{Results}
\label{sec:results}

\begin{figure}\begin{center}
   \begin{tabular}{c} \includegraphics[height=12.0cm,trim={0cm 0cm 0cm 0cm}]{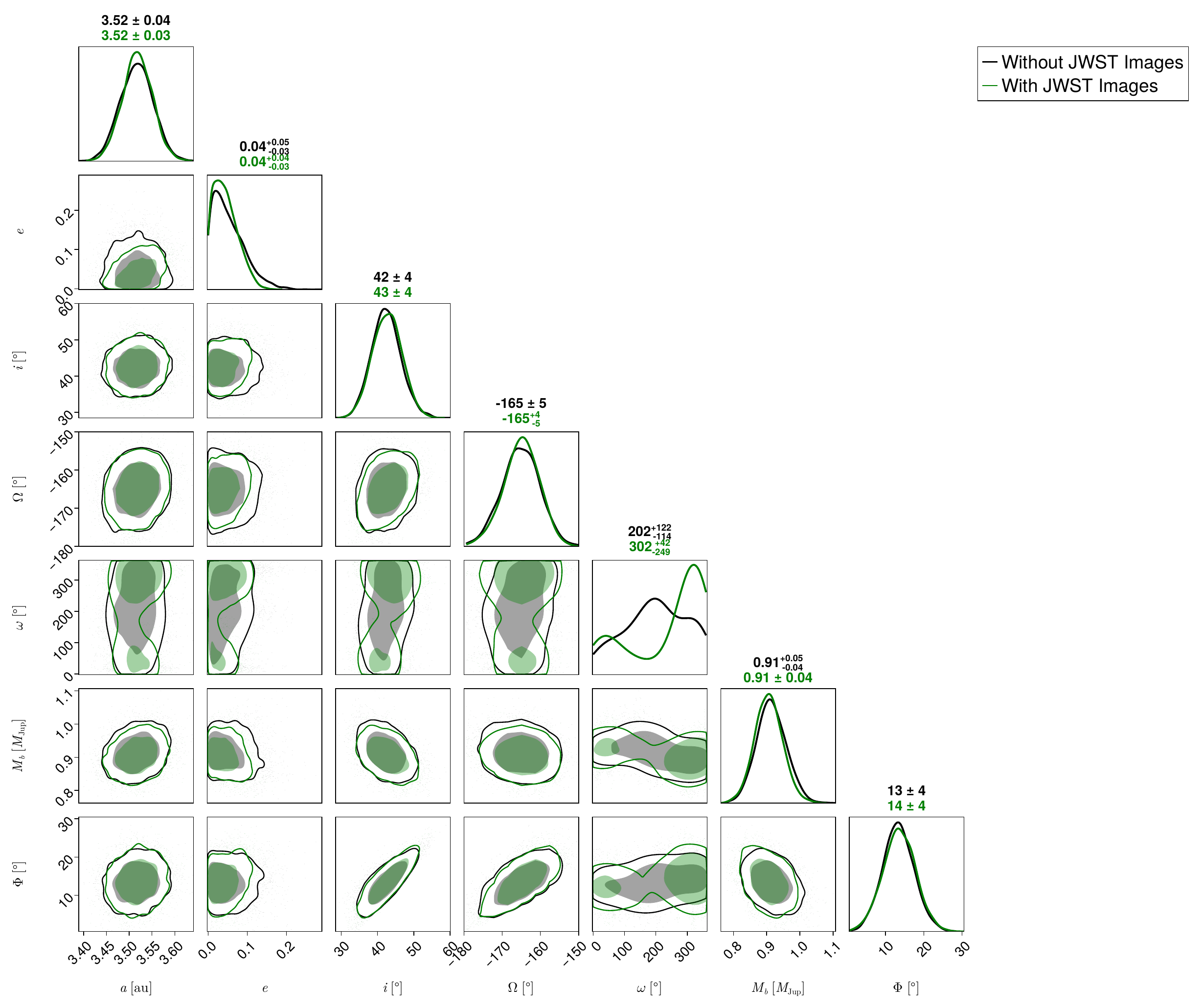}
   \end{tabular}
   \end{center}
   \caption{Corner plot for the orbital parameters and mass of \epseri~b with and without the JWST/NIRCam F444W images in the data. The fit does not include an atmopheric model; the flux is \textit{free-floating} in the two NIRCam observation epochs. Even though the walkers tend to \textit{cling} to the low-SNR feature present in the second epoch, the orbit is practically identical. The argument of the periastron changes but the orbit is largely circular so it does not change the model in any significant way.
   }
   \label{fig:cornerplot_compare_jwst_images}
\end{figure}

We present our planet model fits for \epseri\ b: (1) for the free-floating flux model, (2) the clear atmosphere assumption models, and (3) the cloudy atmosphere assumption models. Figure~\ref{fig:cornerplot_compare_jwst_images} shows the joint posterior on the orbital elements and the dynamical mass obtained with the free-floating flux method, with and without the JWST/NIRCam imaging included. The main changes with respect to T25 are two: the G23H treatment of the Gaia astrometry, and the disk-planet mutual inclination prior of Sec.~\ref{sec:mutual_inclination_prior}; together they account for the bulk of the shifts between our posteriors and theirs. The addition of the JWST data does not affect the orbit and mass significantly. The only apparent change is the argument of the periastron, $\omega$, which has little effect on the actual shape of the orbit given that it is largely circular.

The case without JWST data is the direct analog to T25's orbit and mass solution. With respect to their planet model, there are no major changes to the orbit; all orbital parameters fall within $1\sigma$ of their posteriors. The mass from the new model is likewise within $1\sigma$ of the T25 mass posteriors, shifting from $1.00 \pm 0.1\,M_{\rm Jup}$ to {$0.91 \pm$ 0.04$\,M_{\rm Jup}$}. These changes to the mass and orbital parameters can be attributed to the methodological differences described in Sec.~\ref{sec:planet_model_fit}, as well as to internal changes in \texttt{Octofitter}. To assess the validity of our \texttt{Octofitter} infrastructure, we performed MCMC runs replicating the {T25} setup. The results were almost identical; they differed slightly more only for the case in which all the data were used (black posteriors in their Fig.~2). In any case, all differences were well within $1\sigma$ and are likely attributable to small internal changes in \texttt{Octofitter}.

\begin{figure}\begin{center}
   \begin{tabular}{c} \includegraphics[height=7.0cm,trim={0cm 0cm 0cm 0cm}]{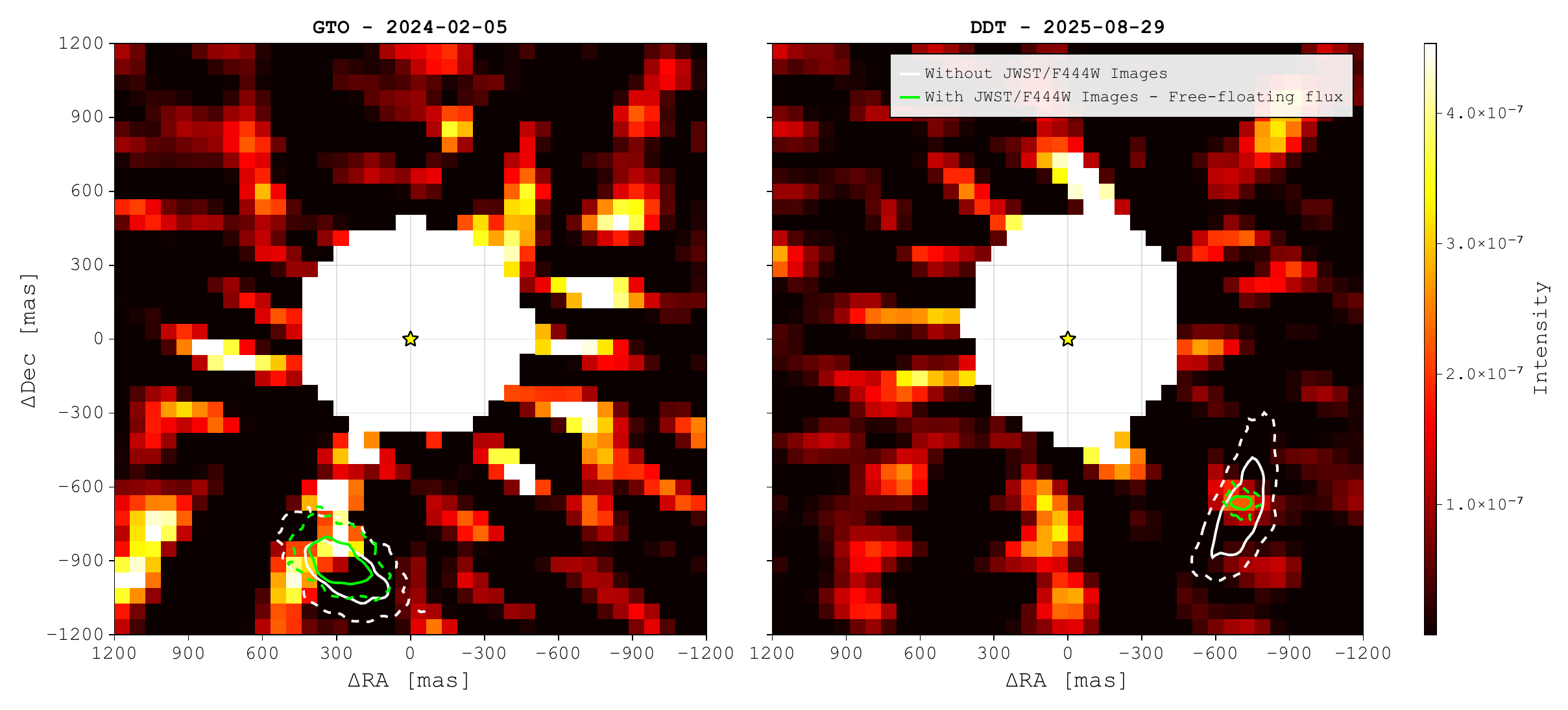}
   \end{tabular}
   \end{center}
   \caption{Post-processed NIRCam images for the two epochs, with 1- and 2-$\sigma$ contours of the expected position, computed with the posteriors of the \octofitter~results: in \textit{white} without including the images in the fit, and in \textit{green} including the images, but without making a connection between flux and atmospheric model. The postprocessing method is described in Sec.~\ref{sec:imaging_data}. When including the images, some walkers tend towards the feature whose location coincides with the expected position of planet~b. 
   \label{fig:gto_ddt_images}
   }  
\end{figure}

Figure~\ref{fig:gto_ddt_images} shows the post-processed flux maps, with the with- and without-JWST orbit posteriors overplotted as contours. The free-floating flux fit converges on a faint feature in the DDT epoch. The feature is of low significance and we claim no detection with meaningful statistical evidence; the flux posterior peaks at $2.6\pm0.7$  $\times 10^{-7}$, corresponding to a SNR of only $\sim 3$. Notably, the feature lands essentially on top of where the orbit, constrained by the RV and astrometry alone, already places the planet. As a result, including the imaging leaves the orbit almost unchanged from the no-JWST case, as seen in Fig.~\ref{fig:cornerplot_compare_jwst_images}; the data nudge the fit toward a location it was already going to favor. We leave the detailed discussion of this feature for Sec.~\ref{sec:lowsnr_feature}.

\begin{figure}\begin{center}
   \begin{tabular}{c} \includegraphics[height=13.0cm,trim={0cm 0cm 0cm 0cm}]{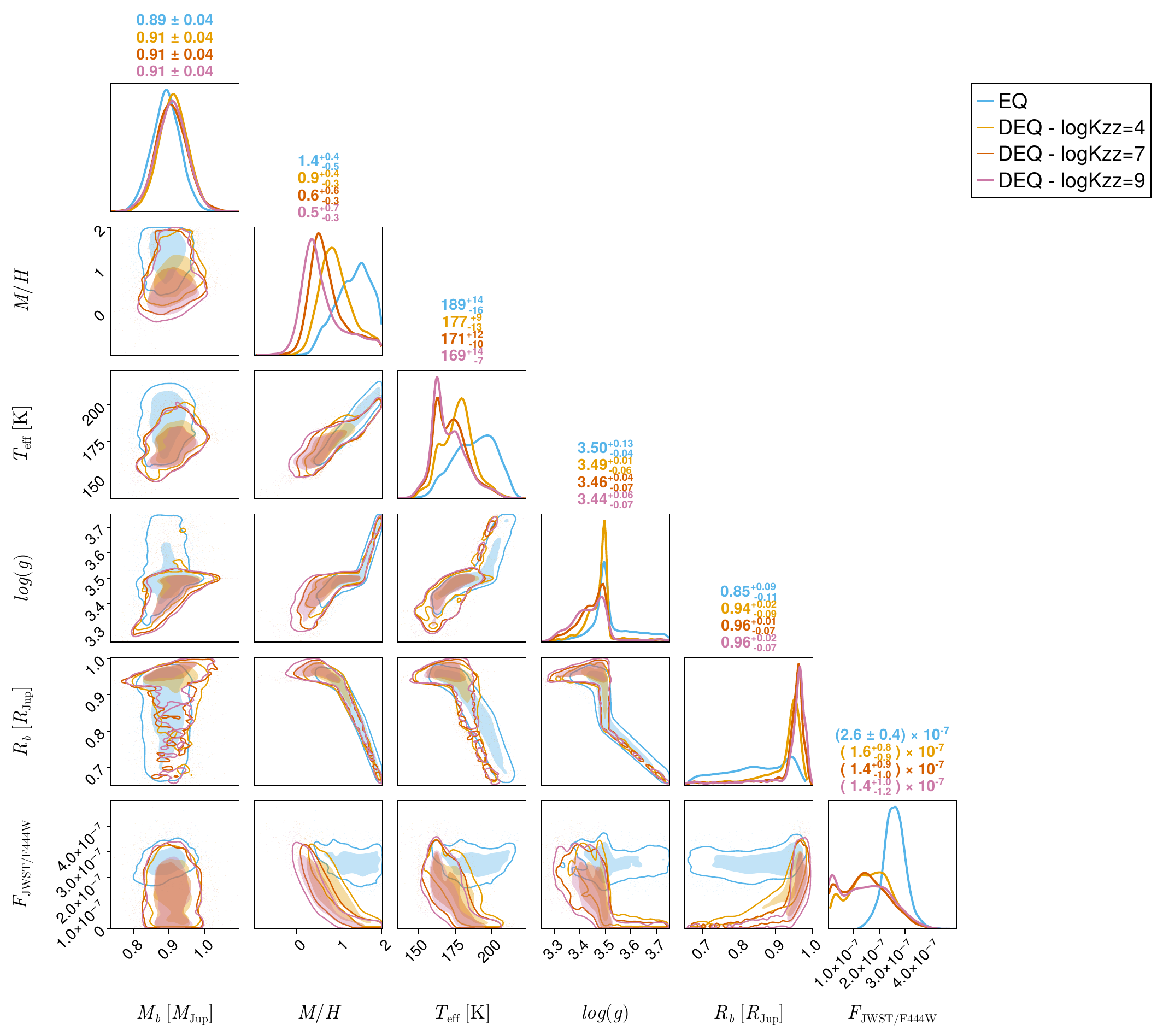}
   \end{tabular}
   \end{center}
   \caption{Corner plot for the atmospheric model fit with the clear atmosphere assumption using the Sonora Flame Skimmer model \citep[][in prep.]{Mang2026flame-skimmer}. Since the mass and age are well determined, the metallicity and effective temperature are constrained to a relatively narrow track, which forces some models to prefer non-zero F444W fluxes from the NIRCam data, in particular, the low-SNR feature shown in Fig.~\ref{fig:gto_ddt_images}. 
{The chemical equilibrium case has no levers to keep the 4~$\mu$m flux low, which makes the F444W flux of the feature more statistically plausible under this assumption; this is a consequence of the model assumption rather than evidence for the feature.}
   On the other hand, the chemical disequilibrium case has more ways to shape the 4 $\mu$m peak, which makes some samples prefer the image feature, and some others tend towards zero.
   \label{fig:atmo_corner_clear}
   }  
\end{figure}

The fit to the data with the clear atmosphere assumption yields the posteriors shown in Figure~\ref{fig:atmo_corner_clear}. We show the effective temperature $T_\mathrm{eff}$, the metallicity [M/H], the surface gravity $\log g$, and the radius, for both the equilibrium-chemistry (EQ) and disequilibrium-chemistry (DEQ) fits. We leave out the mass from this corner plot, since it is unaffected by the addition of the NIRCam images and it shows no correlations with the other evolutionary or atmospheric parameters. In other words, the mass is solely determined by the RV and astrometry data. The fit has a strong preference for the enhanced metallicity cases. This is in part due to preference of some samples to \textit{cling} to the low-SNR feature in the DDT epoch, and non-zero F444W flux solutions. We discuss this further in Sec.~\ref{sec:discussion}. The $\mathrm{C/O}$ is not shown in the corner plots because the posteriors are rather flat across the prior range of 0.5 to 2.5, and show no correlations with the other parameters. 

\begin{figure}\begin{center}
   \begin{tabular}{c} \includegraphics[height=13.0cm,trim={0cm 0cm 0cm 0cm}]{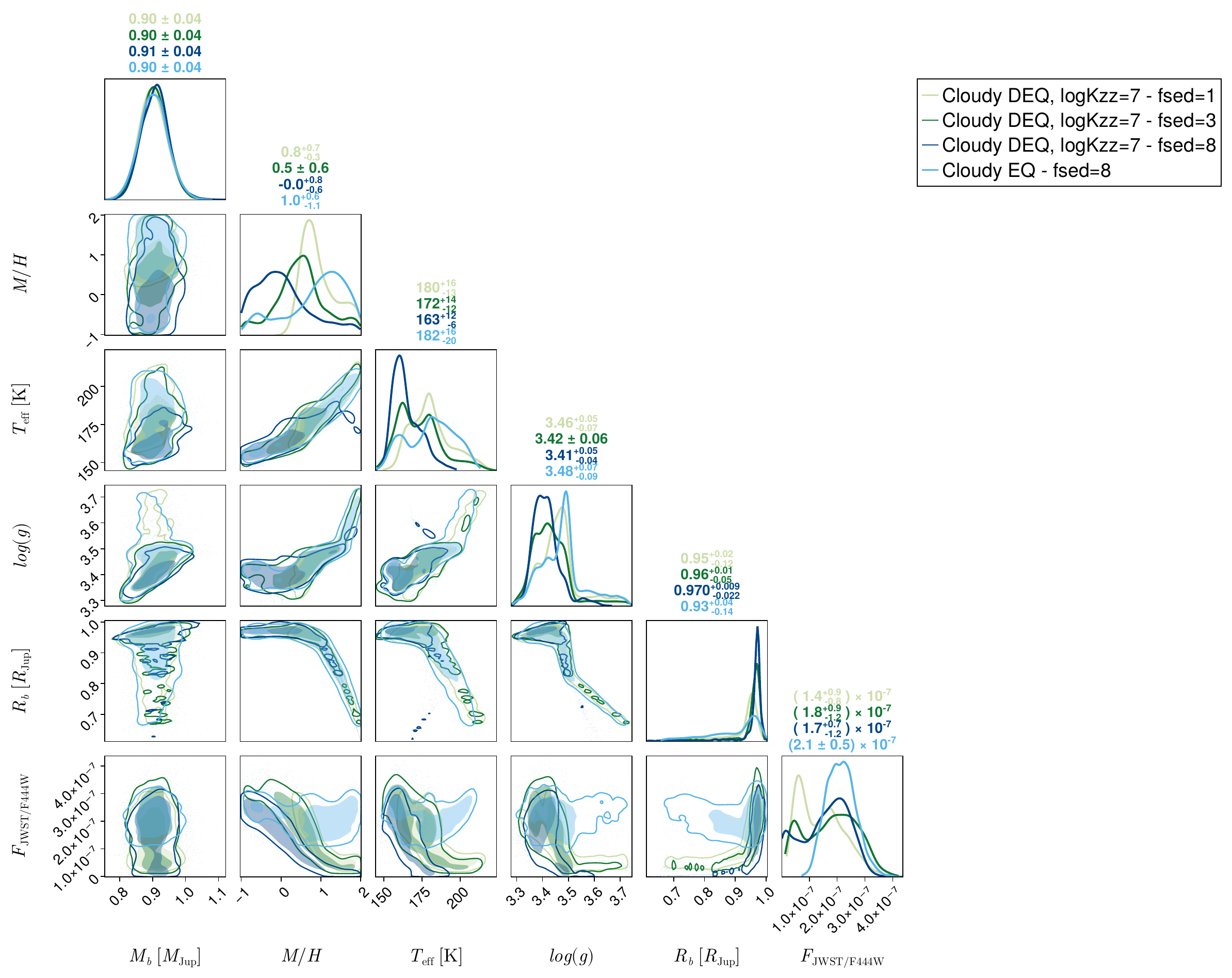}
   \end{tabular}
   \end{center}
   \caption{Same as Fig.~\ref{fig:atmo_corner_clear} for the cloudy atmosphere models. The atmospheric grids of fluxes for the evolutionary and atmospheric parameters are computed with PICASO, as explained in Sec.~\ref{sec:cloudy_picaso}. 
   \label{fig:atmo_corner_cloudy}
   }  
\end{figure}

Fig.~\ref{fig:atmo_corner_cloudy} shows the analogous corner plot for the cloudy atmosphere assumption. These fits also have many non-zero F444W solutions, but the resulting metallicity solutions are less extreme.

\subsection{4 $\mu$m flux constraints}\label{sec:res_flux}
The 4 $\mu$m flux constraint is, as expected, model-dependent. Moreover, non-uniformity of the intensity in the image, and in particular, the low-SNR feature consistent with the expected position in the DDT epoch, complicates the reporting of the flux constraints.
For instance, the flux posteriors from the free-floating flux method, shown in Figs.~\ref{fig:gto_ddt_images}, show a slight preference for the low-SNR feature present in the NIRCam images; the samples tend to gather around it rather than spreading toward zero. For this case, the posterior does not behave as an upper limit, but rather as a representation of the flux associated with that low-SNR feature given the data.

\begin{figure}\begin{center}
   \begin{tabular}{c} \includegraphics[height=13cm,trim={0cm 0cm 0cm 0cm}]{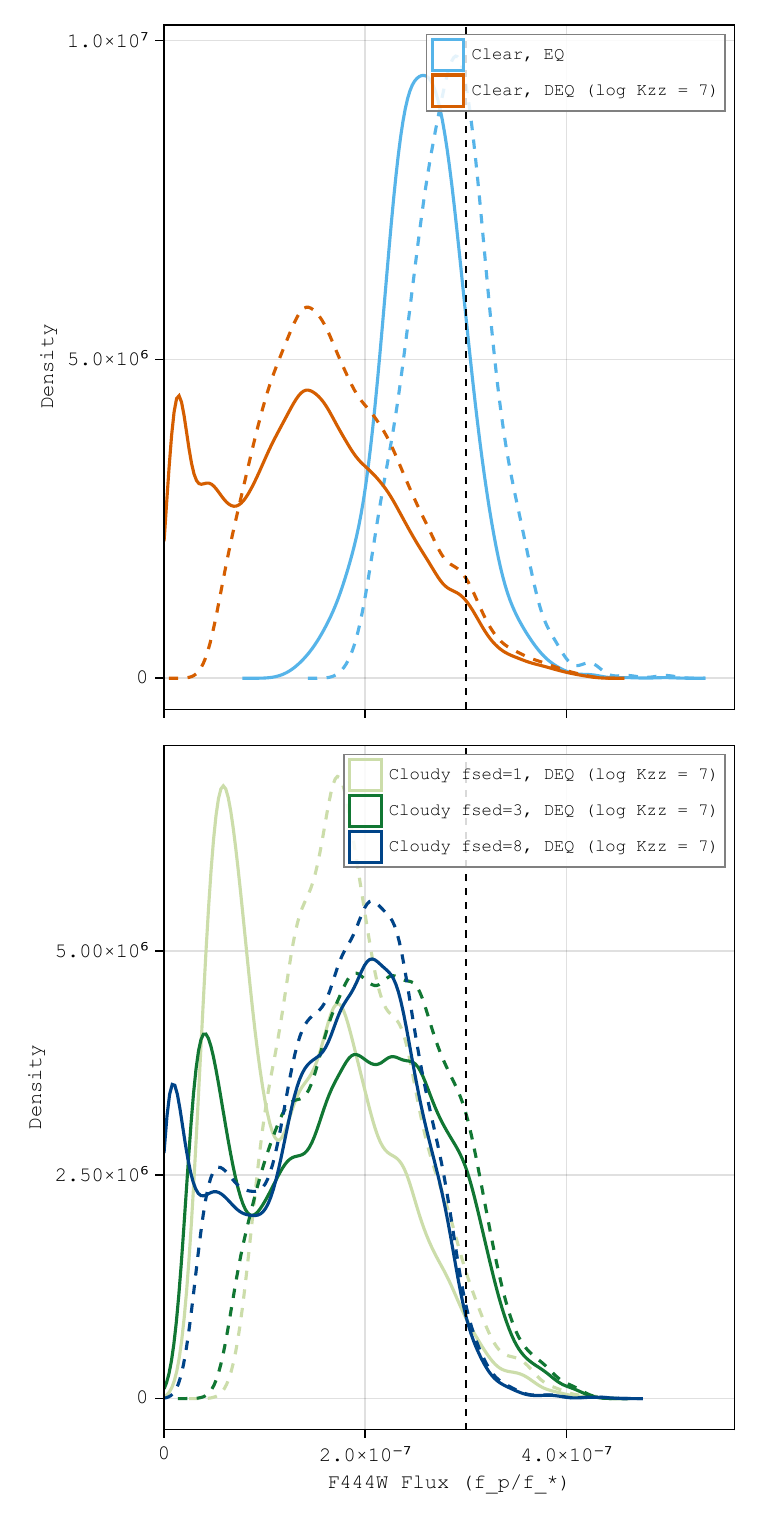}
   \end{tabular}
   \end{center}
   \caption{Posterior density distribution for the F444W flux from the \octofitter~results. The solid curves contain all the samples, the dashed lines contain metallicities under 1, M/H$<$1. This is to illustrate that the very enhanced metallicity solutions are the ones allowing the flux to go to zero. The physical interpretation is given in the text. 
   The dashed vertical line indicates \citet{Sanghi2026} 5-$\sigma$ sensitivity limit. 
   \label{fig:f444w_post}
   }  
\end{figure}

Similarly, in the chemical equilibrium fit the MCMC walkers drift toward the low-SNR feature, {the flux implied by the data under this model rather than a bound on it}. On the other hand, the disequilibrium chemistry models do not. These treat the imaging flux closer to an upper limit, though not a strict one. Indeed, a good fraction of the walkers still settle on a non-zero flux, and the posteriors come out bimodal (Figs.~\ref{fig:f444w_post}): some walkers collapse toward zero, or pile up near it, while others prefer a low flux consistent with the NIRCam images. 

We discuss the significance of the low-SNR feature in the Discussion section. For a conventional upper-limit value, the 5$\sigma$ upper limit reported by {S26} is still the appropriate one to cite, \sensitivitySanghi; the posteriors shown in Figs.~\ref{fig:atmo_corner_clear} and \ref{fig:atmo_corner_cloudy} being the more nuanced, less caveat-free story to the flux constraints.

The correlations between flux and other parameters in Figs.~\ref{fig:atmo_corner_clear} and \ref{fig:atmo_corner_cloudy} contain interesting correlations. 
In both clear and cloudy atmospheres, the disequilibrium cases reach the lowest F444W flux levels. The dominant driver is chemical: stronger vertical mixing quenches the CO/CH$_4$ ratio at deeper, hotter levels than the photosphere, so at the pressures probed by F444W the CH$_4$ abundance decreases and the CO abundance increases relative to equilibrium. The enhanced CO produces absorption between 4.5 and 4.8~$\mu$m that suppresses the flux across the F444W bandpass. In the
equilibrium cases the CO abundance stays lower, because CH$_4$ is favored at the colder, lower-pressure photosphere, and the F444W window remains comparatively open.

Some of the values for both surface gravity and metallicity in these low flux models are rather extreme, in Fig.~\ref{fig:f444w_post} we plot the posteriors with the same posteriors overplotted but discarding the samples with M/H$>$1 (dashed lines). This plots illustrate how, if we discard these more extreme cases, the flux posteriors tend to forego the low flux solutions. Interestingly, most pile around the 3-$\sigma$ sensitivity from S26, $\sim$1.8$\times$\tentos, meaning that observing this system more epochs with the same NIRCam configuration should discard many of the models in our posteriors, assuming a similar sensitivity is reached.

\subsection{Evolutionary and Atmospheric Model Constraints}
\label{sec:atmo_constraints}
As with the flux case, there is no single, one-size-fits-all value for the atmospheric and evolutionary parameters. In both clear and cloudy atmosphere cases, even for each independent chemical equilibrium assumption the non-uniformity of the intensity in the images complicates the story. Since many of the samples latch onto the non-zero flux data, the posteriors for all evolutionary and atmospheric model parameters reflect that pull. Both equilibrium and disequilibrium models are affected by the non-uniformity: the posteriors for the F444W flux exhibit a bi-modal distribution due to some walkers preferring the feature while others tend towards the zero-flux solution, and this is reflected in the evolutionary and atmospheric model parameters posteriors. 

A helpful way to analyze these results is to start by looking at the metallicity-effective temperature correlations. These follow a positive linear correlation between the allowed values because the age of the system and the mass are well constrained. The age is treated as a prior of 1.1$\pm$0.1 Gyr, taken from the revised analysis in S26, and the mass is well constrained by the RV and astrometry. Neither changes in any significant way with the addition of the images. For a well constrained age and mass, the evolutionary models impose a narrow, positive linear correlation between the allowed effective temperatures and metallicities. For instance, for high-metallicity, only high-\teff s are allowed, and similarly, for low-metallicities, only low-\teff s are available to the MCMC walkers. These \teff-M/H traces are present for all models. What changes is what combinations of \teff, M/H and gravity yield a F444W flux that fits well with the NIRCam images. For instance, in the clear atmosphere case, an atmosphere in chemical equilibrium favors a higher M/H and \teff\ combination, and correspondingly a brighter F444W flux, than an atmosphere in chemical disequilibrium. Vertical mixing under chemical disequilibrium enhances CO at the expense of CH$_4$, deepening absorption in the 4--5 $\mu$m CO band and suppressing the emergent flux; the disequilibrium models therefore reproduce the NIRCam photometry at cooler \teff\ and lower metallicity.

The same process happens in the cloudy atmospheres, where the 4 $\mu$m peak flux comes from the thermal emission that gets past through the cloud holes.
In this case, for chemical disequilibrium, the lower \fsed~prefers a higher metallicity and higher effective temperature. This is also due to the samples in the MCMC fit preferring non-zero F444W solutions, which are allowed at different metallicity/temperature levels for the different \fsed~values. Looking at the Fig.~\ref{fig:atmo_corner_cloudy} bottom left corner panel, the \fsed=1 case needs a higher the metallicity to match the $\sim$2$\times$\tentos~flux when compared to the \fsed=8 case. {Intuitively, more metals are needed in a high cloud deck case for the flux from the interior to make it through the cloud holes.}

\begin{figure}\begin{center}
   \begin{tabular}{c} \includegraphics[height=13cm,trim={0cm 0cm 0cm 0cm}]{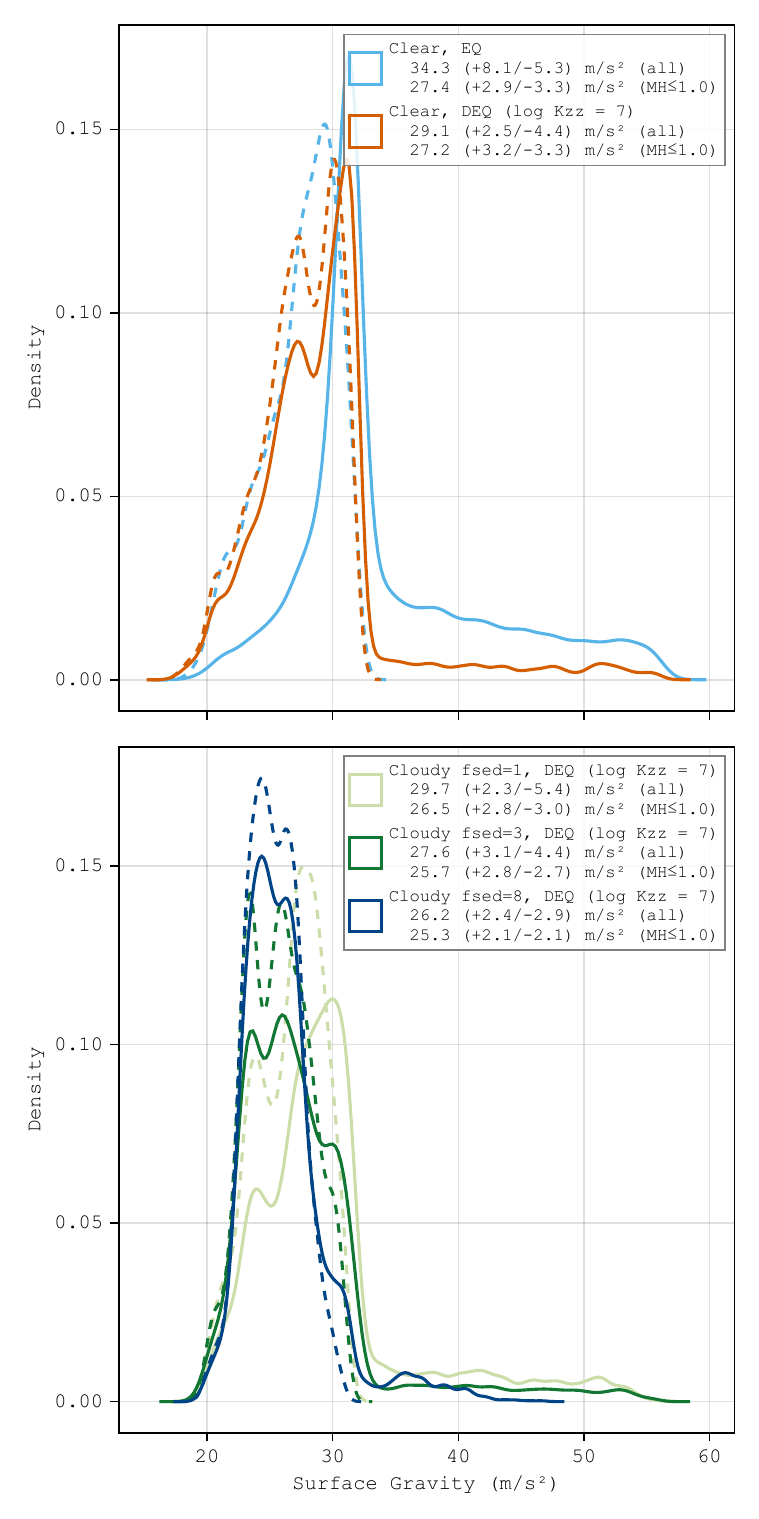}
   \end{tabular}
   \end{center}
   \caption{Same as Fig.~\ref{fig:f444w_post} but for the surface gravity. The posteriors in which the very high metallicity (M/H$>$1) are discarded (dashed), are the ones with very high surface gravity. The physical interpretation is discussed in the text.
   \label{fig:surf_grav}
   }  
\end{figure}

The story for the surface gravity posteriors is more complex. From the corner plots we see that for M/H spanning from -1 to +1, the gravities allowed are roughly \logg~= 3.3 to 3.5; then for higher metallicities, the \logg~shoots up linearly to 3.7. In order to see how this skews the surface gravity result, in Fig.~\ref{fig:surf_grav} we plot the posteriors for the surface gravity and overplot the same posteriors but cropping out the samples for which M/H$>$1 (dashed lines). These leave out the tails in the surface gravity posteriors with values beyond $\sim$32 m/s$^2$. {We stress that these posteriors are conditional on the fixed chemistry and cloud assumption of each run (Sec.~\ref{sec:imaging_likelihood}).}

\section{Discussion} \label{sec:discussion}

\subsection{The low-SNR feature in the second epoch}
\label{sec:lowsnr_feature}
The low-SNR feature in the second epoch coincidental in position with the expected astrometry of planet b (see Fig.~\ref{fig:gto_ddt_images}), deserves a closer look. 
The free-floating flux method, described in Sec.~\ref{sec:planet_model_fit}, lets the orbit fit assign a flux to each epoch without imposing an atmospheric model. Applied to the data including the JWST images, the corner plot places the low-SNR feature recovered in the image (Sec.~\ref{sec:results}) at the center of the positional posterior predicted for the planet. The feature is of low statistical significance, at an SNR of 2 to 3, yet the MCMC walkers consistently assign it as the planet; they do so because it coincides with where the posteriors predicted the planet is, and because it has some, while small, statistical significance. Another angle to make sense of why is this feature preferred in the MCMC fit is that the RV and astrometry data are also rather noisy, i.e. the planet is not detected with great confidence in any individual dataset (see Fig. 2 of T25). Therefore, the walkers are naturally drawn to another relatively low-SNR datapoint to cling to, in this case, the image feature, especially if it's very consistent with the rest of the data. 
Taking the flux posteriors, dividing the MAP flux by its error bar gives an SNR of $\sim$2.5. Hence the feature remains of low statistical significance even though the walkers tend toward it.

It is worth noting that a different data reduction method for these images will yield different results. A similar feature appears in S26 image, which was analyzed with an independent reduction. The SNR is likely different, which means that in the case of the free-floating flux \octofitter~run, the resulting flux posterior distribution would be different. A smaller SNR for the feature would yield a less tight posterior distribution for he flux. As for the flux, photometry for both reductions is comparable, which in the case of the atmospheric model fits means that the parameters retrieved would be also comparable. The SNR of the feature would drive the uncertainty in those parameters. 

\subsubsection{Why the equilibrium case clings to the feature and the disequilibrium case does not as much}
\label{sec:disc_eq_vs_deq}

The two model classes do not treat the feature in the same way: the equilibrium model clings to it, whereas the disequilibrium models do not. During the Octofitter MCMC fit, the walkers settle on the parameter combinations that are statistically more probable given the data, the model parameters, and the discrete options available in the spectral grid. The number of such combinations differs between the two model classes, and that difference is what decides whether the walkers are driven onto the feature.

In the equilibrium case, the parameter space offers more statistically favorable ways for the walkers to settle on the low-SNR feature. 
The low-metallicity, low-temperature end is excluded in the equilibrium case, because the evolutionary model would then require a surface gravity too low, i.e. too \textit{puffy}, making the planet too bright
at 4~$\mu$m. Only the high-metallicity, high-temperature combination remains, and it yields a surface gravity and a 4~$\mu$m flux that coincide with the JWST image. Because the equilibrium model produces a single spectral shape for a given temperature, metallicity, and surface gravity, it has no freedom to avoid the JWST flux and treat it as an upper limit for most of the sampled parameters.

The disequilibrium models admit a wider range of allowable combinations of surface gravity and 4~$\mu$m flux, so the walkers are not driven onto the low-SNR feature as much as for the equilibrium case. The effective temperature and metallicity still follow the same correlation as in
the equilibrium case, running from low-low to high-high. In the disequilibrium models, however, the parameters may also settle into an intermediate temperature and metallicity range, over which the allowed surface gravities are considerably more spread out. The strong-disequilibrium models give the spectrum additional freedom near 4~$\mu$m, admitting a range of peak shapes and CO/CO$_2$ combinations. These broader allowances, with the several possible shapes of the 4~$\mu$m peak, give the walkers enough flexibility to avoid the low-SNR feature. There are still parameter combinations that cling to the low-SNR feature, but many don't.

\subsubsection{Is the low-SNR feature the planet?}
Given the data and its uncertainties, and assuming the chemical equilibrium model as given by the Flame Skimmer grids, the feature in the DDT epoch is the planet with a decently high degree of confidence. Defining the SNR as the median of the flux posterior divided by its error bars yields a value of $\sim$6. This is not to claim that we have found the planet.  The choice of expanding the metallicity to enhanced levels up to M/H = 2 allowed the walkers to find more solutions consistent with the F444W flux in the images. Is an enhanced metallicity, say M/H$>$1.5, as likely as a more modest one? The MCMC setup gives equal weight to these cases, so if one can argue that M/H$>$1.5 is unlikely for this planet, then the walkers have less space to cover. Indeed, as S26 stated, and because the age and mass are relatively well constrained, the equilibrium case is only allowed in the enhanced metallicity hypothesis.

The flux posterior we report is also conditioned on our choice of parameters to marginalize over. For instance, we did not marginalize over $\log K_{zz}$, with zero being the equilibrium case. Had we done so, the posteriors would likely sample those combinations among the available solutions for a non-detection, and the feature would not be as significant. 
Indeed, we have no information on whether the planet is in equilibrium or in disequilibrium, so we cannot say which model carries more weight in terms of what is physically possible for \epseri~b. The same holds for the question of clouds: is the atmosphere cloudy or clear? At $\sim$1.1~Gyr we would expect some clouds to be present, but with what degree of confidence? With clouds, the 4$\mu$m peak is more suppressed, so as discussed in S26, the NIRCam fluxes are most likely upper limits, similarly to the disequilibrium case.
The only satisfying answer to this question would be to observe the system again with JWST/NIRCam. Two more epochs at a comparable SNR would confirm whether this feature is the planet. If the feature is not there, then we can finally rule out that the planet is in equilibrium with a clear atmosphere.

{Within the disequilibrium and cloudy cases, however, we have no empirical means to favor a given mixing strength or sedimentation efficiency. \epseri~b occupies a region of parameter space ($\sim$1~$M_{\rm J}$, $\sim$1~Gyr, $\sim$150--200~K) with essentially no characterized analogs; Jupiter itself is the closest one. We therefore present results for each assumption and regard the spread across them as a measure of the atmospheric-model uncertainty, rather than weighting one over the others.}

\subsection{Surface gravity posteriors and correlations}
\label{sec:disc_logg}

The surface gravity displays interesting correlations with the flux and other parameters that deserves further dicussion. Much as we did to explain the low-SNR feature in the equilibrium case, it is instructive to follow the trace of effective temperature against metallicity. Given a well-determined mass and age, metallicity and effective temperature fall along a positive, roughly linear correlation. The position along this trace is what gates the surface gravity: depending on where the samples land in the effective temperature and metallicity plane, the allowable values of \logg~become more or less restricted. In other words, the radius and log(g) structure is inherited from where the walkers are permitted to sit on the \teff–M/H relation.

Across all of the atmospheric modeling cases we consider, a surface gravity of approximately \logg~$\approx$ 3.5 is the preferred value. This preference is least pronounced for the high-mixing disequilibrium model with high log Kzz. In that case the shape of the 4-micron spectrum has considerably more freedom and can take on a wider variety of forms; consequently, for different surface gravities the resulting fluxes vary much more than they do in the other cases, and the fit no longer pulls as sharply toward a single \logg.
The equilibrium case behaves differently. Here the MCMC walkers prefer a higher metallicity, and as a result the upper tail of the posterior samples higher surface gravities. Even though \logg~$\approx$ 3.5 remains the peak, the distribution is noticeably broader toward the high end. This skew propagates directly into the radius: medium metallicities correspond predominantly to \logg~$\approx$ 3.5, whereas the high-metallicity samples open up the higher \logg~values. The net effect is that the equilibrium case admits a wider range of allowable radii, specifically those tied to the higher-metallicity, higher-\logg~scenarios.

As shown in Fig.~\ref{fig:surf_grav}, when discarding the very high metallicity cases, the surface gravity posteriors take more reasonable values. These are the planet models with relatively larger F444W fluxes, or less skewed to the zero flux (see Fig.~\ref{fig:f444w_post}). As explained in Sec.~\ref{sec:res_flux}, this is a complex combination of effects. The very high metallicity forces the \logg~to high values, which corresponds to smaller planets; this is has a negative effect on the overall thermal emission flux. The \teff~for these combinations of metallicities and \logg~are on the high side, which has a positive effect on the emission flux. The high \teff for the chemical disequilibrium cases, drives the abundances of CH$_4$ up which has a negative effect on the 4 $\mu$m emission. The net effect is overall lower emission at the 4 $\mu$m peak. 

\subsection{Mutual Inclination with the disk}
\begin{figure}\begin{center}
   \begin{tabular}{c} \includegraphics[height=8.5cm,trim={0cm 0cm 0cm 0cm}]{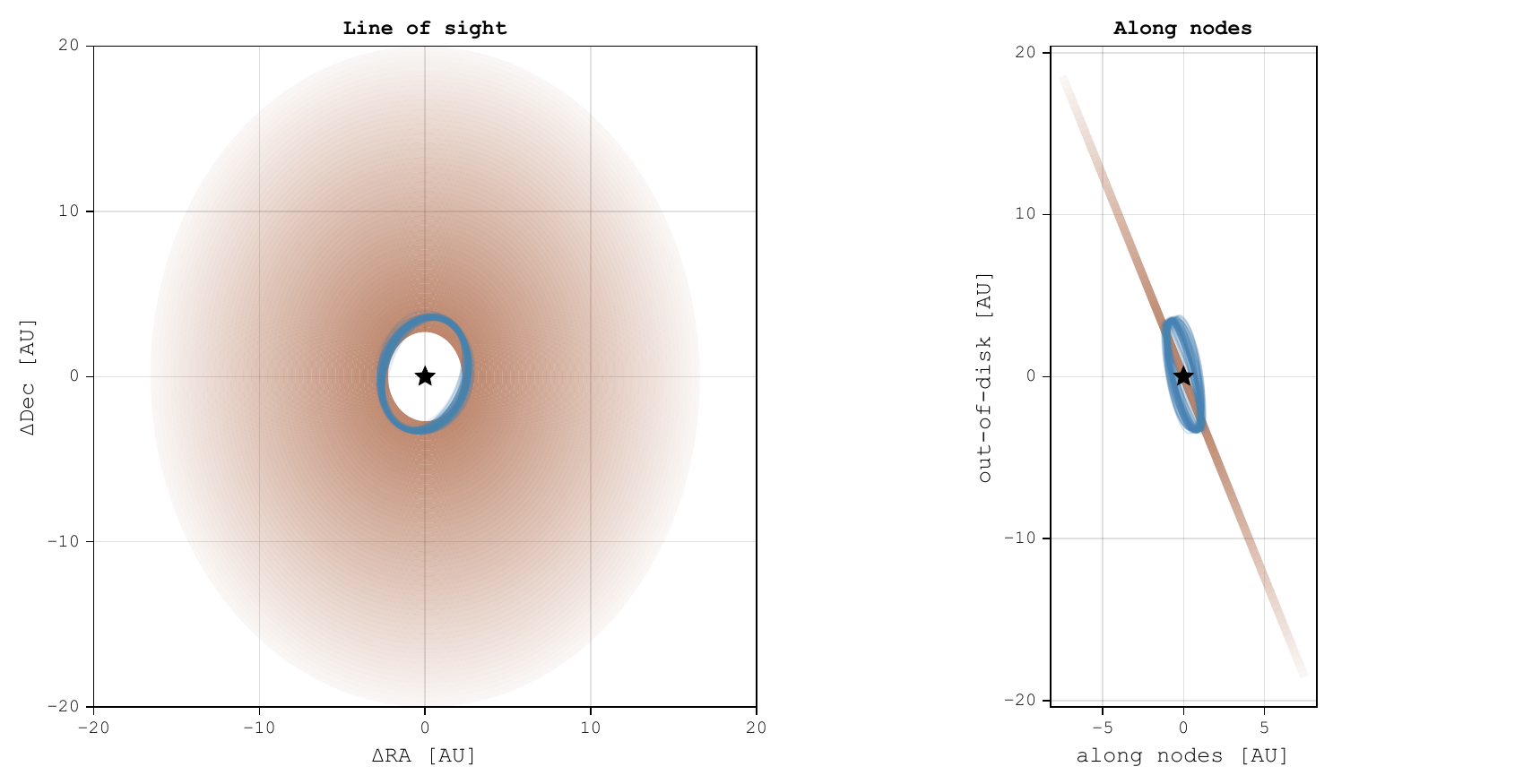}
   \end{tabular}
   \end{center}
   \caption{  Visualization of the MIRI disk \citep{Wolff2025} and planet orbit relative positions. The mutual inclination resulting from our fits is $\Phi$=14$\pm$4$^\circ$, the inclination of the disk is $\sim$34$^\circ$, and the planet's, 43$\pm$4$^\circ$. The inner edge of the displayed disk image is 2.7 au, where the MIRI PSF core saturates at 25.5 $\mu$m.
   \label{fig:mutual_inclination}
   }  
\end{figure}

The resulting mutual inclination between the planet and the disk from our model fits is $\Phi$=14$\pm$4$^\circ$; Fig.~\ref{fig:mutual_inclination} show a visual representation of the MIRI disk \citep{Wolff2025} and our 3D orbits. Although co-planarity is not fully discarded, our result mostly favors moderate mutual inclinations. How do we reconcile these with the smoothness of the disk revealed in the MIRI observations? The disk presents no visible structure in the inner region, even within the planet's orbit. Here are two pathways to reconcile the mutual inclination and the flat MIRI disk:
\begin{itemize}
    \item The structure imprinted by the presence of the planet could be buried in diffraction effects near the MIRI image core. The inner region for massive disks in MIRI images is notoriously hard to treat \cite{Gaspar2023,Millar-Blanchaer2025}. \item The planet is unable to imprint enough gravitational torque for the disk particles, and these flow mostly undisturbed, or disturbed in such a way that is unreachable for MIRI. \end{itemize}

To differentiate between these instrumental and dynamical scenarios, one must consider the competition between dust transport and gravitational perturbation timescales. The mid-infrared emission captured by MIRI is primarily sensitive to small, micron-sized grains. In the environment of \epseri, where the stellar wind is exceptionally potent, these small grains migrate inward rapidly via wind-driven and Poynting-Robertson drag. If the migration timescale ($t_{\text{drag}}$) through the planet's vicinity is significantly shorter than the secular nodal precession timescale ($t_{\text{sec}}$) forced by a $\sim 1 \, M_{\text{Jup}}$ planet, the dust particles will stream past $\epsilon$ Eri b before their orbital planes can be significantly torqued. This would validate the second scenario, yielding an apparently undisturbed disk morphology despite the planet's geometric tilt. 
Conversely, if $t_{\text{sec}} \lesssim t_{\text{drag}}$, the $\Phi \sim 14^\circ$ inclination must inevitably induce a localized warp or a vertical scale-height inflation as the grain orbits precess around the planet's misaligned angular momentum vector. If such physical distortions are real, the lack of visible structures in \citet{Wolff2025} strongly points toward our first scenario: these geometric features are simply buried within the aggressive point spread function (PSF) subtractions and complex diffraction artifacts. Distinguishing between a truly drag-dominated, unperturbed material flux and a physically warped inner disk will ultimately require dedicated $N$-body simulations that couple stellar wind drag with secular gravitational dynamics.

\subsection{Albedo computed from posteriors} \label{sec:albedo}
\begin{figure}\begin{center}
   \begin{tabular}{c} \includegraphics[height=6cm,trim={0cm 0cm 0cm 0cm}]{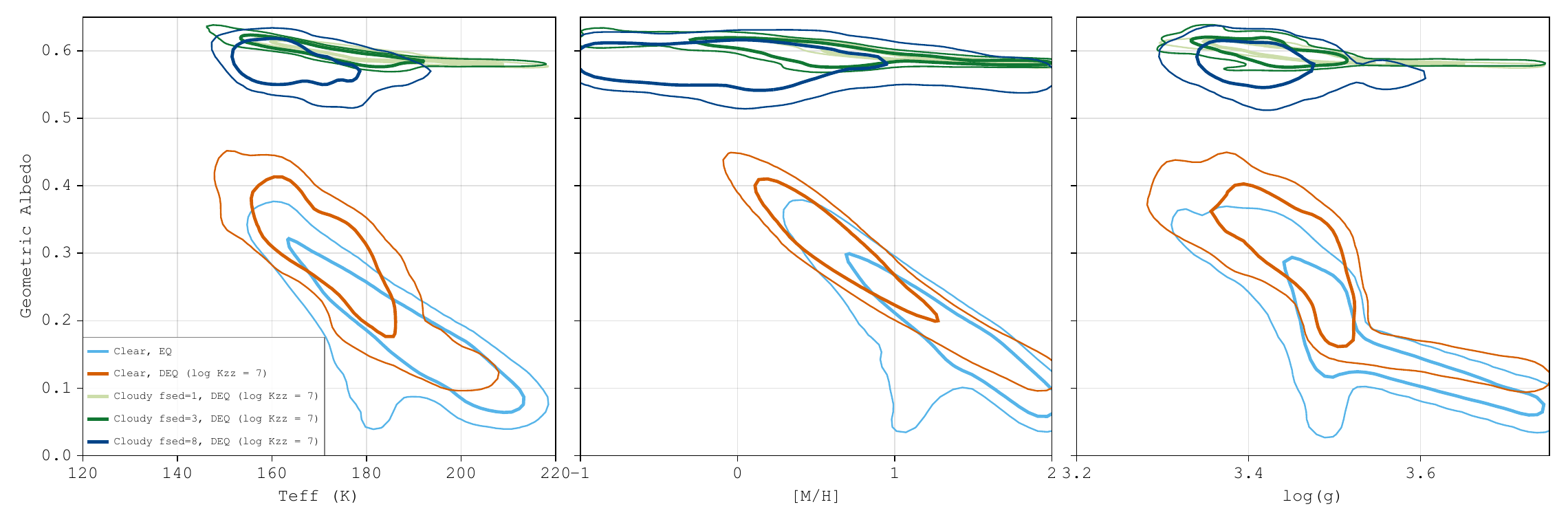}
   \end{tabular}
   \end{center}
   \caption{1- and 2-$\sigma$ contours for the geometric albedo computed with PICASO based on the posteriors from the \octofitter~runs presented in Sec.~\ref{sec:results}. The physical interpretation can be found in the text.
   \label{fig:albedo_contours}
   }  
\end{figure}

With the posteriors from the \octofitter~runs we can compute the geometric albedo of all planet models with PICASO. In Fig.~\ref{fig:albedo_contours} we show the 1- and 2-$\sigma$ contours for the clear and cloudy, and chemical equilibrium and disequilibrium run assumptions plotted against \teff, M/H, and \logg. These are computed for the Roman Coronagraph, Band 1 bandpass (10\% bandwidth at 575nm) over which the albedo is averaged. The clear-atmosphere geometric albedos decline monotonically with \teff, $\log g$, and metallicity, but these are not independent dependencies since each traces the methane column above the scattering photosphere. For a clear $\sim$150--200~K giant, $A_g$ is set by the competition between Rayleigh scattering, whose optical depth accumulates as $\tau_{\rm ray} \propto (P^2/g)\,\sigma_{\rm ray}$, and $CH_4$ absorption, which scales as $n(CH_4)\,(P/g)$. Higher metallicity adds methane per unit $H_2$ column while leaving the Rayleigh cross-section fixed; higher $\log g$ thins both columns but pushes the $\tau_{\rm scat} = 1$ level to higher pressure faster than it removes absorber; higher \teff\ expands the scale height and drives condensation levels deeper, lengthening the absorbing path. Radius shows no trend because it enters only through $g = GM/R^2$: geometric albedo is an intensive property of the atmosphere, and radius merely converts it into a flux ratio. In the cloudy models the dependence largely vanishes, since the scattering photosphere sits at the cloud top above most of the methane, and $A_g$ flattens near $0.55$--$0.62$. The disequilibrium models sit $\sim$0.05--0.1 above their equilibrium counterparts in the clear panels because vertical mixing at $\log K_{zz} = 7$ quenches $CH_4$ at a deeper, hotter level and transports that lower abundance upward; the offset narrows toward high \teff, where CO is already the favored carbon carrier. 

The near-degeneracy between the $f_{\rm sed} = 1$ and $f_{\rm sed} = 3$ models is expected. Once a condensate deck is optically thick and composed of conservative scatterers, the geometric albedo becomes insensitive to its optical depth: reflectance climbs with $\tau_{\rm cl}$ only until $\tau_{\rm cl} \sim$ a few, beyond which additional condensate redistributes photons within the deck without altering what escapes. At the \teff\ and $K_{zz}$ values sampled here, both models clear that threshold, and both converge on the same asymptote set by the particle single-scattering albedo and phase function rather than by $f_{\rm sed}$. A compensating effect narrows the difference further: the $f_{\rm sed} = 1$ deck is vertically extended and composed of small, more isotropically scattering particles, which reflect efficiently per unit optical depth but admit more $CH_4$ into the scattering layer. The $f_{\rm sed} = 3$ deck is compact and composed of larger, more forward-scattering grains, which reflect less efficiently per unit optical depth but place a sharper scattering surface above most of the absorbing column. The two effects act in opposition and largely cancel. The degeneracy breaks only when the deck ceases to be optically thick or conservative, which is what the $f_{\rm sed} = 8$ models show: their albedos lie $\sim$0.03 lower and are considerably more sensitive to \teff\ and $\log g$, because gas absorption below the deck begins to contribute to the emergent reflectance and the cloud-top pressure tracks the condensation level. The same mechanism produces the mild downturn in the $f_{\rm sed} = 1$ contours at high \teff, where the deck sinks below part of the methane column. Separating $f_{\rm sed} = 1$ from $f_{\rm sed} = 3$ therefore requires an observable other than broadband geometric albedo at $\alpha = 0$, which is close to the least informative choice available. Either wavelength dependence; the particle-size contrast alters the optical spectral slope, since the Mie scattering efficiency turns over at different $\lambda/r$ (or phase-curve coverage), where the forward-scattering asymmetry of the larger $f_{\rm sed} = 3$ grains yields a measurably different brightness at small phase angles, would be needed to constrain the sedimentation efficiency.

\subsection{Gaia Data Release 4}
Gaia's Data Release 4 (DR4) will publish the intermediate astrometric data (IAD) for \epseri, which will further constrain the orbit and the dynamical mass of the planet. Although \epseri~is bright enough to saturate the Gaia detectors, we still expect usable IAD for this system, albeit at degraded astrometric precision relative to the fainter, unsaturated regime. A tighter mass narrows the region of the evolutionary and atmospheric grids that the sampler needs to search. A tighter orbit sharpens the predicted position of the planet at our observing epochs; consequently, the walkers draw from fewer pixels in the NIRCam images. \octofitter~has the ability to perform planet model fits with DR4 data, we thus perform a fit with mock \epseri~DR4 IAD data to assess how the orbit, mass and atmospheric parameters would change when DR4 is released. 

\begin{figure}\begin{center}
   \begin{tabular}{c} \includegraphics[height=7cm,trim={0cm 0cm 0cm 0cm}]{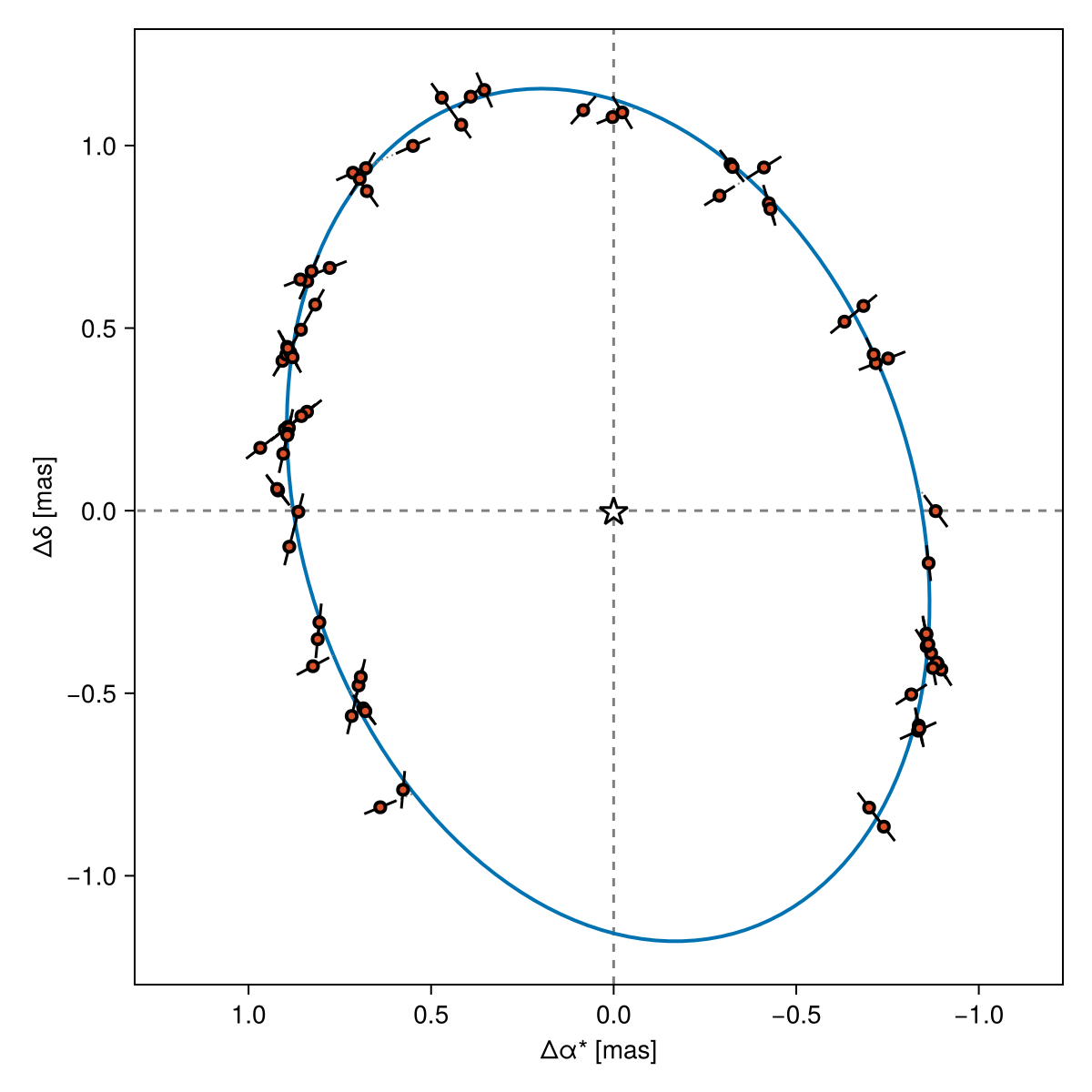}
   \end{tabular}
   \end{center}
   \caption{\epseri~b orbit selected from our \octofitter~posteriors with mock Gaia DR4 IAD data overplotted. The ``along scan'' uncertainty selected is 50 $\mu$as, which is what is expected for a birght star \citep{Lindegren2018}.
   \label{fig:dr4_orbit}
   }  
\end{figure}

We produce a mock DR4 data following the \octofitter~documentation tutorial, and the available pre-released data from the Gaia BH3 black hole \citep{GaiaPrerelease} as reference. We select, the orbit from our posteriors with the highest log probability and simulate Gaia scans with an ``along scan'' uncertainty. We simulate for 50 and 100 $\mu$as, which are the expected errors for a bright star such as \epseri~\citep{Lindegren2018}.
In Fig.~\ref{fig:dr4_orbit} the orbit of planet b with the Gaia scans and data points from our generated mock data.

\begin{figure}\begin{center}
   \begin{tabular}{c} \includegraphics[height=7cm,trim={0cm 0cm 0cm 0cm}]{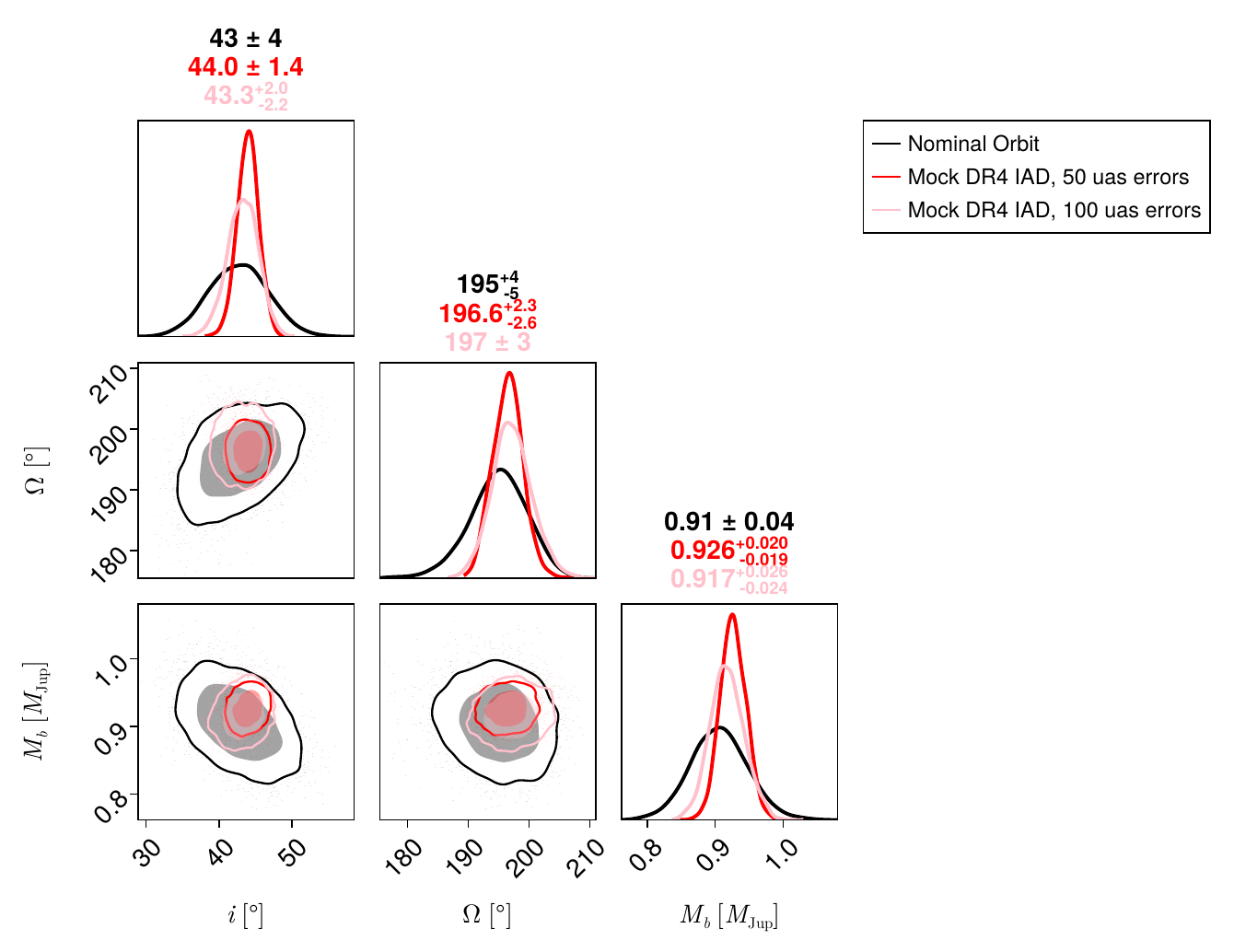}
   \end{tabular}
   \end{center}
   \caption{Corner plot comparing \octofitter~runs using the mock astrometry Gaia DR4 data with our nominal results. The resulting orbit and mass are better determined thanks to the precision of the DR4 measurements; the error bars are improved by a factor of $\sim$2. The nominal run is the \textit{free-floating flux} fit presented in Sec.~\ref{sec:results}.
   \label{fig:dr4_corner_orbit}
   }  
\end{figure}

We use the new \octofitter~likelihood object \texttt{GaiaDR4AstromObs} to fit a new planet b model to this mock data. We use the RV and JWST images data in the joint fit, but drop all other astrometry data: the G23H object (Hipparcos, Gaia DR2 and DR3), and HST/FGS. To assess the constraining power over the atmopheric parameters, we fit an atmospheric model for a representative case of a cloudy atmosphere: \fsed=3, in chemical disequilibrium, $\log$K$_{zz}$=7.
The resulting orbit is better constrained, as well as the mass; in Fig.~\ref{fig:dr4_corner_orbit} we show the corner plot comparing some relevant orbital parameters and the mass with our nominal orbit from the \textit{free-floating flux} fit. The error bars are reduced by a factor of $\sim$2. 

\begin{figure}\begin{center}
   \begin{tabular}{c} \includegraphics[height=7cm,trim={0cm 0cm 0cm 0cm}]{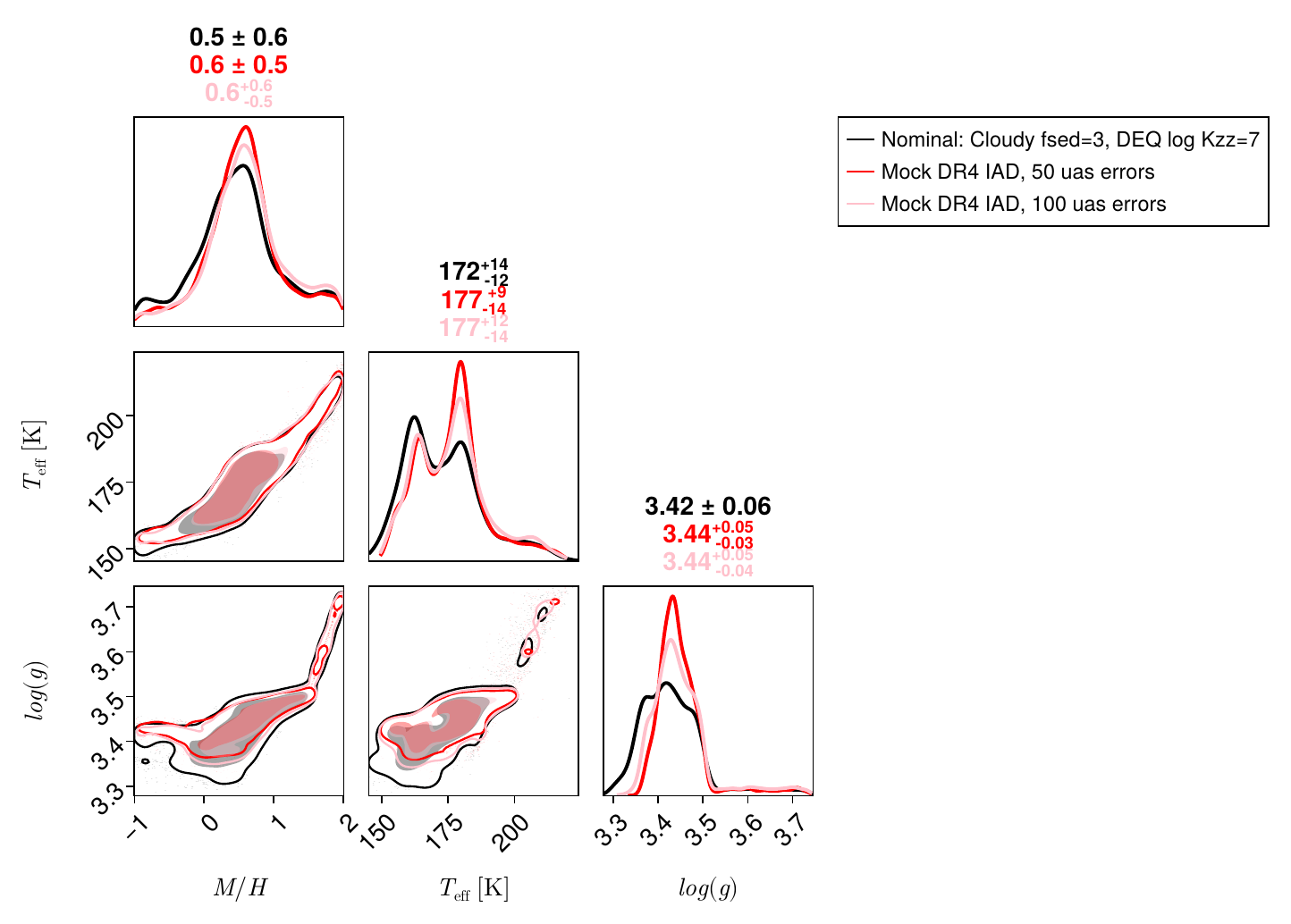}
   \end{tabular}
   \end{center}
   \caption{Same as Fig.~\ref{fig:dr4_corner_orbit} but for the {atmospheric} parameters. Here we compare to a representative run from Sec.~\ref{sec:results}: a cloudy model wit \fsed=3 in chemical disequilibrium ($\log$K$_{zz}$=7). Although the constraints on the orbit and mass are significantly improved, this does not translate to comparable improvements in the atmospheric models. We discuss this in the text.
   \label{fig:dr4_corner_atm}
   }  
\end{figure}

The atmospheric parameters are not as well constrained, as shown in Fig.~\ref{fig:dr4_corner_atm}. This is because these parameters remain dependent on the NIRCam flux. Although the better-determined orbit helps the MCMC walkers constrain the expected position to a smaller area on the NIRCam detector, the flux still varies across the image since the FWHM in the F444W band is $\sim$130 mas. More importantly, the uncertainty in the flux is relatively high.
 
\section{Upcoming Observations with the Roman Coronagraph}
\label{sec:roman}
The Nancy Grace Roman Space Telescope {was recently launched by} NASA. While most of its time will be devoted to wide-field near-infrared surveys, Roman will also carry the Coronagraph Instrument, which will demonstrate space-based, visible-band coronagraphy with active wavefront sensing and control for the first time \citep{Bailey2023}. The instrument relies on a pair of deformable mirrors operating in concert, a series of high-precision coronagraphic masks, and photon-counting EMCCD detectors; together these are expected to deliver small-separation contrast ratios better than \tentos, and potentially down to \tenton~levels, in reflected light. This regime is precisely the one that makes \epseri~b a compelling target: as we will show, the planet-to-star flux ratio and angular separation predicted by our model fall within reach of the Coronagraph's expected sensitivity, making it one of the most favorable known giant planets for a reflected-light detection.

{\epseri~is among the systems that the  Roman Coronagraph Community Participation Program (CPP) is considering as observing targets~\citep{Wolff2026}. }
The Roman CPP solicited white papers from the worldwide science community to define potential technology demonstration and scientific observations during the technology demonstration phase; of the submissions received, a handful proposed \epseri~b as an observing target.\footnote{The submitted white papers are publicly available \citep{10.5281/zenodo.21203363}.} These observations were also discussed by S26, who also laid out the case for \epseri{}~b as a Roman Coronagraph target; in this work we build on that foundation, leveraging our orbit fits (Sec.~\ref{sec:results}) to deliver more precise predictions of the Roman Coronagraph's sensitivity to this planet. As we detail in the following sections, our predictions point to three conclusions: (1) the target should be observed soon, both because of its current orbital phase and because it may move behind the outer working angle (OWA) within the operational window, a point already raised in one of the white papers; (2) the deeper the achievable contrast, the more our model can be constrained, so that even a non-detection becomes informative about the planet's properties; and (3) multi-epoch observations would not only improve the flux sensitivity through combination, but could also recover the planet in cases where a single epoch yields only a low-significance signal.

\subsection{Roman coronagraph photometry and astrometry predictions from our results}
\begin{figure}\begin{center}
   \begin{tabular}{c} \includegraphics[height=8.5cm,trim={0cm 0cm 0cm 0cm}]{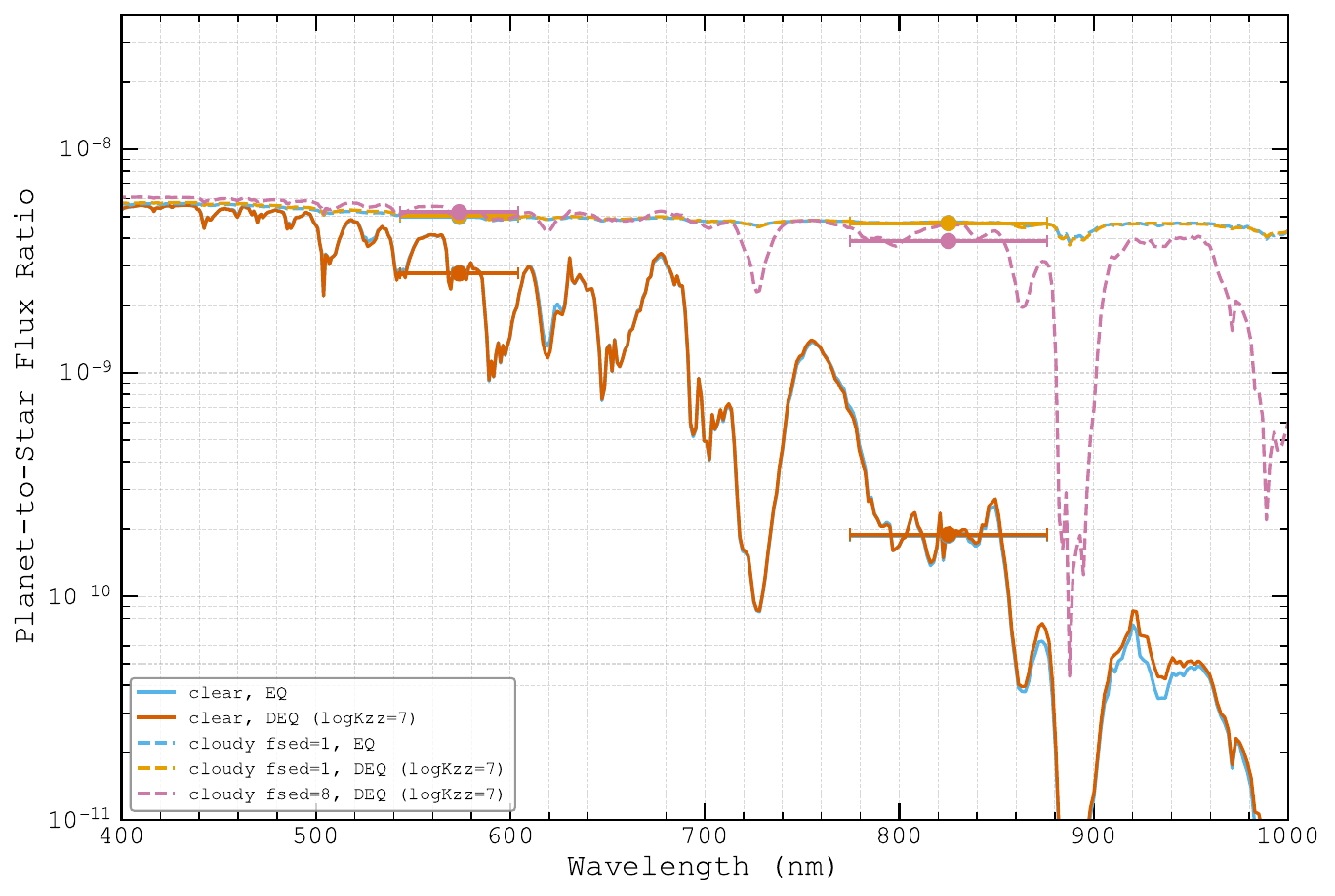}
   \end{tabular}
   \end{center}
   \caption{Planet-to-star flux ratio for some representative models of \epseri~b. Shaded area represent the Roman Coronagraph's Band 1 and 4 bandpasses. This illustrates the constraining power of the Roman Coronagraph observations of this system: just a Band 1 detection has some constraining power over different parameters. Including Band 4 observations would help constrain the cloud abundance and properties.
   \label{fig:picaso_reflect}
   }  
\end{figure}

\begin{figure}\begin{center}
   \begin{tabular}{c} \includegraphics[height=7.5cm,trim={0cm 0cm 0cm 0cm}]{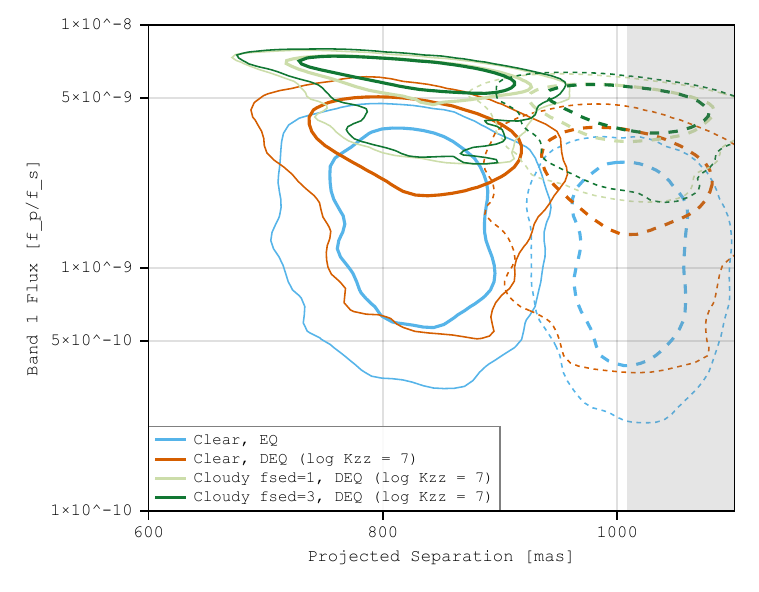}
   \end{tabular}
   \end{center}
   \caption{Flux contours as a function of separation computed with the posteriors from the \octofitter~runs. Solid line contours are for a January 1st, 2027 observation, and the dashed contours are for October 1st, 2027. This illustrates the constraining power of Roman Coronagraph observations; a final contrast of $\sim$2$\times$\tenton~would reach all the cloudy planets according to our \octofitter~results. The cloudy models with different sedimentation efficiencies have similar expected  values due to the posteriors: the \fsed=3 posteriors have more samples covering \textit{puffier} solutions, which compensates the lower reflectivity with respect to \fsed=1 models. The grey region indicates the area outside the field of view of the coronagraph mode considered (wide field of view, WFOV, Band 1).
   \label{fig:RCflux_vs_projsep}
   }  
\end{figure}

We present the reflected light flux prediction for the Roman Coronagraph Band 1 based on our results. We use PICASO to compute the grid of reflected light fluxes for a set of evolutionary and atmospheric model parameters in the vicinity of planet b's expected parameters. The process is analogous to the one explained in Sec.~\ref{sec:cloudy_picaso}; Fig.~\ref{fig:picaso_reflect} shows some example reflected light spectra computed with PICASO, with Band 1 and Band 4 fluxes indicated for each model. Using the posteriors presented earlier (see Sec.~\ref{sec:results}) we compute the expected fluxes for all the models. These are shown in Fig.~\ref{fig:RCflux_vs_projsep} as a function of projected separation. We show the expected fluxes for two epochs: early after launch, in January 2027, and in October 2027. The former yields a very favorable projected separation where the planet should be within the FOV of teh instrument within 2-sigma. In October 2027, the planet is expected to be close to the outer working angle of the wide field of view (WFOV) mode, in which Band 1 and Band 4 have the most potential to constrain cloud abundances and properties.

The degeneracy in albedos between \fsed=1 and 3 was explained in Sec.~\ref{sec:albedo}. Since Fig.~\ref{fig:RCflux_vs_projsep} shows the reflected light flux, the radius of the planets considered also comes into play. For both \fsed=1 and 3, the flux is such that the walkers are choosing to sample the combinations of parameters that allow them to \textit{cling} to the low-SNR feature, or simply non-zero flux solutions. In particular, they're choosing higher gravity values for \fsed=1 with respect to \fsed=3. This translates in puffier planets for the latter, with higher planet surface to reflect light from, thus resulting in higher reflected flux in the visible. 

\begin{figure}\begin{center}
   \begin{tabular}{c} \includegraphics[height=7.5cm,trim={0cm 0cm 0cm 0cm}]{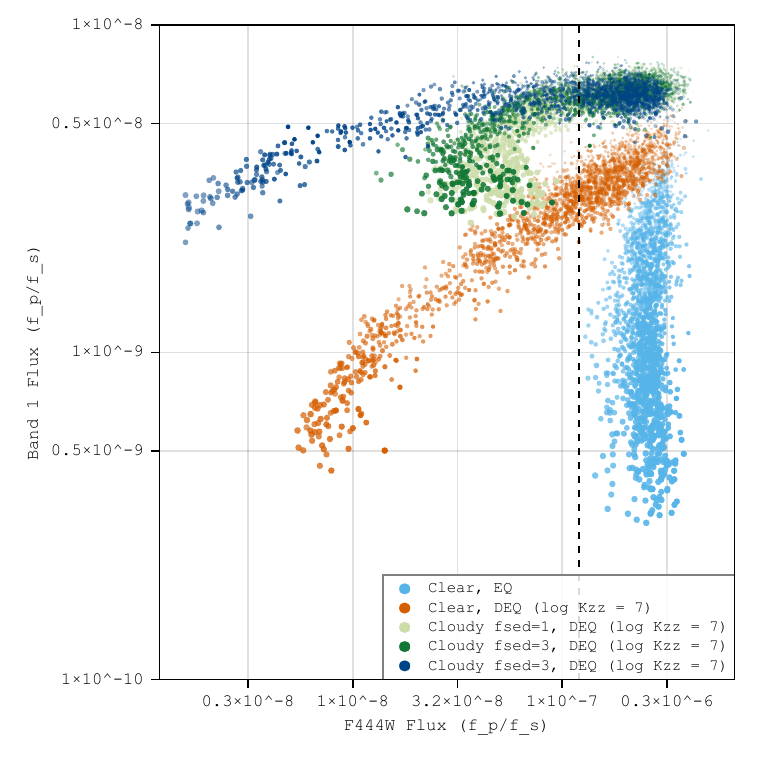}
   \end{tabular}
   \end{center}
   \caption{Roman Coronagraph Band 1 flux versus JWST/NIRCam F444W flux for the samples in the posteriors from our \octofitter~results. This illustrates the combined power of Roman Coronagraph and NIRCam coronagraph observations. A Roman Coronagraph Band 1 non-detection at a sensitivity limit of $\sim$2$\times$\tenton~would discard most planet assumptions except clear atmospheres in chemical disequilibrium. The dashed vertical line indicates the NIRCam 3-$\sigma$ upper limit from \citet{Sanghi2026}; any point to its right can be discarded or confirmed with additional NIRCam followup observations.
   \label{fig:cloud_points_rc_f444wlimit}
   }  
\end{figure}

The clear atmosphere case is not as favorable in the visible due to the lack os scattering particles in the top of the atmosphere. Many samples from the posteriors yield flux levels below \tenton. However, many of these parameter combinations that yield low reflected light flux can be discarded with the already proven sensitivity of JWST/NIRCam corongraph. In Fig.~\ref{fig:cloud_points_rc_f444wlimit} we show half of the samples from the last \texttt{Pigeons Octofitter} fit round in a Roman Corongraph Band 1 versus F444W flux distribution. If the Roman Corongraph observations are able to reach a sensitivity limit of, say, {2}$\times$\tenton, and no detection is confirmed, that would discard most of the planet models, except the clear atmosphere in chemical disequilibrium. A contrast sensitivity, after postprocessing, of \tenton~is not unrealistic; a raw contrast of 5$\times$\tenton~combined with a postprocessing gain factor of 5$\times$ would suffice. With better sensitivity, the planet has little combinations of evolutionary and atmospheric parameters possible, given our model assumptions. For reference, the instrument results from the TVAC tests yielded a raw contrast (coherent) of 1.42$\times$\tentoe \citep{Cady2025}; making the 2$\times$\tenton~final sensitivity rather reasonable. The plot shows a vertical line indicating the 3-$\sigma$ limit of S26; this is to indicate at which F444W contrast the planet would technically detectable with NIRCam with reasonable number of observations. In other words, any point at the right of that line can be discarded or confirmed with new NIRCam observations. 

\begin{figure}\begin{center}
   \begin{tabular}{c} \includegraphics[height=7.5cm,trim={0cm 0cm 0cm 0cm}]{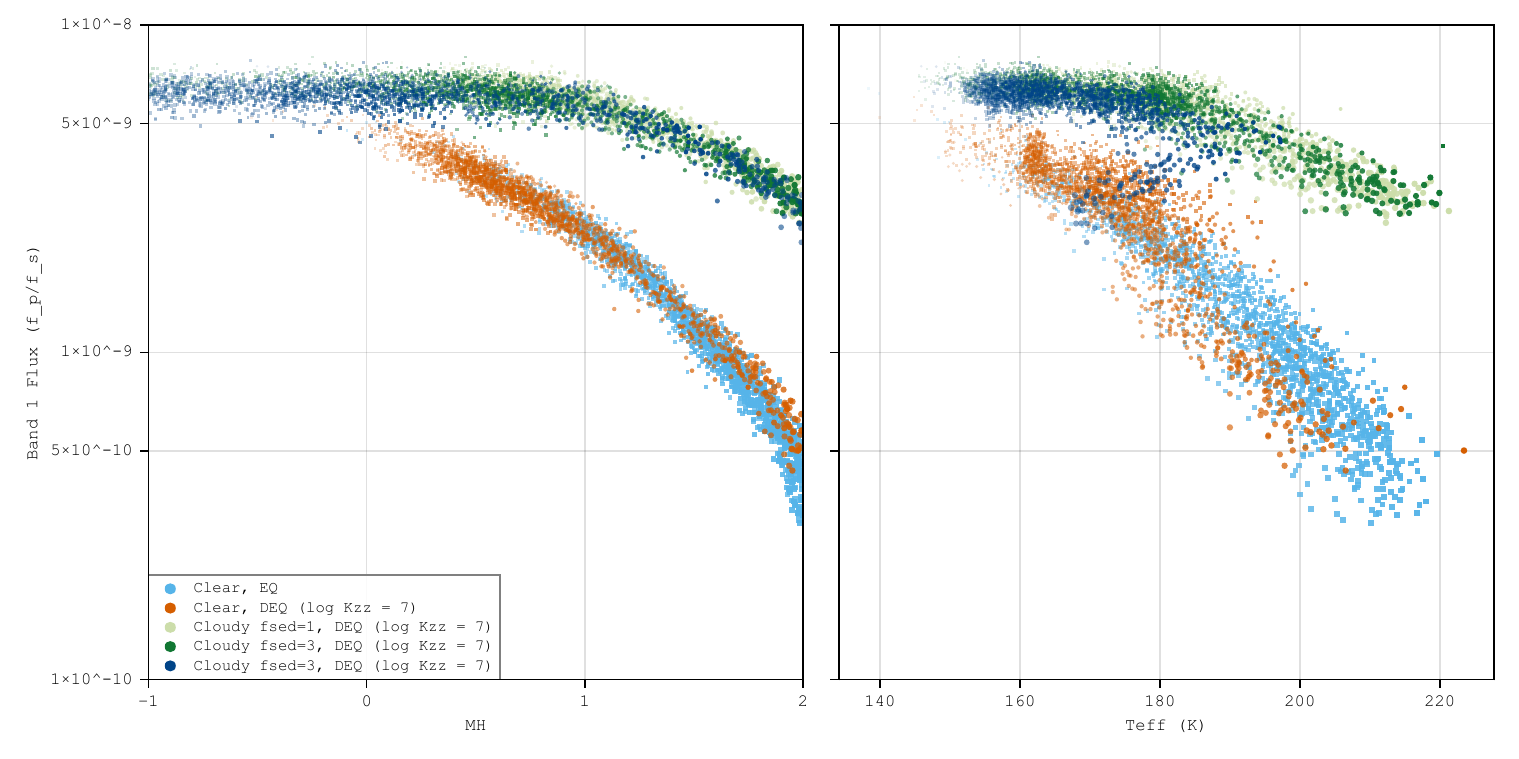}
   \end{tabular}
   \end{center}
   \caption{Roman Coronagraph Band 1 fluxes computed from the \octofitter~result posteriors versus metallicity (\textit{left}) and effective temperature (\textit{right})). This illustrates the constraining power of the Roman Coronagraph observation: a non-detection with a final sensitivity $\sim$2$\times$\tenton~would force the planet models to fall into very high metallicity solutions, and rather high effective temperatures. 
   \label{fig:cloud_points_rc}
   }  
\end{figure}

Fig.~\ref{fig:cloud_points_rc} shows a similar plot with M/H and \teff, both of which have clear correlations with the reflected flux. Our results show how a contrast sensitivity, after post-processing, of 5$\times$\tenton~would already discard most of the possible cloudy M/Hs and \teff s. {In this case, only the very high metallicity and high \teff~solutions would remain given our JWST data.}
A contrast sensitivity of 2$\times$\tenton, given the contrast limits in the F444W flux, would discard all cloudy planets under our model assumptions. {At worse sensitivities, $>$7$\times$\tenton, the Roman Coronagraph data would not be very constraining, since the posteriors for M/Hs of -1 to 1, and \teff of 150 to 185, all pile up under this contrast level. A contrast sensitivity of $>$1$\times$\tentoe, would provide no meaningful constraints. However, given the TVAC raw contrast results of 1.42$\times$\tentoe, this scenario would mean that the performance on sky would have degraded with respect to TVAC}.

\begin{figure}\begin{center}
   \begin{tabular}{c} \includegraphics[height=7.5cm,trim={0cm 0cm 0cm 0cm}]{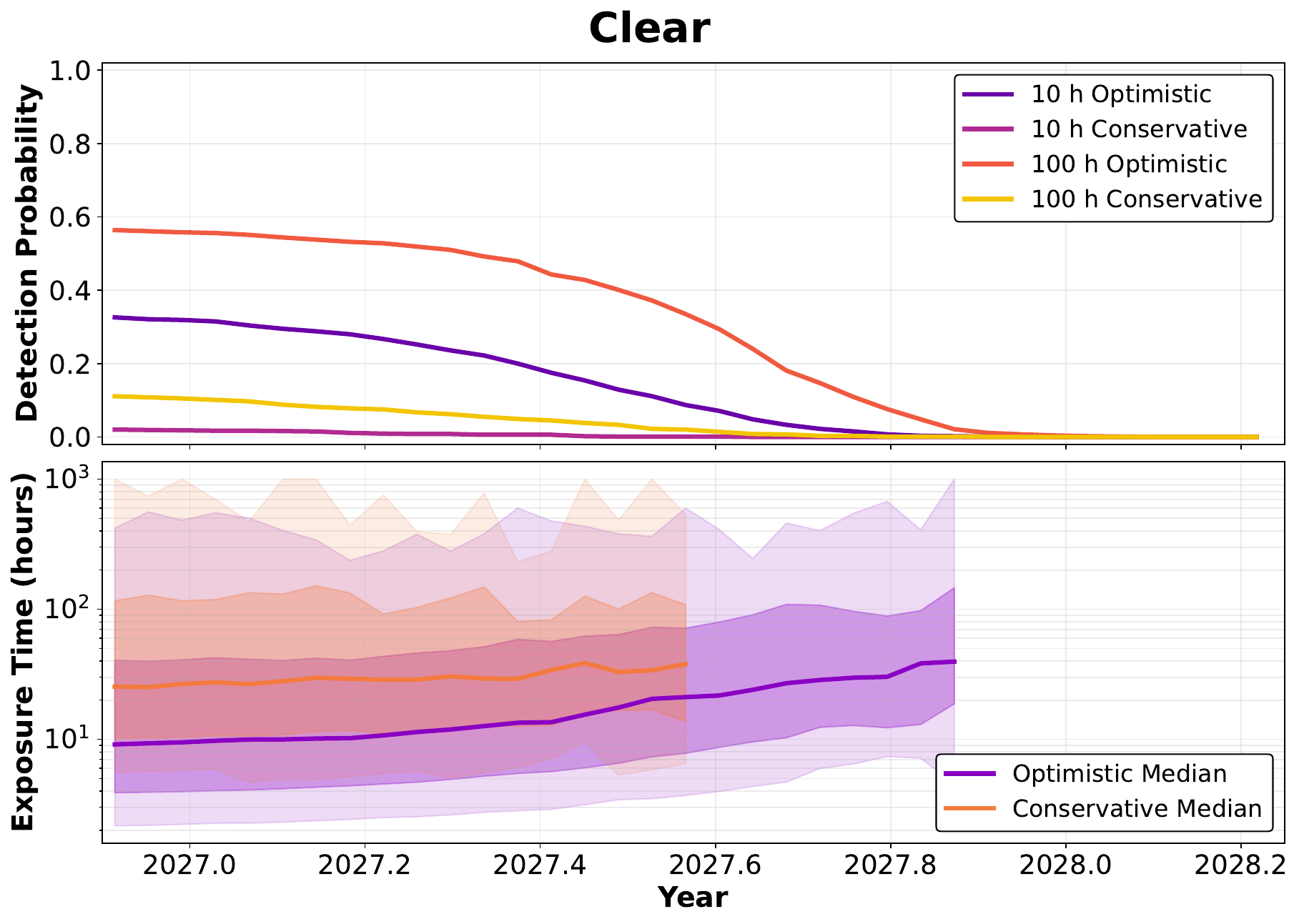}
   \end{tabular}
   \end{center}
   \caption{ Predictions of the detection probability and exposure time (to SNR=5) for Roman Coronagraph observations of \epseri. These are computed with \texttt{corgietc}, the  Roman Coronagraph ETC. Here we show the predictions for the worst case from our model assumptions for the Roman Coronagraph: the clear atmosphere in chemical {equilibrium}.
   \label{fig:rc_clarissa_clear}
   }  
\end{figure}

\begin{figure}\begin{center}
   \begin{tabular}{c} \includegraphics[height=7.5cm,trim={0cm 0cm 0cm 0cm}]{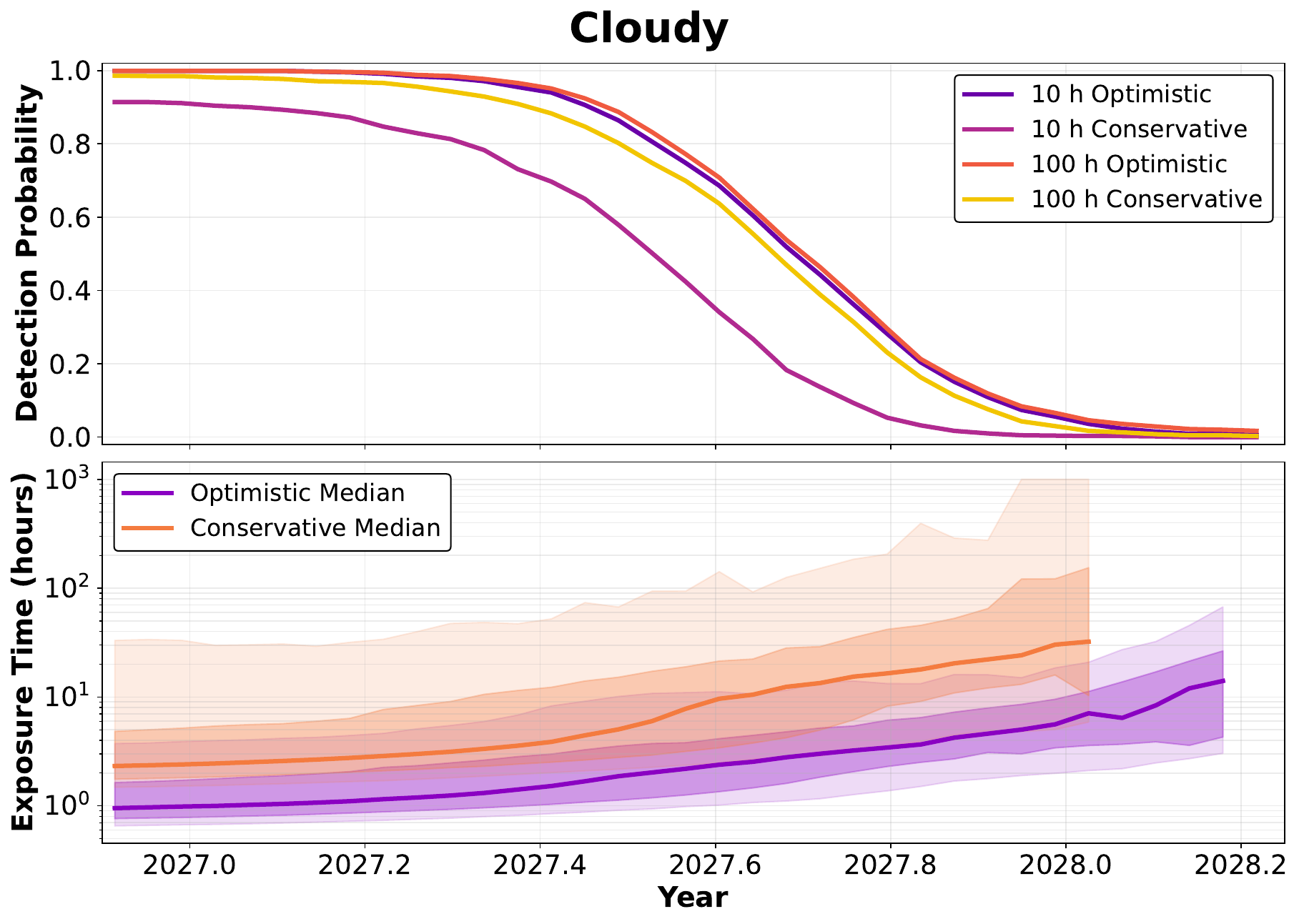}
   \end{tabular}
   \end{center}
   \caption{Same as Fig.~\ref{fig:rc_clarissa_clear} but for a cloudy case: \fsed=3 and chemical disequilibrium ($\log$K$_{zz}$=7). The cloudy atmosphere scenario is significantly more favorable than the clear one for Roman Coronagraph observations of \epseri~b.
   \label{fig:rc_clarissa_cloudy}
   }  
\end{figure}

{Using the Roman Coronagraph exposure time calculator (ETC), \texttt{corgietc}, we compute the exposure time required to detect \epseri~b for each posterior sample at each propagated Roman epoch. The required exposure time depends primarily on the planet-star flux contrast and angular separation, both of which vary across the posterior as the planet moves along its orbit. From these exposure time distributions, we compute a detection probability: for a fixed integration time budget, this is the fraction of posterior samples that could be detected by the Roman Coronagraph with SNR = 5 within that time.
We evaluate two representative atmospheric scenarios from the posteriors: a cloudy disequilibrium model with $f_{\rm sed}=3$ and $\log K_{zz}=7$, and a clear equilibrium chemistry model. The resulting detection probabilities and exposure time distributions are shown in Figs.~\ref{fig:rc_clarissa_clear} and~\ref{fig:rc_clarissa_cloudy}, assuming a 300 zodi exozodiacal dust level for \epseri~\citep{Ertel2020}. In both cases, the planet is most favorable for Roman observations before the end of 2027, when the combination of contrast and projected separation yields the shortest required exposure times. This emphasizes the importance of observing \epseri~b early in the mission.}

\subsection{What would a high-sensitivity non-detection mean}
Figs.~\ref{fig:cloud_points_rc} and ~\ref{fig:cloud_points_rc_f444wlimit} eloquently show the constraining power of the upcoming observations with the Roman Coronagraph. A final contrast sensitivity of 5$\times$\tentot~leaves little room for the planet to hide, in terms of atmospheric and evolutionary parameters. We take this to the extreme with a full model fit with fake Roman Coronagraph images that simulate a contrast sensitivity of 5$\times$\tentot~and 1$\times$\tentot. {Although an observing scenario where the Roman Coronagraph reaches such deep levels of final contrast is rather optimistic, it is not impossible. For instance, the "optimistic" scenarios simulated with \texttt{cgisim} offer solutions where the average contrast in the dark hole is $\sim$2$\times$\tenton~\citep{Krist2023}. A factor of $\times$20 separates this contrast level to a final contrast of 1$\times$\tentot. Post-processing gain factors for this instrument are uncertain at this point, but other studies assume or find factors of this order of magnitude, e.g. \citet{Stark2019,Ygouf2025}.  }

\begin{figure}\begin{center}
   \begin{tabular}{c} \includegraphics[height=9.5cm,trim={3cm 5cm 0cm 4cm}]{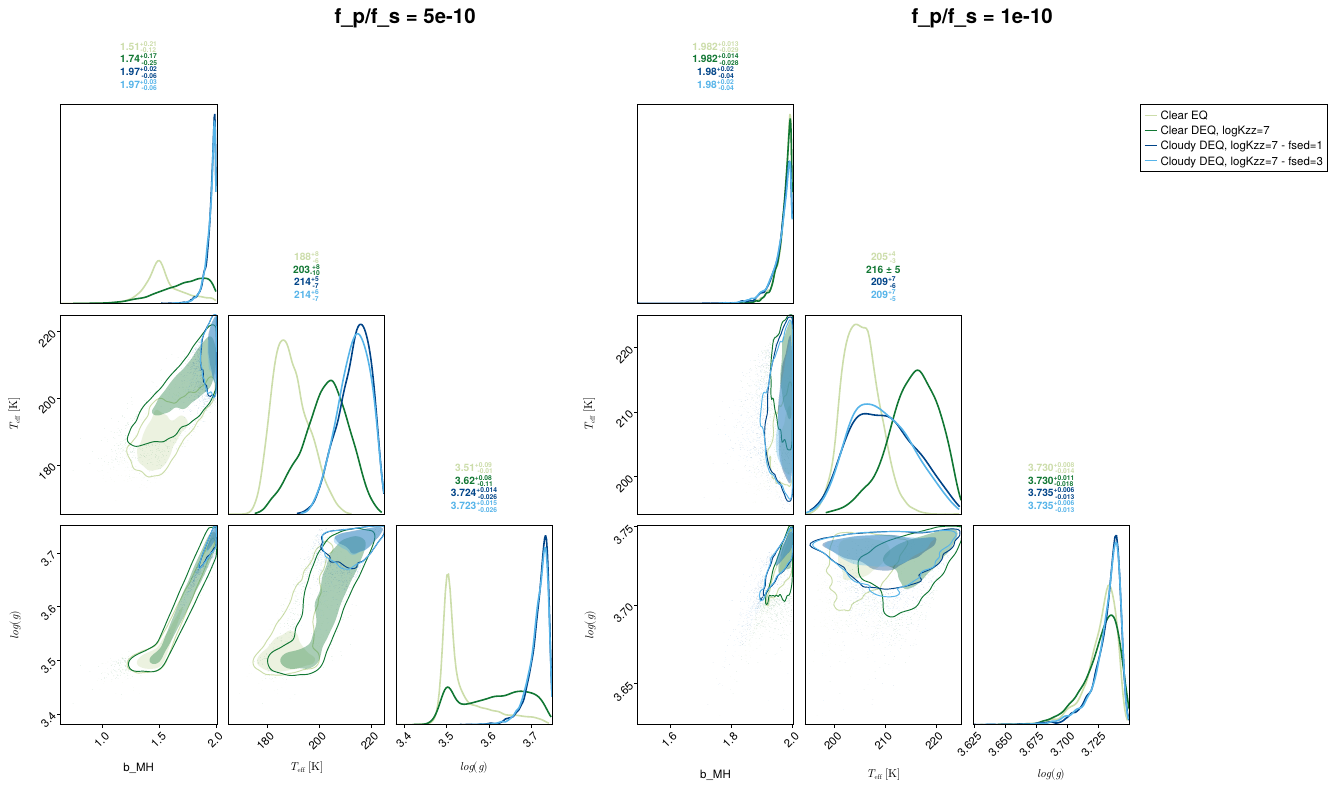}
   \end{tabular}
   \end{center}
   \caption{\octofitter~runs with synthetic Roman Coronagraph images assuming non detections at very high sensitivities to illustrate what happens to the atmospheric parameters if the planet is not detected with very strict upper limits in Band 1. Assuming a contrast sensitivity of 1$\times$\tentot~yields extreme best-fit parameters: enhanced metallicity, elevated \teff, and high surface gravity. A non-detection in this contrast regime would challenge current atmospheric models or cast doubt on the planet's existence.
   \label{fig:corner_nondetection}
   }  
\end{figure}

We feed \octofitter~with images of the same contrast and flux, corresponding to 1$\times \sigma$ of 5$\times$\tentot~and 1$\times$\tentot, so that the walkers see an equally low likelihood at all positions. We do this to avoid field dependent solutions that would skew the orbit and mass. We perform the fit with these images and the NIRCam images to assess what would happen in these two high sensitivity observations with the Roman Coronagraph, we show the results in a corner plot in Fig.~\ref{fig:corner_nondetection}.
The case of 5$\times$\tentot~still allows rather reasonable, albeit very high metallicity solutions for the clear atmosphere case. On the other hand, constraining the model fit with an assumed contrast sensitivity of 1$\times$\tentot~produces highly anomalous parameter values, specifically enhanced metallicity, high \teff, and high surface gravity. Consequently, a non-detection at this contrast limit would contradict existing atmospheric models or suggest the planet does not exist.

\subsection{Low-SNR images of the planet could still yield a confident detection}
\label{sec:lowsnr}
 
A planet that is too faint to register as a confident point source in any single Coronagraph image can still be recovered if it is observed at several epochs, because its motion between epochs is not arbitrary: it must trace a Keplerian orbit. The orbital model we have constrained with the radial velocity and astrometric data predicts where \epseri~b should sit at any future date, up to the uncertainties of the
fit. One can therefore tie together the flux at the predicted position in each epoch under the constraint that all of those positions belong to the same orbit. This shifts the search from the image plane into the orbital domain, where information from epochs spread across months or years can be combined coherently even when no individual frame contains a significant detection \citep{LeCoroller2020,Nowak2018}. \octofitter~\citep{Thompson2023} adopts the image likelihood of \citet{Ruffio2017} within a Bayesian orbit fit, so that the Keplerian constraints supplied by the RV and astrometry act directly on the recovery of a faint imaging signal.
 
We tested how well this works for the specific case of \epseri~b and the Roman Coronagraph using end-to-end simulations of the instrument. We generated simulated Coronagraph observations with \texttt{corosims} \citep{Llop-Sayson2025corosims}, which wraps the \texttt{cgisim} diffraction model of the instrument built on \texttt{PROPER} \citep{Krist2007,Krist2023}, and adds
photon-counting EMCCD detector noise through \texttt{emccd\_detect}. The wavefront errors driving the speckle field were taken from the time series of the Observing Scenario~11 (OS11), so that each simulated frame carries a
realistic, time-varying speckle structure rather than a single static realization. For each simulated epoch we produced a two-roll observation of \epseri\ together with a reference-star observation, and reduced the data with both roll subtraction and reference subtraction; that is, angular and reference differential imaging (ADI and RDI). These images include a
synthetic planet at a grid of single-epoch signal-to-noise ratios, placing it at the sky positions consistent with a single Keplerian orbit drawn from our posterior. We then recovered the injected signal with the image likelihood of \citet{Ruffio2017} in an orbit fit, as explained in Sec.~\ref{sec:planet_model_fit} with \octofitter. The fit is performing a blind search by construction, meaning that, although operating under the assumption that the planet exists, it doesn't have any information regarding the fake planet injection in the images. The recovered SNR for the grid of planets injected is shown in Fig.~\ref{fig:lowsnr}. The SNR is computed from the marginal photometry posterior, defined as the mean of that posterior divided by its standard deviation, following the convention of \citet{Thompson2023}.
\begin{figure}\begin{center}
   \begin{tabular}{c} \includegraphics[height=9.5cm,trim={0cm 0cm 0cm 0cm}]{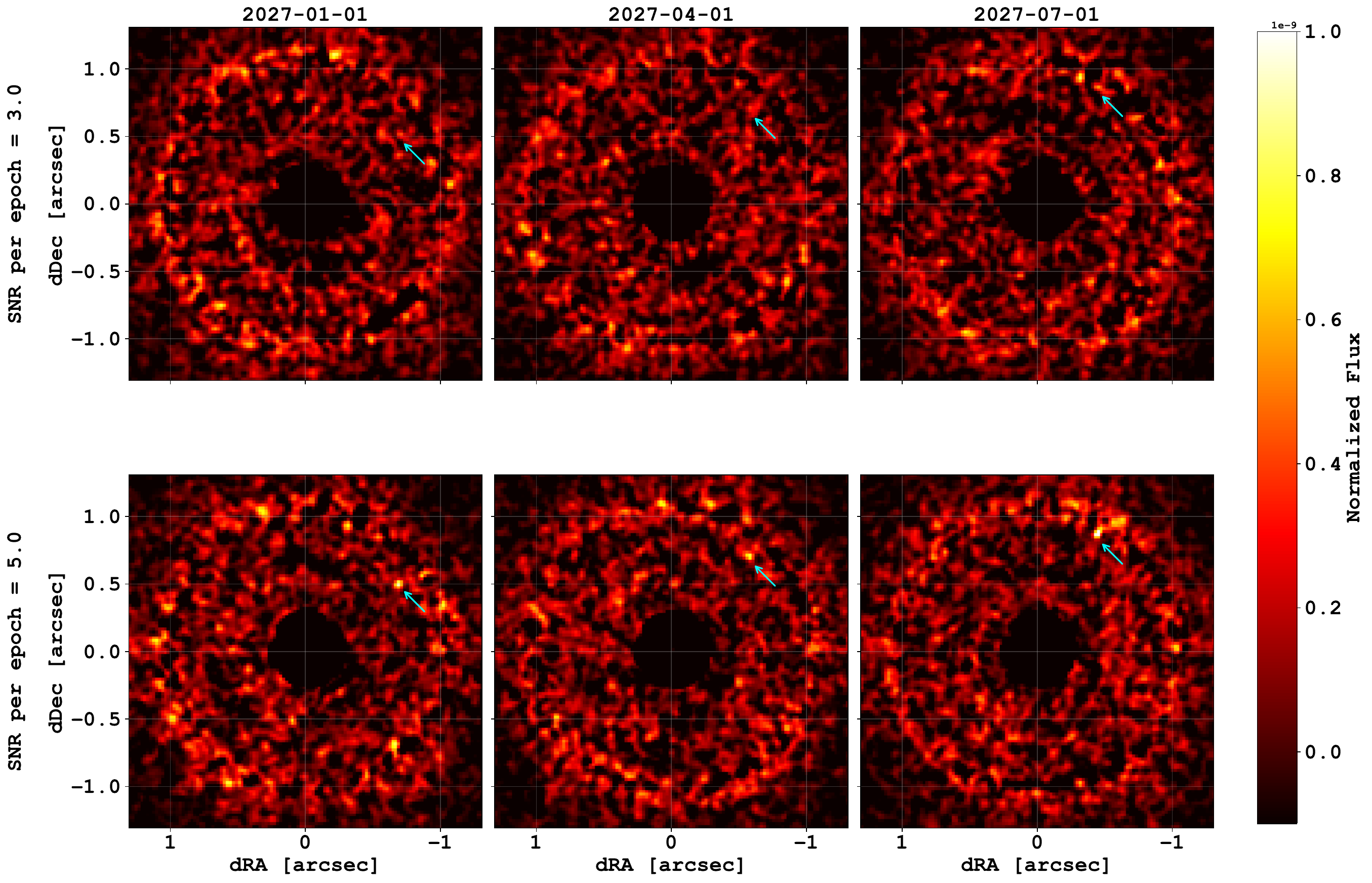}
   \end{tabular}
   \end{center}
   \caption{ Roman Coronagraph simulations using \texttt{cgisim} and \texttt{emccd\_detect} with the \texttt{corosims} wrapper; these images represent the post processed images after reference and roll subtraction (ADI + RDI). \textit{Top row:} planet injected at three epochs with an SNR=3 per epoch; \textit{bottom row:} planet injected with an SNR=5. Although the low SNR planet is barely visible in the top images, a full model fit can retrieve it at high statistical significance with enough epochs thanks to the constraining power of the Keplerian motion.}
   \label{fig:cgisim}
     
\end{figure} 

\begin{figure}\begin{center}
   \begin{tabular}{c} \includegraphics[height=7.5cm,trim={0cm 0cm 0cm 0cm}]{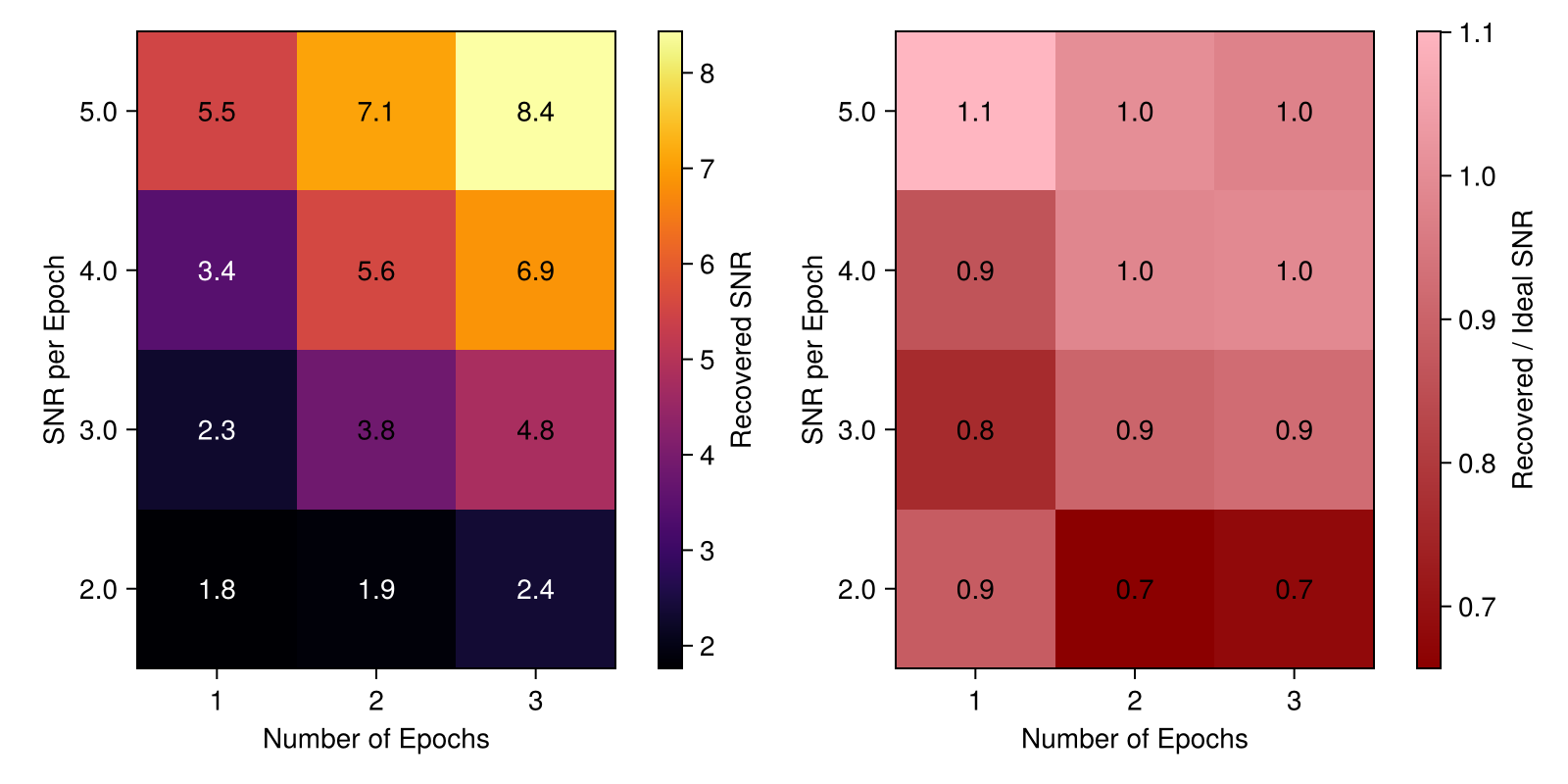}
   \end{tabular}
   \end{center}
   \caption{ \textit{Left:} Recovered photometric SNR as a function of injected planet SNR per epoch and number of epochs in model fit. \textit{Right:} Comparison of the recovered SNR to the ideal SNR =  $\sqrt{N_\mathrm{epochs}}\times\mathrm{SNR}_\mathrm{epoch}$. This test is analogous to the one performened in \citet{Thompson2023}, with similar results: the Keplerian motion is so constraining that even with low-SNR images, the planet can be detected with high confidence with enough epochs. 
   \label{fig:lowsnr}
   }  
\end{figure} 

\begin{figure}\begin{center}
   \begin{tabular}{c} \includegraphics[height=7.5cm,trim={0cm 0cm 0cm 0cm}]{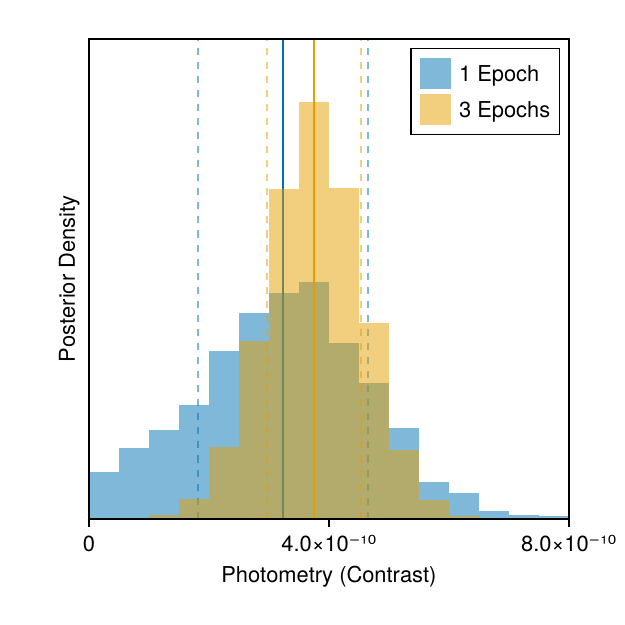}
   \end{tabular}
   \end{center}
   \caption{ Roman Coronagraph Band 1 photometry posteriors for the case of an injected planet at SNR=3 per epoch, one epoch fit (\textit{blue}), and three epochs fit (\textit{orange}). The simulated Roman Coronagraph can be seen in Fig.~\ref{fig:cgisim}. Although the injected planet is not detected with enough statistical significance at any one epoch ---the blue photometry posteriors are rather spread---the combination of multiple epochs can yield a confident detection---illustrated by the tightening of the orange posteriors.
   \label{fig:lowsnr_post}
   }  
\end{figure}

Our results closely reproduce the behavior found by \citet{Thompson2023} for ground-based imaging. Fig.~\ref{fig:lowsnr} shows the recovered
signal-to-noise ratio as a function of the per-epoch signal-to-noise and the number of epochs, in the same form as their Figure~9. We find that the combined significance grows essentially as $\sqrt{N_\mathrm{epochs}}\times\mathrm{SNR}_\mathrm{epoch}$, the ideal scaling expected for stacking independent measurements of a stationary source in uncorrelated noise.  Here it is recovered thanks to the Keplerian motion of the planet, which ties the flux measurements at the predicted positions across epochs into a single coherent constraint. In practice this means that a handful of epochs, each carrying a signal well below the canonical detection threshold, can together place \epseri~b above it. The benefit of multi-epoch observations is thus twofold: more epochs deepen the effective flux sensitivity, and they do so while simultaneously tightening the orbit.
Fig.~\ref{fig:lowsnr_post} illustrates the tightening of the posterior distribution for the case of SNR=3 in individual epochs when combining three epochs.  

The recovered significances quoted above are computed from the marginal photometry posterior of the orbit fit, which already folds in the RV and astrometric constraints: the fit marginalizes over the credible orbits, so the planet flux marginal posterior weights most heavily the positions at each epoch allowed by the other data. It is worth noting that this is specifically the significance of detecting non-zero photometry -- the planet is already detected with extreme significance from the RV and astrometry data. Because of this, we suggest that the $5\sigma$ threshold usually applied in direct imaging is not appropriate here. That threshold is set high to account for (A) noise that is not strictly gaussian (still applies here) (B) small sample statistics (still applies here) and most importantly, (C), as a \textit{de facto} correction for performing multiple comparisons (not applicable here). A blind search must admit a candidate at any azimuth, since a speckle of a given height is roughly equally likely to appear anywhere around the star at a fixed separation; the false positive fraction is therefore set by how often such a peak occurs across the full annulus. The orbit, by contrast, predicts the planet's position angle to within a narrow range at any given epoch. Essentially, only a single comparison is made: ``is the data consistent with non-zero photometry along this orbit?'' rather than ``are there any locations in any of the images that are consistent with non-zero photometry?''.

\section{Conclusions}
\label{sec:conclusions}

We have presented an updated model of \epseri~b that brings its orbit, mass, and atmosphere into a single framework. Building on the model fit work of T25 and the atmospheric upper limits of S26, we fit the radial velocity record, the absolute astrometry, and the two JWST/NIRCam F444W epochs jointly within \octofitter, with the predicted planet flux supplied by a grid of evolutionary and atmospheric models. We then applied the resulting model to the prospects for imaging \epseri~b in reflected light with the Roman Coronagraph.

The orbit and mass are largely consistent with T25. All orbital parameters fall within $1\sigma$ of the T25 posteriors, and the mass shifts only slightly, from $1.00\pm0.10\,M_{\rm Jup}$ to {$0.91\pm$0.04$\,M_{\rm Jup}$}; the difference is attributable to the G23H treatment of the Gaia astrometry \citep{Thompson2026}, the disk-planet mutual-inclination prior, the new NEID RV data, and small internal changes in \octofitter, rather than to the imaging. Indeed, adding the JWST images barely moves the orbit or the mass. The fit recovers a faint feature in the DDT epoch whose flux posterior peaks at $2.6\pm0.7\times10^{-7}$, a photometric SNR of only $\sim$3. We also find the planet's orbit to be mildly inclined with respect to the cold debris belt, with a mutual inclination of $\Phi = 14\pm4^\circ$.

Because the mass and the system age are both well constrained, the evolutionary models tie effective temperature and metallicity along a narrow track, which in turn restricts the surface gravity and the predicted 4~\mum\ flux; a surface gravity near $\log g \approx 3.5$ is preferred across all of the cases we consider, although rather unconstrained for some cases. Under the clear-atmosphere assumption this track leaves no freedom to keep the 4~\mum\ flux below the NIRCam limit except at very enhanced metallicity. In other words, a clear atmosphere in chemical equilibrium is viable only if the planet is strongly metal-enriched; if such an enhancement (M/H~$\gtrsim 0.5$) is judged unlikely for \epseri~b, then a cloudy atmosphere becomes the favored explanation, consistent with the suppression of the 4~\mum\ peak expected from water clouds at $\sim$1.1~Gyr. The cloudy and disequilibrium cases treat the NIRCam flux closer to an upper limit, and their metallicity solutions are correspondingly less extreme.

Resolving this ambiguity cleanly will require returning to the system with JWST. The low-SNR feature in the DDT epoch is consistent with the planet under the equilibrium assumption, but its significance is too low to settle the question on its own. Two further NIRCam epochs at comparable SNR would confirm whether the feature is real; if it is absent, the clear-atmosphere, equilibrium-chemistry hypothesis can be ruled out, and the orbital and atmospheric solutions that currently rely on it discarded.

Applied to the Roman Coronagraph, our model reiterates \epseri~b as a particularly favorable target for a reflected-light detection. Even a non-detection would carry considerable constraining power, and the deeper the contrast the more it constrains. Working down our model grid, a post-processed sensitivity of 5$\times$\tenton\ would already discard most of the allowed cloudy metallicities and effective temperatures; 2$\times$\tenton\ would discard all cloudy atmospheres under our assumptions and  would leave only the clear, disequilibrium case standing. Such a limit is not unrealistic: a raw contrast of 5$\times$\tenton\ combined with a post-processing gain of 5$\times$ would suffice. Pushing further, a final sensitivity, after post-processing, of 5$\times$\tentot\ leaves the planet almost nowhere to hide, and a non-detection at 1$\times$\tentot\ would drive the fit to anomalous parameters (very high metallicity, elevated effective temperature, and high surface gravity), which would either contradict current atmospheric models or call the planet's existence into question.

We therefore make three recommendations for the Coronagraph observations of \epseri~b. (1) Reach the deepest contrast the instrument allows, since sensitivity translates directly into constraints on the atmosphere whether or not the planet is detected. (2) Observe at multiple epochs: combination deepens the effective flux sensitivity, and, because the orbit predicts the planet's position at every date, the Keplerian motion can knit together several low-significance epochs into a confident detection, as demonstrated for ground-based imaging by \citet{Thompson2023}. (3) Observe in both Band~1 and Band~4, whose predicted flux ratios separate the cloudy and clear cases and would discriminate between the two atmospheric hypotheses that the JWST data alone cannot. As the nearest Jupiter analog, \epseri~b sits at the intersection of these two facilities; together, further JWST/NIRCam epochs and the upcoming Roman Coronagraph are poised to deliver the first direct characterization of its atmosphere.

\begin{acknowledgments}
We thank Samuel Halverson and the NEID team for providing the NEID radial velocity data used in this work. 
NIRCam development and use at the University of Arizona is supported through NASA Contract NAS5-02105. Part of this work was carried out at the Jet Propulsion Laboratory, California Institute of Technology, under a contract with the National Aeronautics and Space Administration (80NM0018D0004). The High Performance Computing resources used in this investigation were provided by funding from the JPL Enterprise IT Division. 
The work of A.G., and S.W. was partially supported by NASA grants NNX13AD82G and 1255094. This material is based upon work supported by the National Science Foundation Graduate Research Fellowship under Grant No.~2139433. We are grateful for support from NASA for the JWST NIRCam project though contract number NAS5-02105 (M. Rieke, University of Arizona, PI). 

Artificial intelligence tools were used to assist with drafting, code development, and figure preparation, in the same way as any other computational instrument. The authors verified all results and take full responsibility for the contents of this article.

\end{acknowledgments}

\begin{contribution}
J. Ll-S.: Observation planning, formal analysis, writing original draft. A. F.: various \octofitter~runs, and \octofitter~assistance. J. M.: atmospheric modeling, and assistance with \texttt{PICASO}. W. T.: \octofitter~assistance. C. D. \'O., I. H. and L. P.: Roman Coronagraph observations. A. S., C. B., G. B., and M. Y.: Observation planning. A. G. and S. W.: disk observations and planet-disk interactions. All authors have contributed to the observation analysis and assisted during the writing of this manuscript.

\end{contribution}

\facilities{JWST/NIRCam}

\software{astropy \citep{astropy},  
\texttt{pyKLIP} \citep{Wang2015},
\octofitter~\citep{Thompson2023},
PICASO~\citep{Batalha2019,Mang2026picaso},
\texttt{corgietc}, 
\texttt{corosims} \citep{Llop-Sayson2025corosims},  \texttt{cgisim} \citep{Krist2023},
\texttt{PROPER} \citep{Krist2007},
\texttt{jwst} \citep{jwst2022},
\texttt{STPSF} \citep{Perrin2014STPSF}, and
\texttt{webbpsf\_ext} \citep{Leisenring2025}.
}

\appendix

\section{Other tables}

\startlongtable
\begin{deluxetable}{ccc}
\tablecaption{Radial Velocities \label{tab:rvs}}
\tablehead{
    \colhead{Time} & \colhead{RV} & \colhead{RV Unc.} \\
    \colhead{(JD)} & \colhead{($\rm m~s^{-1}$)} & \colhead{($\rm m~s^{-1}$)}
}
\startdata
2459475.97600 & 96.51 & 0.1628 \\
2459478.99554 & 100.28 & 0.1533 \\
2459480.00188 & 102.30 & 0.1796 \\
2459482.00503 & 110.63 & 0.2936 \\
2459502.93589 & 101.55 & 0.2461 \\
2459504.88147 & 109.22 & 0.1609 \\
2459505.95270 & 118.08 & 0.2229 \\
2459517.89002 & 114.84 & 0.2057 \\
2459523.82790 & 106.55 & 0.1428 \\
2459528.90584 & 114.61 & 0.2214 \\
... & ... & ... \\
\enddata
\tablecomments{The full version of this table is available in machine-readable format in the online journal. A portion is shown here for guidance regarding its form and content.}
\end{deluxetable}

\startlongtable
\begin{deluxetable}{lcc}
\tablecaption{With and without NIRCam imaging data, for the free-floating flux method \label{tab:posteriors1}}
\tablehead{
    \colhead{Parameter} & \colhead{No Imaging Data} & \colhead{NIRCam Images, Free-Floating Flux} \\
}
\startdata
    M\_pri & $0.820_{-0.019}^{+0.020}$ & $0.819_{-0.019}^{+0.020}$ \\
    M\_sec & $0.913_{-0.041}^{+0.046}$ & $0.905_{-0.040}^{+0.041}$ \\
    b\_e & $0.044_{-0.032}^{+0.049}$ & $0.040_{-0.028}^{+0.037}$ \\
    b\_{$\omega$} & $3.530_{-1.982}^{+2.124}$ & $5.267_{-4.352}^{+0.735}$ \\
    b\_{$\tau$} & $0.562_{-0.417}^{+0.250}$ & $0.195_{-0.116}^{+0.302}$ \\
    b\_P & $7.278_{-0.067}^{+0.071}$ & $7.287_{-0.051}^{+0.057}$ \\
    b\_a & $3.517_{-0.037}^{+0.035}$ & $3.518_{-0.033}^{+0.033}$ \\
    b\_tp & $58885.030_{-1107.676}^{+663.397}$ & $57906.994_{-306.365}^{+792.314}$ \\
    b\_mass & $0.913_{-0.041}^{+0.046}$ & $0.905_{-0.040}^{+0.041}$ \\
    b\_i & $0.738_{-0.068}^{+0.071}$ & $0.745_{-0.070}^{+0.070}$ \\
    b\_{$\Omega$} & $-2.882_{-0.087}^{+0.083}$ & $-2.877_{-0.081}^{+0.078}$ \\
    b\_JWST\_flux & \nodata & $(1.953_{-0.760}^{+0.753})\times10^{-7}$ \\
    ... & ... & ... \\
\enddata
\tablecomments{The full version of this table is available in machine-readable format in the online journal. A portion is shown here for guidance regarding its form and content. Values represent the median of the posterior distribution, with errors denoting the 16th and 84th percentiles.}
\end{deluxetable}

\startlongtable
\begin{deluxetable}{lcccc}
\tablecaption{Clear atmosphere assumption \label{tab:posteriors2}}
\tablehead{
    \colhead{} & \colhead{Clear Atmosphere} & \colhead{Clear Atmosphere} & \colhead{Clear Atmosphere} & \colhead{Clear Atmosphere} \\
    \colhead{Parameter} & \colhead{EQ} & \colhead{DEQ - log Kzz=4} & \colhead{DEQ - log Kzz=7} & \colhead{DEQ - log Kzz=9} \\
}
\startdata
    M\_pri & $0.816_{-0.019}^{+0.020}$ & $0.818_{-0.020}^{+0.019}$ & $0.819_{-0.020}^{+0.019}$ & $0.818_{-0.019}^{+0.020}$ \\
    M\_sec & $0.892_{-0.040}^{+0.038}$ & $0.912_{-0.039}^{+0.040}$ & $0.906_{-0.040}^{+0.043}$ & $0.908_{-0.043}^{+0.041}$ \\
    b\_e & $0.043_{-0.029}^{+0.035}$ & $0.036_{-0.025}^{+0.036}$ & $0.039_{-0.027}^{+0.037}$ & $0.039_{-0.027}^{+0.038}$ \\ 
    b\_{$\omega$} & $5.167_{-4.106}^{+0.756}$ & $3.295_{-2.662}^{+2.609}$ & $4.633_{-3.923}^{+1.008}$ & $4.622_{-4.130}^{+1.150}$ \\
    b\_{$\tau$} & $0.178_{-0.121}^{+0.327}$ & $0.250_{-0.137}^{+0.314}$ & $0.274_{-0.200}^{+0.615}$ & $0.244_{-0.163}^{+0.419}$ \\
    b\_P & $7.288_{-0.050}^{+0.056}$ & $7.279_{-0.053}^{+0.057}$ & $7.281_{-0.051}^{+0.055}$ & $7.283_{-0.051}^{+0.057}$ \\
    b\_a & $3.514_{-0.033}^{+0.033}$ & $3.514_{-0.033}^{+0.033}$ & $3.515_{-0.033}^{+0.033}$ & $3.516_{-0.032}^{+0.032}$ \\
    b\_tp & $57865.048_{-322.744}^{+864.654}$ & $58054.144_{-363.418}^{+824.872}$ & $58118.978_{-534.278}^{+1634.552}$ & $58038.068_{-433.879}^{+1106.643}$ \\
    b\_mass & $0.892_{-0.040}^{+0.038}$ & $0.912_{-0.039}^{+0.040}$ & $0.906_{-0.040}^{+0.043}$ & $0.908_{-0.043}^{+0.041}$ \\
    b\_i & $0.755_{-0.065}^{+0.068}$ & $0.733_{-0.064}^{+0.069}$ & $0.745_{-0.071}^{+0.070}$ & $0.742_{-0.069}^{+0.071}$ \\
    b\_{$\Omega$} & $-2.872_{-0.077}^{+0.069}$ & $-2.873_{-0.077}^{+0.078}$ & $-2.873_{-0.078}^{+0.077}$ & $-2.871_{-0.086}^{+0.080}$ \\
    b\_JWST\_Age & $1.150_{-0.085}^{+0.087}$ & $1.101_{-0.094}^{+0.098}$ & $1.106_{-0.101}^{+0.103}$ & $1.104_{-0.101}^{+0.099}$ \\
    b\_JWST\_MH & $1.371_{-0.488}^{+0.383}$ & $0.888_{-0.320}^{+0.447}$ & $0.626_{-0.275}^{+0.564}$ & $0.463_{-0.283}^{+0.705}$ \\
    b\_JWST\_CO & $1.486_{-0.665}^{+0.661}$ & $1.502_{-0.695}^{+0.671}$ & $1.491_{-0.671}^{+0.673}$ & $1.492_{-0.670}^{+0.687}$ \\
    b\_JWST\_logKzz & $0.000_{-0.000}^{+0.000}$ & $4.000_{-0.000}^{+0.000}$ & $7.000_{-0.000}^{+0.000}$ & $9.000_{-0.000}^{+0.000}$ \\
    b\_JWST\_flux & $(2.595_{-0.394}^{+0.403})\times10^{-7}$ & $(1.641_{-0.930}^{+0.785})\times10^{-7}$ & $(1.412_{-0.998}^{+0.934})\times10^{-7}$ & $(1.415_{-1.169}^{+1.027})\times10^{-7}$ \\
    b\_JWST\_Teff & $189.298_{-16.088}^{+13.529}$ & $177.249_{-12.569}^{+9.395}$ & $171.361_{-9.594}^{+11.983}$ & $168.700_{-7.416}^{+13.894}$ \\
    b\_JWST\_Radius & $0.852_{-0.111}^{+0.089}$ & $0.944_{-0.091}^{+0.018}$ & $0.956_{-0.068}^{+0.014}$ & $0.962_{-0.068}^{+0.016}$ \\
    b\_JWST\_logg & $3.500_{-0.037}^{+0.127}$ & $3.488_{-0.055}^{+0.012}$ & $3.461_{-0.069}^{+0.039}$ & $3.445_{-0.071}^{+0.055}$ \\
    ... & ... & ... & ... & ... \\
\enddata
\tablecomments{The full version of this table is available in machine-readable format in the online journal. A portion is shown here for guidance regarding its form and content. Values represent the median of the posterior distribution, with errors denoting the 16th and 84th percentiles.}
\end{deluxetable}

\startlongtable
\begin{deluxetable}{lcccc}
\tablecaption{Cloudy atmosphere assumption \label{tab:posteriors3}}
\tablehead{
    \colhead{} & \colhead{Cloudy Atmosphere} & \colhead{Cloudy Atmosphere} & \colhead{Cloudy Atmosphere} & \colhead{Cloudy Atmosphere} \\
    \colhead{} & \colhead{fsed=8} & \colhead{fsed=1} & \colhead{fsed=3} & \colhead{fsed=8} \\
    \colhead{Parameter} & \colhead{EQ} & \colhead{DEQ, log Kzz=7} & \colhead{DEQ, log Kzz=7} & \colhead{DEQ, log Kzz=7} \\
}
\startdata
    M\_pri & $0.817_{-0.020}^{+0.020}$ & $0.818_{-0.020}^{+0.020}$ & $0.818_{-0.019}^{+0.020}$ & $0.817_{-0.019}^{+0.020}$ \\
    M\_sec & $0.903_{-0.041}^{+0.042}$ & $0.900_{-0.039}^{+0.043}$ & $0.904_{-0.040}^{+0.041}$ & $0.909_{-0.042}^{+0.038}$ \\
    b\_e & $0.037_{-0.026}^{+0.035}$ & $0.039_{-0.028}^{+0.039}$ & $0.039_{-0.027}^{+0.037}$ & $0.036_{-0.026}^{+0.038}$ \\ 
    $b_{\omega}$ & $4.343_{-3.488}^{+1.268}$ & $4.854_{-2.035}^{+0.943}$ & $4.902_{-4.219}^{+0.985}$ & $4.424_{-3.252}^{+1.393}$ \\
    $b_{\tau}$ & $0.353_{-0.254}^{+0.556}$ & $0.238_{-0.167}^{+0.651}$ & $0.220_{-0.135}^{+0.432}$ & $0.288_{-0.179}^{+0.511}$ \\
    b\_P & $7.272_{-0.048}^{+0.057}$ & $7.277_{-0.052}^{+0.064}$ & $7.282_{-0.051}^{+0.057}$ & $7.277_{-0.053}^{+0.060}$ \\
    b\_Age & $1.115_{-0.091}^{+0.091}$ & $1.100_{-0.098}^{+0.099}$ & $1.111_{-0.102}^{+0.102}$ & $1.107_{-0.097}^{+0.098}$ \\
    b\_MH & $1.049_{-1.099}^{+0.556}$ & $0.834_{-0.293}^{+0.651}$ & $0.491_{-0.570}^{+0.623}$ & $-0.049_{-0.600}^{+0.812}$ \\
    b\_CO & $1.295_{-0.574}^{+0.832}$ & $1.462_{-0.648}^{+0.694}$ & $1.467_{-0.669}^{+0.702}$ & $1.497_{-0.667}^{+0.688}$ \\
    b\_a & $3.511_{-0.032}^{+0.034}$ & $3.514_{-0.034}^{+0.034}$ & $3.516_{-0.034}^{+0.034}$ & $3.513_{-0.033}^{+0.033}$ \\
    b\_tp & $58325.481_{-670.959}^{+1455.025}$ & $58029.176_{-451.367}^{+1716.632}$ & $57976.475_{-360.653}^{+1126.364}$ & $58152.878_{-474.504}^{+1358.759}$ \\
    b\_mass & $0.903_{-0.041}^{+0.042}$ & $0.900_{-0.039}^{+0.043}$ & $0.904_{-0.040}^{+0.041}$ & $0.909_{-0.042}^{+0.038}$ \\
    b\_i & $0.746_{-0.072}^{+0.072}$ & $0.751_{-0.068}^{+0.069}$ & $0.744_{-0.068}^{+0.072}$ & $0.744_{-0.070}^{+0.068}$ \\
    $b_{\Omega}$ & $-2.875_{-0.081}^{+0.074}$ & $-2.876_{-0.085}^{+0.076}$ & $-2.872_{-0.082}^{+0.076}$ & $-2.870_{-0.081}^{+0.078}$ \\
    b\_logKzz & $0.000_{-0.000}^{+0.000}$ & $7.000_{-0.000}^{+0.000}$ & $7.000_{-0.000}^{+0.000}$ & $7.000_{-0.000}^{+0.000}$ \\
    $b_{\alpha\_1}$ & $115.419_{-4.069}^{+4.546}$ & $115.233_{-4.444}^{+5.040}$ & $116.004_{-4.207}^{+4.940}$ & $115.499_{-4.224}^{+4.901}$ \\
    $b_{\alpha\_2}$ & $64.159_{-3.244}^{+3.149}$ & $63.765_{-3.235}^{+3.543}$ & $64.433_{-3.369}^{+3.259}$ & $64.507_{-3.314}^{+3.609}$ \\
    b\_JWST\_flux & $(2.123_{-0.518}^{+0.534})\times10^{-7}$ & $(1.359_{-0.788}^{+0.936})\times10^{-7}$ & $(1.847_{-1.246}^{+0.921})\times10^{-7}$ & $(1.706_{-1.217}^{+0.720})\times10^{-7}$ \\
    b\_JWST\_Teff & $181.690_{-19.944}^{+15.558}$ & $179.525_{-12.613}^{+15.641}$ & $171.850_{-12.210}^{+14.108}$ & $162.524_{-6.220}^{+12.260}$ \\
    b\_JWST\_Radius & $0.928_{-0.139}^{+0.042}$ & $0.947_{-0.120}^{+0.016}$ & $0.964_{-0.053}^{+0.013}$ & $0.970_{-0.022}^{+0.009}$ \\
    b\_JWST\_logg & $3.478_{-0.086}^{+0.074}$ & $3.458_{-0.072}^{+0.047}$ & $3.424_{-0.057}^{+0.063}$ & $3.408_{-0.040}^{+0.049}$ \\
    ... & ... & ... & ... & ... \\
 \enddata
\tablecomments{The full version of this table is available in machine-readable format in the online journal. A portion is shown here for guidance regarding its form and content. Values represent the median of the posterior distribution, with errors denoting the 16th and 84th percentiles.}
\end{deluxetable}

\clearpage
\bibliography{biblio-jorgellop.bib}{}

@ARTICLE{Backman1993,
    author = {{Backman}, D.~E. and {Paresce}, F.},
    title = "{Main-sequence stars with circumstellar solid material - The VEGA phenomenon}",
    journal = {Protostars and Planets III},
    year = {1993},
    pages = {1253}
}

@ARTICLE{Booth2017,
    author = {{Booth}, M. and {Dent}, W.~R.~F. and {Jord{\'a}n}, A. and others},
    title = "{Resolving the planetesimal belt of HR 8799 with ALMA}",
    journal = {\mnras},
    year = {2017},
    volume = {469},
    pages = {3200},
    doi = {10.1093/mnras/stx1072}
}

@MISC{jwst2022,
    author = {{Bushouse}, H. and {Eisenhamer}, J. and {Dencheva}, N. and others},
    title = "{JWST Calibration Pipeline}",
    year = {2022},
    publisher = {Zenodo},
    doi = {10.5281/ZENODO.7038885},
    url = {https://zenodo.org/record/7038885}
}

@INPROCEEDINGS{Gillett1986,
    author = {{Gillett}, F.~C.},
    title = "{IRAS observations of cool excess around main sequence stars}",
    booktitle = {Light on Dark Matter},
    year = {1986},
    series = {Astrophysics and Space Science Library},
    volume = {124},
    editor = {{Israel}, F.~P.},
    pages = {61-69}
}

@ARTICLE{Hatzes2000,
    author = {{Hatzes}, A.~P. and {Cochran}, W.~D. and {McArthur}, B. and others},
    title = "{Evidence for a Long-Period Planet Orbiting epsilon Eridani}",
    journal = {\apjl},
    year = {2000},
    volume = {544},
    pages = {L145},
    doi = {10.1086/317319}
}

@INPROCEEDINGS{Kammerer2022,
    author = {{Kammerer}, J. and {Girard}, J. and {Carter}, A.~L. and others},
    title = "{Performance of JWST NIRCam's coronagraphs in flight}",
    booktitle = {Space Telescopes and Instrumentation 2022: Optical, Infrared, and Millimeter Wave},
    year = {2022},
    series = {Society of Photo-Optical Instrumentation Engineers (SPIE) Conference Series},
    volume = {12180},
    editor = {{Coyle}, L.~E. and {Matsuura}, S. and {Perrin}, M.~D.},
    pages = {121803N}
}

@ARTICLE{Llop-Sayson2021,
    author = {{Llop-Sayson}, J. and {Wang}, J.~J. and {Ruffio}, J.-B. and others},
    title = "{Constraining the Orbit and Mass of epsilon Eridani b with Radial Velocities, Hipparcos IAD-Gaia DR2 Astrometry, and Multiepoch Vortex Coronagraphy Upper Limits}",
    journal = {\aj},
    year = {2021},
    volume = {162},
    pages = {181},
    doi = {10.3847/1538-3881/ac134a}
}

@ARTICLE{Mawet2014,
    author = {{Mawet}, D. and {Milli}, J. and {Wahhaj}, Z. and others},
    title = "{Fundamental limitations of high contrast imaging set by small space angle scattering}",
    journal = {\apj},
    year = {2014},
    volume = {792},
    pages = {97},
    doi = {10.1088/0004-637X/792/2/97}
}

@ARTICLE{Mawet2019,
    author = {{Mawet}, D. and {Hirsch}, L. and {Lee}, E.~J. and others},
    title = "{Deep Exploration of the Planets HR 8799 b, c, and d with Alternate Strategies for High-contrast Imaging}",
    journal = {\aj},
    year = {2019},
    volume = {157},
    pages = {33},
    doi = {10.3847/1538-3881/aaef8a}
}

@INPROCEEDINGS{Perrin2014STPSF,
       author = {{Perrin}, Marshall D. and {Sivaramakrishnan}, Anand and {Lajoie}, Charles-Philippe and {Elliott}, Erin and {Pueyo}, Laurent and {Ravindranath}, Swara and {Albert}, Lo{\"\i}c.},
        title = "{Updated point spread function simulations for JWST with WebbPSF}",
    booktitle = {Space Telescopes and Instrumentation 2014: Optical, Infrared, and Millimeter Wave},
         year = 2014,
       editor = {{Oschmann}, Jr., Jacobus M. and {Clampin}, Mark and {Fazio}, Giovanni G. and {MacEwen}, Howard A.},
       series = {Society of Photo-Optical Instrumentation Engineers (SPIE) Conference Series},
       volume = {9143},
        month = aug,
          eid = {91433X},
        pages = {91433X},
          doi = {10.1117/12.2056689},
       adsurl = {https://ui.adsabs.harvard.edu/abs/2014SPIE.9143E..3XP}
}

@ARTICLE{Su2017,
    author = {{Su}, K.~Y.~L. and {De Buizer}, J.~M. and {Rieke}, G.~H. and others},
    title = "{The Inner Ring of epsilon Eridani}",
    journal = {\aj},
    year = {2017},
    volume = {153},
    pages = {226},
    doi = {10.3847/1538-3881/aa696b}
}

@MISC{Wang2015,
    author = {{Wang}, J.~J. and {Ruffio}, J.-B. and {De Rosa}, R.~J. and others},
    title = "{pyKLIP: PSF Subtraction for Exoplanets and Disks}",
    year = {2015},
    howpublished = {Astrophysics Source Code Library, record ascl:1506.001}
}

@ARTICLE{Booth2023,
    author = {{Booth}, M. and {Pearce}, T.~D. and {Krivov}, A.~V. and {Wyatt}, M.~C. and {Dent}, W.~R.~F. and {Hales}, A.~S. and {Lestrade}, J.-F. and {Cruz-S{\'a}enz de Miera}, F. and {Faramaz}, V.~C. and {L{\"o}hne}, T. and {Chavez-Dagostino}, M.},
    title = "{Title missing in bibitem}",
    journal = {\mnras},
    year = {2023},
    volume = {521},
    pages = {4},
    doi = {10.1093/mnras/stad938}
}

@ARTICLE{Greenbaum2023,
    author = {{Greenbaum}, A.~Z. and {Llop-Sayson}, J. and {Lew}, B.~W.~P. and others},
    title = "{First Observations of the Brown Dwarf HD 19467 B with JWST}",
    journal = {\apj},
    year = {2023},
    volume = {945},
    pages = {126},
    doi = {10.3847/1538-4357/acb68b}
}

@ARTICLE{Ygouf2024,
    author = {{Ygouf}, M. and {Beichman}, C.~A. and {Llop-Sayson}, J. and others},
    title = "{Title missing in bibitem}",
    journal = {\aj},
    year = {2024},
    volume = {167},
    pages = {26},
    doi = {10.3847/1538-3881/ad08c8}
}

@ARTICLE{Carter2023,
    author = {{Carter}, A.~L. and {Hinkley}, S. and {Kammerer}, J. and others},
    title = "{The JWST Early Release Science Program for Direct Observations of Exoplanetary Systems I: High-contrast Imaging of the Exoplanet HIP 65426 b from 2 to 16 \ensuremath{\mu}m}",
    journal = {\apjl},
    year = {2023},
    volume = {951},
    pages = {L20},
    doi = {10.3847/2041-8213/acd93e}
}

@ARTICLE{Mamajek2024,
    author = {{Mamajek}, E. and {Stapelfeldt}, K.},
    title = "{NASA Exoplanet Exploration Program (ExEP) Mission Star List for the Habitable Worlds Observatory (2023)}",
    journal = {arXiv e-prints},
    year = {2024},
    eprint = {2402.12414},
    doi = {10.48550/arXiv.2402.12414}
}

@ARTICLE{Thompson2025,
    author = {{Thompson}, W. and {Nielsen}, E. and {Ruffio}, J.-B. and others},
    title = "{Revised Mass and Orbit of $\varepsilon$ Eridani b: A 1 Jupiter-Mass Planet on a Near-Circular Orbit}",
    journal = {arXiv e-prints},
    year = {2025},
    eprint = {2502.20561},
    doi = {10.48550/arXiv.2502.20561}
}

@ARTICLE{Backman2009,
    author = {{Backman}, D. and {Marengo}, M. and {Stapelfeldt}, K. and others},
    title = "{Epsilon Eridani's Planetary Debris Disk: Structure and Dynamics based on Spitzer and CSO Observations}",
    journal = {\apj},
    year = {2009},
    volume = {690},
    pages = {1522},
    doi = {10.1088/0004-637X/690/2/1522}
}

@ARTICLE{Gray2003,
    author = {{Gray}, R.~O. and {Corbally}, C.~J. and {Garrison}, R.~F. and others},
    title = "{Contributions to the Nearby Stars (NStars) Project: Spectroscopy of Stars Earlier than M0 within 40 Parsecs: The Northern Sample. I.}",
    journal = {\aj},
    year = {2003},
    volume = {126},
    pages = {2048},
    doi = {10.1086/378365}
}

@ARTICLE{Tschudi2024,
       author = {{Tschudi}, C. and {Schmid}, H.~M. and {Nowak}, M. and others},
        title = "{SPHERE RefPlanets: Search for \varepsilon Eridani b and warm dust}",
      journal = {\aap},
         year = {2024},
       volume = {687},
        pages = {A74},
          doi = {10.1051/0004-6361/202449442}
}

@ARTICLE{Gaspar2023,
       author = {{G{\'a}sp{\'a}r}, Andr{\'a}s and {Wolff}, Schuyler Grace and {Rieke}, George H. and others},
        title = "{Spatially resolved imaging of the inner Fomalhaut disk using JWST/MIRI}",
      journal = {Nature Astronomy},
         year = {2023},
       volume = {7},
        pages = {790-798},
          doi = {10.1038/s41550-023-01962-6}
}

@ARTICLE{Anglada-Escude2012,
       author = {{Anglada-Escud{\'e}}, Guillem and {Butler}, R. Paul},
        title = "{The HARPS-TERRA Project. I. Description of the Algorithms, Performance, and New Measurements on a Few Remarkable Stars Observed by HARPS}",
      journal = {\apjs},
         year = {2012},
       volume = {200},
        pages = {15},
          doi = {10.1088/0067-0049/200/2/15}
}

@ARTICLE{Llop-Sayson2025,
       author = {{Llop-Sayson}, Jorge and {Beichman}, Charles and {Bryden}, Geoffrey and others},
        title = "{Searching for Planets Orbiting \epsilon Eridani with JWST/NIRCam}",
      journal = {\aj},
         year = {2025},
       volume = {170},
        pages = {229},
          doi = {10.3847/1538-3881/adf727}
}

@ARTICLE{Thompson2023,
       author = {{Thompson}, William and {Lawrence}, Jensen and {Blakely}, Dori and others},
        title = "{Octofitter: Fast, Flexible, and Accurate Orbit Modeling to Detect Exoplanets}",
      journal = {\aj},
         year = {2023},
       volume = {166},
        pages = {164},
          doi = {10.3847/1538-3881/acf5cc}
}

@ARTICLE{LeCoroller2020,
       author = {{Le Coroller}, H. and {Nowak}, M. and {Delorme}, P. and others},
        title = "{K-Stacker: an algorithm to hack the orbital parameters of planets hidden in high-contrast imaging. First applications to VLT/SPHERE multi-epoch observations}",
      journal = {\aap},
         year = {2020},
       volume = {639},
        pages = {A113},
          doi = {10.1051/0004-6361/202037605}
}

@ARTICLE{Ruffio2017,
       author = {{Ruffio}, Jean-Baptiste and {Macintosh}, Bruce and {Wang}, Jason J. and others},
        title = "{Improving and Assessing Planet Sensitivity of the GPI Exoplanet Survey with a Forward Model Matched Filter}",
      journal = {\apj},
         year = {2017},
       volume = {842},
        pages = {14},
          doi = {10.3847/1538-4357/aa72dd}
}

@ARTICLE{Wolff2025,
       author = {{Wolff}, Schuyler G. and {G{\'a}sp{\'a}r}, Andr{\'a}s and {Rieke}, George and others},
        title = "{JWST/MIRI Imaging of the Warm Dust Component of the Epsilon Eridani Debris Disk}",
      journal = {arXiv e-prints},
         year = {2025},
       eprint = {2509.24976}
}

@ARTICLE{Campbell1988,
       author = {{Campbell}, Bruce and {Walker}, G.~A.~H. and {Yang}, S.},
        title = "{A Search for Substellar Companions to Solar-type Stars}",
      journal = {\apj},
         year = 1988,
        month = aug,
       volume = {331},
        pages = {902},
          doi = {10.1086/166608},
       adsurl = {https://ui.adsabs.harvard.edu/abs/1988ApJ...331..902C}
}

@ARTICLE{Fischer2014,
       author = {{Fischer}, Debra A. and {Marcy}, Geoffrey W. and {Spronck}, Julien F.~P.},
        title = "{The Twenty-five Year Lick Planet Search}",
      journal = {\apjs},
         year = 2014,
        month = jan,
       volume = {210},
       number = {1},
          eid = {5},
        pages = {5},
          doi = {10.1088/0067-0049/210/1/5},
archivePrefix = {arXiv},
       eprint = {1310.7315},
 primaryClass = {astro-ph.EP},
       adsurl = {https://ui.adsabs.harvard.edu/abs/2014ApJS..210....5F}
}

@ARTICLE{Zechmeister2013,
       author = {{Zechmeister}, M. and {K{\"u}rster}, M. and {Endl}, M. and {Lo Curto}, G. and {Hartman}, H. and {Nilsson}, H. and {Henning}, T. and {Hatzes}, A.~P. and {Cochran}, W.~D.},
        title = "{The planet search programme at the ESO CES and HARPS. IV. The search for Jupiter analogues around solar-like stars}",
      journal = {\aap},
         year = 2013,
        month = apr,
       volume = {552},
          eid = {A78},
        pages = {A78},
          doi = {10.1051/0004-6361/201116551},
archivePrefix = {arXiv},
       eprint = {1211.7263},
 primaryClass = {astro-ph.EP},
       adsurl = {https://ui.adsabs.harvard.edu/abs/2013A&A...552A..78Z}
}

@ARTICLE{Giguere2016,
       author = {{Giguere}, Matthew J. and {Fischer}, Debra A. and {Zhang}, Cyril X.~Y. and {Matthews}, Jaymie M. and {Cameron}, Chris and {Henry}, Gregory W.},
        title = "{A Combined Spectroscopic and Photometric Stellar Activity Study of Epsilon Eridani}",
      journal = {\apj},
         year = 2016,
        month = jun,
       volume = {824},
       number = {2},
          eid = {150},
        pages = {150},
          doi = {10.3847/0004-637X/824/2/150},
archivePrefix = {arXiv},
       eprint = {1606.08553},
 primaryClass = {astro-ph.SR},
       adsurl = {https://ui.adsabs.harvard.edu/abs/2016ApJ...824..150G}
}

@ARTICLE{Roettenbacher2022,
       author = {{Roettenbacher}, Rachael M. and {Cabot}, Samuel H.~C. and {Fischer}, Debra A. and {Monnier}, John D. and {Henry}, Gregory W. and {Harmon}, Robert O. and {Korhonen}, Heidi and {Brewer}, John M. and {Llama}, Joe and {Petersburg}, Ryan R. and {Zhao}, Lily L. and {Kraus}, Stefan and {Le Bouquin}, Jean-Baptiste and {Anugu}, Narsireddy and {Davies}, Claire L. and {Gardner}, Tyler and {Lanthermann}, Cyprien and {Schaefer}, Gail and {Setterholm}, Benjamin and {Clark}, Catherine A. and {Jorstad}, Svetlana G. and {Kuehn}, Kyler and {Levine}, Stephen},
        title = "{EXPRES. III. Revealing the Stellar Activity Radial Velocity Signature of ϵ Eridani with Photometry and Interferometry}",
      journal = {\aj},
         year = 2022,
        month = jan,
       volume = {163},
       number = {1},
          eid = {19},
        pages = {19},
          doi = {10.3847/1538-3881/ac3235},
archivePrefix = {arXiv},
       eprint = {2110.10643},
 primaryClass = {astro-ph.EP},
       adsurl = {https://ui.adsabs.harvard.edu/abs/2022AJ....163...19R}
}

@ARTICLE{Foreman-Mackey2017,
       author = {{Foreman-Mackey}, Daniel and {Agol}, Eric and {Ambikasaran}, Sivaram and {Angus}, Ruth},
        title = "{Fast and Scalable Gaussian Process Modeling with Applications to Astronomical Time Series}",
      journal = {\aj},
         year = 2017,
        month = dec,
       volume = {154},
       number = {6},
          eid = {220},
        pages = {220},
          doi = {10.3847/1538-3881/aa9332},
archivePrefix = {arXiv},
       eprint = {1703.09710},
 primaryClass = {astro-ph.IM},
       adsurl = {https://ui.adsabs.harvard.edu/abs/2017AJ....154..220F}
}

@INPROCEEDINGS{Schwab2016,
       author = {{Schwab}, C. and {Rakich}, A. and {Gong}, Q. and {Mahadevan}, S. and {Halverson}, S.~P. and {Roy}, A. and {Terrien}, R.~C. and {Robertson}, P.~M. and {Hearty}, F.~R. and {Levi}, E.~I. and {Monson}, A.~J. and {Wright}, J.~T. and {McElwain}, M.~W. and {Bender}, C.~F. and {Blake}, C.~H. and {St{\"u}rmer}, J. and {Gurevich}, Y.~V. and {Chakraborty}, A. and {Ramsey}, L.~W.},
        title = "{Design of NEID, an extreme precision Doppler spectrograph for WIYN}",
    booktitle = {Ground-based and Airborne Instrumentation for Astronomy VI},
         year = 2016,
       editor = {{Evans}, Christopher J. and {Simard}, Luc and {Takami}, Hideki},
       series = {Society of Photo-Optical Instrumentation Engineers (SPIE) Conference Series},
       volume = {9908},
        month = aug,
          eid = {99087H},
        pages = {99087H},
          doi = {10.1117/12.2234411},
       adsurl = {https://ui.adsabs.harvard.edu/abs/2016SPIE.9908E..7HS}
}

@ARTICLE{Jiang2024,
       author = {{Jiang}, Sarah and {Roy}, Arpita and {Halverson}, Samuel and {Bender}, Chad F. and {Selgas}, Carlos and {Otor}, O. Justin and {Mahadevan}, Suvrath and {Stef{\'a}nsson}, Gu{\dj}mundur and {Terrien}, Ryan C. and {Schwab}, Christian},
        title = "{Revisiting ϵ Eridani with NEID: Identifying New Activity-sensitive Lines in a Young K Dwarf Star}",
      journal = {\aj},
         year = 2024,
        month = jan,
       volume = {167},
       number = {1},
          eid = {9},
        pages = {9},
          doi = {10.3847/1538-3881/ad0b0b},
archivePrefix = {arXiv},
       eprint = {2311.10677},
 primaryClass = {astro-ph.EP},
       adsurl = {https://ui.adsabs.harvard.edu/abs/2024AJ....167....9J}
}

@BOOK{vanLeeuwen2007,
       author = {{van Leeuwen}, Floor},
        title = "{Hipparcos, the New Reduction of the Raw Data}",
         year = 2007,
       volume = {350},
          doi = {10.1007/978-1-4020-6342-8},
       adsurl = {https://ui.adsabs.harvard.edu/abs/2007ASSL..350.....V}
}

@article{Benedict2006,
	Author = {G. Fritz Benedict and Barbara E. McArthur and George Gatewood and Edmund Nelan and William D. Cochran and Artie Hatzes and Michael Endl and Robert Wittenmyer and Sallie L. Baliunas and Gordon A. H. Walker and Stephenson Yang and Martin K{\"u}rster and Sebastian Els and Diane B. Paulson},
	Doi = {10.1086/508323},
	Journal = {The Astronomical Journal},
	Month = {oct},
	Number = {5},
	Pages = {2206--2218},
	Publisher = {American Astronomical Society},
	Title = {The Extrasolar Planet $\upepsilon$ Eridani b: Orbit and Mass},
	Url = {https://doi.org/10.1086/508323},
	Volume = {132},
	Year = 2006}

@ARTICLE{Thompson2026,
       author = {{Thompson}, William and {Blakely}, Dori and {Xuan}, Jerry W. and {Blouin}, Simon and {Zhang}, Jingwen and {Johnstone}, Doug and {Ruffio}, Jean-Baptiste and {Nielsen}, Eric and {Speedie}, Jessica and {Bowler}, Brendan P. and {Bouchard-C{\^o}t{\'e}}, Alexandre and {Franson}, Kyle and {Blunt}, Sarah and {Roberson}, William and {Cloutier}, Ryan and {Fogal}, Andre and {Hessel}, Kaitlyn and {Marois}, Christian and {Rochon}, Alexandra},
        title = "{Detecting and Characterizing Companions with a Calibrated Gaia DR2, DR3, and Hipparcos Catalog (G23H)}",
      journal = {arXiv e-prints},
         year = 2026,
        month = jan,
          eid = {arXiv:2602.00235},
        pages = {arXiv:2602.00235},
          doi = {10.48550/arXiv.2602.00235},
archivePrefix = {arXiv},
       eprint = {2602.00235},
 primaryClass = {astro-ph.EP},
       adsurl = {https://ui.adsabs.harvard.edu/abs/2026arXiv260200235T}
}

@ARTICLE{Nielsen2020,
       author = {{Nielsen}, Eric L. and {De Rosa}, Robert J. and {Wang}, Jason J. and {Sahlmann}, Johannes and {Kalas}, Paul and {Duch{\^e}ne}, Gaspard and {Rameau}, Julien and {Marley}, Mark S. and {Saumon}, Didier and {Macintosh}, Bruce and {Millar-Blanchaer}, Maxwell A. and {Nguyen}, Meiji M. and {Ammons}, S. Mark and {Bailey}, Vanessa P. and {Barman}, Travis and {Bulger}, Joanna and {Chilcote}, Jeffrey and {Cotten}, Tara and {Doyon}, Rene and {Esposito}, Thomas M. and {Fitzgerald}, Michael P. and {Follette}, Katherine B. and {Gerard}, Benjamin L. and {Goodsell}, Stephen J. and {Graham}, James R. and {Greenbaum}, Alexandra Z. and {Hibon}, Pascale and {Hung}, Li-Wei and {Ingraham}, Patrick and {Konopacky}, Quinn and {Larkin}, James E. and {Maire}, J{\'e}r{\^o}me and {Marchis}, Franck and {Marois}, Christian and {Metchev}, Stanimir and {Oppenheimer}, Rebecca and {Palmer}, David and {Patience}, Jennifer and {Perrin}, Marshall and {Poyneer}, Lisa and {Pueyo}, Laurent and {Rajan}, Abhijith and {Rantakyr{\"o}}, Fredrik T. and {Ruffio}, Jean-Baptiste and {Savransky}, Dmitry and {Schneider}, Adam C. and {Sivaramakrishnan}, Anand and {Song}, Inseok and {Soummer}, Remi and {Thomas}, Sandrine and {Wallace}, J. Kent and {Ward-Duong}, Kimberly and {Wiktorowicz}, Sloane and {Wolff}, Schuyler},
        title = "{The Gemini Planet Imager Exoplanet Survey: Dynamical Mass of the Exoplanet {\ensuremath{\beta}} Pictoris b from Combined Direct Imaging and Astrometry}",
      journal = {\aj},
         year = 2020,
        month = feb,
       volume = {159},
       number = {2},
          eid = {71},
        pages = {71},
          doi = {10.3847/1538-3881/ab5b92},
archivePrefix = {arXiv},
       eprint = {1911.11273},
 primaryClass = {astro-ph.EP},
       adsurl = {https://ui.adsabs.harvard.edu/abs/2020AJ....159...71N}
}

@ARTICLE{Brandt2021,
       author = {{Brandt}, Timothy D.},
        title = "{The Hipparcos-Gaia Catalog of Accelerations: Gaia EDR3 Edition}",
      journal = {\apjs},
         year = 2021,
        month = jun,
       volume = {254},
       number = {2},
          eid = {42},
        pages = {42},
          doi = {10.3847/1538-4365/abf93c},
archivePrefix = {arXiv},
       eprint = {2105.11662},
 primaryClass = {astro-ph.GA},
       adsurl = {https://ui.adsabs.harvard.edu/abs/2021ApJS..254...42B}
}

@ARTICLE{Chance2025,
       author = {{Chance}, Quadry and {Foreman-Mackey}, Daniel and {Ballard}, Sarah and {Casey}, Andrew R. and {David}, Trevor J. and {Price-Whelan}, Adrian M.},
        title = "{paired: A Statistical Framework for Detecting Stellar Binarity with Gaia RVs. I. Sensitivity to Unresolved Binaries}",
      journal = {\apj},
         year = 2025,
        month = oct,
       volume = {992},
       number = {1},
          eid = {131},
        pages = {131},
          doi = {10.3847/1538-4357/adfb68},
archivePrefix = {arXiv},
       eprint = {2206.11275},
 primaryClass = {astro-ph.EP},
       adsurl = {https://ui.adsabs.harvard.edu/abs/2025ApJ...992..131C}
}

@ARTICLE{Kiefer2025,
       author = {{Kiefer}, F. and {Lagrange}, A.-M. and {Rubini}, P. and {Philipot}, F.},
        title = "{Searching for substellar companion candidates with Gaia: I. Introducing the GaiaPMEX tool}",
      journal = {\aap},
         year = 2025,
        month = oct,
       volume = {702},
          eid = {A76},
        pages = {A76},
          doi = {10.1051/0004-6361/202449335},
archivePrefix = {arXiv},
       eprint = {2409.16992},
 primaryClass = {astro-ph.EP},
       adsurl = {https://ui.adsabs.harvard.edu/abs/2025A&A...702A..76K}
}

@ARTICLE{Lindegren2020,
       author = {{Lindegren}, Lennart},
        title = "{The Gaia reference frame for bright sources examined using VLBI observations of radio stars}",
      journal = {\aap},
         year = 2020,
        month = jan,
       volume = {633},
          eid = {A1},
        pages = {A1},
          doi = {10.1051/0004-6361/201936161},
archivePrefix = {arXiv},
       eprint = {1906.09827},
 primaryClass = {astro-ph.IM},
       adsurl = {https://ui.adsabs.harvard.edu/abs/2020A&A...633A...1L}
}

@ARTICLE{Cantat-Gaudin2021,
       author = {{Cantat-Gaudin}, Tristan and {Brandt}, Timothy D.},
        title = "{Characterizing and correcting the proper motion bias of the bright Gaia EDR3 sources}",
      journal = {\aap},
         year = 2021,
        month = may,
       volume = {649},
          eid = {A124},
        pages = {A124},
          doi = {10.1051/0004-6361/202140807},
archivePrefix = {arXiv},
       eprint = {2103.07432},
 primaryClass = {astro-ph.GA},
       adsurl = {https://ui.adsabs.harvard.edu/abs/2021A&A...649A.124C}
}

@misc{Bushouse2025,
       author = {{Bushouse}, Howard and {Eisenhamer}, Jonathan and {Dencheva}, Nadia and {Davies}, James and {Greenfield}, Perry and {Morrison}, Jane and {Hodge}, Phil and {Simon}, Bernie and {Grumm}, David and {Droettboom}, Michael and {Slavich}, Edward and {Sosey}, Megan and {Pauly}, Tyler and {Miller}, Todd and {Jedrzejewski}, Robert and {Hack}, Warren and {Davis}, David and {Crawford}, Steven and {Law}, David and {Gordon}, Karl and {Regan}, Michael and {Cara}, Mihai and {MacDonald}, Ken and {Bradley}, Larry and {Shanahan}, Clare and {Jamieson}, William and {Teodoro}, Mairan and {Williams}, Thomas and {Pena-Guerrero}, Maria and {Graham}, Brett and {Molter}, Edward and {Brandt}, Timothy and {Hayes}, Christian and {Cooper}, Rachel and {Clarke}, Melanie and {Filippazzo}, Joseph},
        title = "{JWST Calibration Pipeline}",
         year = 2025,
        month = apr,
          eid = {10.5281/zenodo.15178003},
          doi = {10.5281/zenodo.15178003},
      version = {1.18.0},
    publisher = {Zenodo},
       adsurl = {https://ui.adsabs.harvard.edu/abs/2025zndo..15178003B}
}

@ARTICLE{Gagliuffi2025,
       author = {{Bardalez Gagliuffi}, Daniella C. and {Balmer}, William O. and {Pueyo}, Laurent and {Brandt}, Timothy D. and {Giovinazzi}, Mark R. and {Millholland}, Sarah and {Black}, Brennen and {Lu}, Tiger and {Rice}, Malena and {Mang}, James and {Morley}, Caroline and {Lacy}, Brianna and {Girard}, Julien H. and {Matthews}, Elisabeth C. and {Carter}, Aarynn L. and {Bowler}, Brendan P. and {Faherty}, Jacqueline K. and {Fontanive}, Clemence and {Rickman}, Emily},
        title = "{JWST Coronagraphic Images of 14 Her c: A Cold Giant Planet in a Dynamically Hot Multiplanet System}",
      journal = {\apjl},
         year = 2025,
        month = jul,
       volume = {988},
       number = {1},
          eid = {L18},
        pages = {L18},
          doi = {10.3847/2041-8213/ade30f},
archivePrefix = {arXiv},
       eprint = {2506.09201},
 primaryClass = {astro-ph.EP},
       adsurl = {https://ui.adsabs.harvard.edu/abs/2025ApJ...988L..18B}
}

@ARTICLE{Brandt2024a,
       author = {{Brandt}, Timothy D.},
        title = "{Likelihood-based Jump Detection and Cosmic Ray Rejection for Detectors Read Out Up-the-ramp}",
      journal = {\pasp},
         year = 2024,
        month = apr,
       volume = {136},
       number = {4},
          eid = {045005},
        pages = {045005},
          doi = {10.1088/1538-3873/ad38da},
archivePrefix = {arXiv},
       eprint = {2404.01326},
 primaryClass = {astro-ph.IM},
       adsurl = {https://ui.adsabs.harvard.edu/abs/2024PASP..136d5005B}
}

@ARTICLE{Brandt2024b,
       author = {{Brandt}, Timothy D.},
        title = "{Optimal Fitting and Debiasing for Detectors Read Out Up-the-Ramp}",
      journal = {\pasp},
         year = 2024,
        month = apr,
       volume = {136},
       number = {4},
          eid = {045004},
        pages = {045004},
          doi = {10.1088/1538-3873/ad38d9},
archivePrefix = {arXiv},
       eprint = {2309.08753},
 primaryClass = {astro-ph.IM},
       adsurl = {https://ui.adsabs.harvard.edu/abs/2024PASP..136d5004B}
}

@misc{Leisenring2025,
  author       = {Leisenring, Jarron},
  title        = {WebbPSF Extensions},
  month        = mar,
  year         = 2025,
  publisher    = {Zenodo},
  version      = {v2.0.1},
  doi          = {10.5281/zenodo.15086592},
  url          = {https://doi.org/10.5281/zenodo.15086592},
  swhid        = {swh:1:dir:20498e7db6df53d71f1c54eb340533d83538e19e
                   ;origin=https://doi.org/10.5281/zenodo.15033130;vi
                   sit=swh:1:snp:5a9f0ae0aa0e9d8587b5b01f09c438f46d9f
                   e140;anchor=swh:1:rel:5b09fa815d81db2320799999bbbb
                   b580555654ee;path=JarronL-webbpsf\_ext-182a238
                  },
}

@ARTICLE{Sanghi2026,
       author = {{Sanghi}, Aniket and {Mang}, James and {Llop-Sayson}, Jorge and {Mamajek}, Eric E. and {Thompson}, William and {Sur}, Ankan and {Beichman}, Charles and {Bryden}, Geoffrey and {G{\'a}sp{\'a}r}, Andr{\'a}s and {Leisenring}, Jarron and {Mawet}, Dimitri and {Morley}, Caroline V. and {Ruffio}, Jean-Baptiste and {Wolff}, Schuyler G. and {Ygouf}, Marie},
        title = "{Worlds Next Door. III. Indirect Evidence for Enhanced Atmospheric Metallicity and/or the Presence of Water Clouds in the Nearest Jupiter-analog ϵ Eri b}",
      journal = {\aj},
         year = 2026,
        month = apr,
       volume = {171},
       number = {4},
          eid = {225},
        pages = {225},
          doi = {10.3847/1538-3881/ae4909},
archivePrefix = {arXiv},
       eprint = {2602.23423},
 primaryClass = {astro-ph.EP},
       adsurl = {https://ui.adsabs.harvard.edu/abs/2026AJ....171..225S}
}

@ARTICLE{Batalha2019,
       author = {{Batalha}, Natasha E. and {Marley}, Mark S. and {Lewis}, Nikole K. and {Fortney}, Jonathan J.},
        title = "{Exoplanet Reflected-light Spectroscopy with PICASO}",
      journal = {\apj},
         year = 2019,
        month = jun,
       volume = {878},
       number = {1},
          eid = {70},
        pages = {70},
          doi = {10.3847/1538-4357/ab1b51},
archivePrefix = {arXiv},
       eprint = {1904.09355},
 primaryClass = {astro-ph.EP},
       adsurl = {https://ui.adsabs.harvard.edu/abs/2019ApJ...878...70B}
}

@ARTICLE{Mukherjee2023,
       author = {{Mukherjee}, Sagnick and {Batalha}, Natasha E. and {Fortney}, Jonathan J. and {Marley}, Mark S.},
        title = "{PICASO 3.0: A One-dimensional Climate Model for Giant Planets and Brown Dwarfs}",
      journal = {\apj},
         year = 2023,
        month = jan,
       volume = {942},
       number = {2},
          eid = {71},
        pages = {71},
          doi = {10.3847/1538-4357/ac9f48},
archivePrefix = {arXiv},
       eprint = {2208.07836},
 primaryClass = {astro-ph.EP},
       adsurl = {https://ui.adsabs.harvard.edu/abs/2023ApJ...942...71M}
}

@ARTICLE{Batalha2025,
       author = {{Batalha}, Natasha E. and {Rooney}, Caoimhe M. and {Visscher}, Channon and {Moran}, Sarah E. and {Marley}, Mark S. and {Sengupta}, Aditya R. and {Kiefer}, Sven and {Lodge}, Matt G. and {Mang}, James and {Morley}, Caroline V. and {Mukherjee}, Sagnick and {Fortney}, Jonathan J. and {Gao}, Peter and {Lewis}, Nikole K. and {Mayorga}, L.~C. and {Pearce}, Logan A. and {Wakeford}, Hannah R.},
        title = "{Condensation Clouds in Substellar Atmospheres with Virga}",
      journal = {arXiv e-prints},
         year = 2025,
        month = aug,
          eid = {arXiv:2508.15102},
        pages = {arXiv:2508.15102},
          doi = {10.48550/arXiv.2508.15102},
archivePrefix = {arXiv},
       eprint = {2508.15102},
 primaryClass = {astro-ph.EP},
       adsurl = {https://ui.adsabs.harvard.edu/abs/2025arXiv250815102B}
}

@ARTICLE{Moran2025,
       author = {{Moran}, Sarah E. and {Lodge}, Matt G. and {Batalha}, Natasha E. and {Ohno}, Kazumasa and {Vahidinia}, Sanaz and {Marley}, Mark S. and {Wakeford}, Hannah R. and {Leinhardt}, Zo{\"e} M.},
        title = "{Fractal Aggregate Aerosols in the Virga Cloud Code. I. Model Description and Application to a Benchmark Cloudy Exoplanet}",
      journal = {\apj},
         year = 2025,
        month = nov,
       volume = {994},
       number = {1},
          eid = {116},
        pages = {116},
          doi = {10.3847/1538-4357/ae0583},
archivePrefix = {arXiv},
       eprint = {2509.06708},
 primaryClass = {astro-ph.EP},
       adsurl = {https://ui.adsabs.harvard.edu/abs/2025ApJ...994..116M}
}

@ARTICLE{Marley2010,
       author = {{Marley}, Mark S. and {Saumon}, Didier and {Goldblatt}, Colin},
        title = "{A Patchy Cloud Model for the L to T Dwarf Transition}",
      journal = {\apjl},
         year = 2010,
        month = nov,
       volume = {723},
       number = {1},
        pages = {L117-L121},
          doi = {10.1088/2041-8205/723/1/L117},
archivePrefix = {arXiv},
       eprint = {1009.6217},
 primaryClass = {astro-ph.SR},
       adsurl = {https://ui.adsabs.harvard.edu/abs/2010ApJ...723L.117M}
}

@ARTICLE{Morley2014,
       author = {{Morley}, Caroline V. and {Marley}, Mark S. and {Fortney}, Jonathan J. and {Lupu}, Roxana},
        title = "{Spectral Variability from the Patchy Atmospheres of T and Y Dwarfs}",
      journal = {\apjl},
         year = 2014,
        month = jul,
       volume = {789},
       number = {1},
          eid = {L14},
        pages = {L14},
          doi = {10.1088/2041-8205/789/1/L14},
archivePrefix = {arXiv},
       eprint = {1406.0863},
 primaryClass = {astro-ph.SR},
       adsurl = {https://ui.adsabs.harvard.edu/abs/2014ApJ...789L..14M}
}

@article{Surjanovic2023,
  title={Pigeons.jl: {D}istributed sampling from intractable distributions},
  author={Surjanovic, Nikola and Biron-Lattes, Miguel and Tiede, Paul and Syed, Saifuddin and Campbell, Trevor and Bouchard-C{\^o}t{\'e}, Alexandre},
  journal={arXiv:2308.09769},
  year={2023}
}

@ARTICLE{Baines2012,
       author = {{Baines}, Ellyn K. and {Armstrong}, J. Thomas},
        title = "{Confirming Fundamental Properties of the Exoplanet Host Star epsilon Eridani Using the Navy Optical Interferometer}",
      journal = {\apj},
         year = 2012,
        month = jan,
       volume = {744},
       number = {2},
          eid = {138},
        pages = {138},
          doi = {10.1088/0004-637X/744/2/138},
       adsurl = {https://ui.adsabs.harvard.edu/abs/2012ApJ...744..138B}
}

@ARTICLE{Trifonov2020,
       author = {{Trifonov}, Trifon and {Tal-Or}, Lev and {Zechmeister}, Mathias and {Kaminski}, Adrian and {Zucker}, Shay and {Mazeh}, Tsevi},
        title = "{Public HARPS radial velocity database corrected for systematic errors}",
      journal = {\aap},
         year = 2020,
        month = apr,
       volume = {636},
          eid = {A74},
        pages = {A74},
          doi = {10.1051/0004-6361/201936686},
archivePrefix = {arXiv},
       eprint = {2001.05942},
 primaryClass = {astro-ph.EP},
       adsurl = {https://ui.adsabs.harvard.edu/abs/2020A&A...636A..74T}
}

@ARTICLE{MacGregor2015,
       author = {{MacGregor}, Meredith A. and {Wilner}, David J. and {Andrews}, Sean M. and {Lestrade}, Jean-Fran{\c{c}}ois and {Maddison}, Sarah},
        title = "{The Epsilon Eridani System Resolved by Millimeter Interferometry}",
      journal = {\apj},
         year = 2015,
        month = aug,
       volume = {809},
       number = {1},
          eid = {47},
        pages = {47},
          doi = {10.1088/0004-637X/809/1/47},
archivePrefix = {arXiv},
       eprint = {1507.01642},
 primaryClass = {astro-ph.SR},
       adsurl = {https://ui.adsabs.harvard.edu/abs/2015ApJ...809...47M}
}

@inproceedings{Bailey2023,
author = {Vanessa P. Bailey and Eduardo Bendek and Brian Monacelli and Caleb Baker and Gasia Bedrosian and Eric Cady and Ewan S. Douglas and Tyler Groff and Sergi R. Hildebrandt and N. Jeremy Kasdin and John Krist and Bruce Macintosh and Bertrand Mennesson and Patrick Morrissey and Ilya Poberezhskiy and Hari B. Subedi and Jason Rhodes and Aki Roberge and Marie Ygouf and Robert T. Zellem and Feng Zhao and Neil T. Zimmerman},
title = {{Nancy Grace Roman Space Telescope coronagraph instrument overview and status}},
volume = {12680},
booktitle = {Techniques and Instrumentation for Detection of Exoplanets XI},
editor = {Garreth J. Ruane},
organization = {International Society for Optics and Photonics},
publisher = {SPIE},
pages = {126800T},
year = {2023},
doi = {10.1117/12.2679036},
URL = {https://doi.org/10.1117/12.2679036}
}

@ARTICLE{Millar-Blanchaer2025,
       author = {{Millar-Blanchaer}, Maxwell A. and {Choquet}, {\'E}lodie and {Lawson}, Kellen and {Marino}, Sebasti{\'a}n and {Kammerer}, Jens and {Carter}, Aarynn L. and {Rebollido}, Isabel and {Leisenring}, Jarron M. and {Kim}, Minjae and {Kalas}, Paul and {Stapelfeldt}, Karl R. and {Hinkley}, Sasha and {Booth}, Mark and {Grady}, Carol A. and {Matthews}, Elisabeth C. and {Biller}, Beth A. and {Skemer}, Andrew and {Girard}, Julien H. and {Wolff}, Schuyler G. and {Ward-Duong}, Kimberly and {Meyer}, Michael R. and {Boccaletti}, Anthony and {Pantin}, Eric and {Matthews}, Brenda C. and {Metchev}, Stanimir and {Perrin}, Marshall D. and {Chen}, Christine H. and {Crotts}, Katie and {Absil}, Olivier and {Balmer}, William O. and {Calissendorff}, Per and {Cugno}, Gabriele and {Currie}, Thayne and {Danielski}, Camilla and {Hoch}, Kielan K.~W. and {Janson}, Markus and {Manjavacas}, Elena and {Lagage}, Pierre-Olivier and {Ben J. Sutlieff} and {Ray}, Shrishmoy and {Ren}, Bin B. and {Rickman}, Emily and {Su{\'a}rez}, Genaro and {Theissen}, Christopher A. and {Uyama}, Taichi and {Quirrenbach}, Andreas and {Wang}, Jason J. and {Whiteford}, Niall and {Wyatt}, Mark C. and {Zurlo}, Alice and {JWST ERS Collaboration}},
        title = "{The JWST Early Release Science Program for Direct Observations of Exoplanetary Systems. VI. Evidence for Radially Evolving Icy Grains in the HD 141569A Disk via NIRCam Coronagraphic Imaging}",
      journal = {\apj},
         year = 2025,
        month = dec,
       volume = {994},
       number = {2},
          eid = {199},
        pages = {199},
          doi = {10.3847/1538-4357/ae0615},
       adsurl = {https://ui.adsabs.harvard.edu/abs/2025ApJ...994..199M}
}

@ARTICLE{Cady2025,
       author = {{Cady}, Eric and {Bowman}, Nicholas and {Greenbaum}, Alexandra Z. and {Ingalls}, James G. and {Kern}, Brian and {Krist}, John and {Marx}, David and {Poberezhskiy}, Ilya and {Eldorado Riggs}, A.~J. and {Ruane}, Garreth and {Seo}, Byoung-Joon and {Shi}, Fang and {Zhou}, Hanying},
        title = "{High-order wavefront sensing and control for the Roman Coronagraph Instrument (CGI): architecture and measured performance}",
      journal = {Journal of Astronomical Telescopes, Instruments, and Systems},
         year = 2025,
        month = apr,
       volume = {11},
          eid = {021408},
        pages = {021408},
          doi = {10.1117/1.JATIS.11.2.021408},
archivePrefix = {arXiv},
       eprint = {2507.23738},
 primaryClass = {astro-ph.IM},
       adsurl = {https://ui.adsabs.harvard.edu/abs/2025JATIS..11b1408C}
}

@ARTICLE{Wang2020,
       author = {{Wang}, Jason J. and {Ginzburg}, Sivan and {Ren}, Bin and {Wallack}, Nicole and {Gao}, Peter and {Mawet}, Dimitri and {Bond}, Charlotte Z. and {Cetre}, Sylvain and {Wizinowich}, Peter and {De Rosa}, Robert J. and {Ruane}, Garreth and {Liu}, Michael C. and {Absil}, Olivier and {Alvarez}, Carlos and {Baranec}, Christoph and {Choquet}, {\'E}lodie and {Chun}, Mark and {Defr{\`e}re}, Denis and {Delorme}, Jacques-Robert and {Duch{\^e}ne}, Gaspard and {Forsberg}, Pontus and {Ghez}, Andrea and {Guyon}, Olivier and {Hall}, Donald N.~B. and {Huby}, Elsa and {Jolivet}, A{\"\i}ssa and {Jensen-Clem}, Rebecca and {Jovanovic}, Nemanja and {Karlsson}, Mikael and {Lilley}, Scott and {Matthews}, Keith and {M{\'e}nard}, Fran{\c{c}}ois and {Meshkat}, Tiffany and {Millar-Blanchaer}, Maxwell and {Ngo}, Henry and {Orban de Xivry}, Gilles and {Pinte}, Christophe and {Ragland}, Sam and {Serabyn}, Eugene and {Catal{\'a}n}, Ernesto Vargas and {Wang}, Ji and {Wetherell}, Ed and {Williams}, Jonathan P. and {Ygouf}, Marie and {Zuckerman}, Ben},
        title = "{Keck/NIRC2 L'-band Imaging of Jovian-mass Accreting Protoplanets around PDS 70}",
      journal = {\aj},
         year = 2020,
        month = jun,
       volume = {159},
       number = {6},
          eid = {263},
        pages = {263},
          doi = {10.3847/1538-3881/ab8aef},
archivePrefix = {arXiv},
       eprint = {2004.09597},
 primaryClass = {astro-ph.EP},
       adsurl = {https://ui.adsabs.harvard.edu/abs/2020AJ....159..263W}
}

@ARTICLE{Krist2023,
       author = {{Krist}, John E. and {Steeves}, John B. and {Dube}, Brandon D. and {Eldorado Riggs}, A.~J. and {Kern}, Brian D. and {Marx}, David S. and {Cady}, Eric J. and {Zhou}, Hanying and {Poberezhskiy}, Ilya Y. and {Baker}, Caleb W. and {McGuire}, James P. and {Nemati}, Bijan and {Kuan}, Gary M. and {Mennesson}, Bertrand and {Trauger}, John T. and {Saini}, Navtej S. and {Rafels}, Sergi Hildebrandt},
        title = "{End-to-end numerical modeling of the Roman Space Telescope coronagraph}",
      journal = {Journal of Astronomical Telescopes, Instruments, and Systems},
         year = 2023,
        month = oct,
       volume = {9},
          eid = {045002},
        pages = {045002},
          doi = {10.1117/1.JATIS.9.4.045002},
archivePrefix = {arXiv},
       eprint = {2309.16012},
 primaryClass = {astro-ph.IM},
       adsurl = {https://ui.adsabs.harvard.edu/abs/2023JATIS...9d5002K}
}

@INPROCEEDINGS{Krist2007,
       author = {{Krist}, John E.},
        title = "{PROPER: an optical propagation library for IDL}",
    booktitle = {Optical Modeling and Performance Predictions III},
         year = 2007,
       editor = {{Kahan}, Mark A.},
       series = {Society of Photo-Optical Instrumentation Engineers (SPIE) Conference Series},
       volume = {6675},
        month = sep,
          eid = {66750P},
        pages = {66750P},
          doi = {10.1117/12.731179},
       adsurl = {https://ui.adsabs.harvard.edu/abs/2007SPIE.6675E..0PK}
}

@ARTICLE{Llop-Sayson2025corosims,
       author = {{Llop-Sayson}, Jorge and {Bailey}, Vanessa P. and {Hom}, Justin and {Krist}, John and {Mennesson}, Bertrand and {Hasler}, Samantha N. and {Greenbaum}, Alexandra Z. and {Riggs}, A.~J. Eldorado and {Bryden}, Geoffrey},
        title = "{Roman coronagraph simulations of exozodi observations in the presence of wavefront errors}",
      journal = {Journal of Astronomical Telescopes, Instruments, and Systems},
         year = 2025,
        month = oct,
       volume = {11},
          eid = {045009},
        pages = {045009},
          doi = {10.1117/1.JATIS.11.4.045009},
archivePrefix = {arXiv},
       eprint = {2512.03308},
 primaryClass = {astro-ph.EP},
       adsurl = {https://ui.adsabs.harvard.edu/abs/2025JATIS..11d5009L}
}

@ARTICLE{Pueyo2016,
       author = {{Pueyo}, Laurent},
        title = "{Detection and Characterization of Exoplanets using Projections on Karhunen Loeve Eigenimages: Forward Modeling}",
      journal = {\apj},
         year = 2016,
        month = jun,
       volume = {824},
       number = {2},
          eid = {117},
        pages = {117},
          doi = {10.3847/0004-637X/824/2/117},
archivePrefix = {arXiv},
       eprint = {1604.06097},
 primaryClass = {astro-ph.IM},
       adsurl = {https://ui.adsabs.harvard.edu/abs/2016ApJ...824..117P}
}

@ARTICLE{Nowak2018,
       author = {{Nowak}, M. and {Le Coroller}, H. and {Arnold}, L. and {Dohlen}, K. and {Estevez}, D. and {Fusco}, T. and {Sauvage}, J.-F. and {Vigan}, A.},
        title = "{K-Stacker: Keplerian image recombination for the direct detection of exoplanets}",
      journal = {\aap},
         year = 2018,
        month = jul,
       volume = {615},
          eid = {A144},
        pages = {A144},
          doi = {10.1051/0004-6361/201629531},
archivePrefix = {arXiv},
       eprint = {1804.02192},
 primaryClass = {astro-ph.IM},
       adsurl = {https://ui.adsabs.harvard.edu/abs/2018A&A...615A.144N}
}

@ARTICLE{Mang2026flame-skimmer,
       author = {{Mang}, James and {Chachan}, Yayaati and {Morley}, Caroline V. and {Batalha}, Natasha E. and {Wogan}, Nicholas F. and {Mukherjee}, Sagnick and {Fortney}, Jonathan J. and {Marley}, Mark S. and {Visscher}, Channon and {Gharib-Nezhad}, Ehsan},
        title = "{The Sonora Substellar Atmosphere Models. VII. Flame Skimmer: Cloud-free Atmospheric and Evolutionary Models for the Coldest Substellar Objects}",
      journal = {arXiv e-prints},
         year = 2026,
        month = aug,
          eid = {arXiv:2608.06454},
        pages = {arXiv:2608.06454},
          doi = {10.48550/arXiv.2608.06454},
archivePrefix = {arXiv},
       eprint = {2608.06454},
 primaryClass = {astro-ph.EP},
       adsurl = {https://ui.adsabs.harvard.edu/abs/2026arXiv260806454M}
}

@ARTICLE{Mang2026picaso,
       author = {{Mang}, James and {Batalha}, Natasha E. and {Morley}, Caroline V. and {Wogan}, Nicholas F. and {Mukherjee}, Sagnick and {Visscher}, Channon and {Marley}, Mark S. and {Fortney}, Jonathan J. and {Chubb}, Katy L. and {Gao}, Peter and {Malsky}, Isaac},
        title = "{PICASO 4.0: Clouds and Photochemistry in Climate Models of Brown Dwarfs and Exoplanets}",
      journal = {\apj},
         year = 2026,
        month = mar,
       volume = {1000},
       number = {1},
          eid = {98},
        pages = {98},
          doi = {10.3847/1538-4357/ae47ff},
archivePrefix = {arXiv},
       eprint = {2602.22468},
 primaryClass = {astro-ph.EP},
       adsurl = {https://ui.adsabs.harvard.edu/abs/2026ApJ..1000...98M}
}

@unpublished{Wolff2026,
  author = {{Wolff}, Schuyler},
  title = "{Roman Coronagraph CPP at SPIE}",
  note = {submitted},
  year = {2026}
}

@ARTICLE{GaiaPrerelease,
       author = {{Gaia Collaboration} and {Panuzzo}, P. and {Mazeh}, T. and {Arenou}, F. and {Holl}, B. and {Caffau}, E. and {Jorissen}, A. and {Babusiaux}, C. and {Gavras}, P. and {Sahlmann}, J. and {Bastian}, U. and {Wyrzykowski}, {\L}. and {Eyer}, L. and {Leclerc}, N. and {Bauchet}, N. and {Bombrun}, A. and {Mowlavi}, N. and {Seabroke}, G.~M. and {Teyssier}, D. and {Balbinot}, E. and {Helmi}, A. and {Brown}, A.~G.~A. and {Vallenari}, A. and {Prusti}, T. and {de Bruijne}, J.~H.~J. and {Barbier}, A. and {Biermann}, M. and {Creevey}, O.~L. and {Ducourant}, C. and {Evans}, D.~W. and {Guerra}, R. and {Hutton}, A. and {Jordi}, C. and {Klioner}, S.~A. and {Lammers}, U. and {Lindegren}, L. and {Luri}, X. and {Mignard}, F. and {Nicolas}, C. and {Randich}, S. and {Sartoretti}, P. and {Smiljanic}, R. and {Tanga}, P. and {Walton}, N.~A. and {Aerts}, C. and {Bailer-Jones}, C.~A.~L. and {Cropper}, M. and {Drimmel}, R. and {Jansen}, F. and {Katz}, D. and {Lattanzi}, M.~G. and {Soubiran}, C. and {Th{\'e}venin}, F. and {van Leeuwen}, F. and {Andrae}, R. and {Audard}, M. and {Bakker}, J. and {Blomme}, R. and {Casta{\~n}eda}, J. and {De Angeli}, F. and {Fabricius}, C. and {Fouesneau}, M. and {Fr{\'e}mat}, Y. and {Galluccio}, L. and {Guerrier}, A. and {Heiter}, U. and {Masana}, E. and {Messineo}, R. and {Nienartowicz}, K. and {Pailler}, F. and {Riclet}, F. and {Roux}, W. and {Sordo}, R. and {Gracia-Abril}, G. and {Portell}, J. and {Altmann}, M. and {Benson}, K. and {Berthier}, J. and {Burgess}, P.~W. and {Busonero}, D. and {Busso}, G. and {Cacciari}, C. and {C{\'a}novas}, H. and {Carrasco}, J.~M. and {Carry}, B. and {Cellino}, A. and {Cheek}, N. and {Clementini}, G. and {Damerdji}, Y. and {Davidson}, M. and {de Teodoro}, P. and {Delchambre}, L. and {Dell'Oro}, A. and {Fraile Garcia}, E. and {Garabato}, D. and {Garc{\'\i}a-Lario}, P. and {Haigron}, R. and {Hambly}, N.~C. and {Harrison}, D.~L. and {Hatzidimitriou}, D. and {Hern{\'a}ndez}, J. and {Hestroffer}, D. and {Hodgkin}, S.~T. and {Jamal}, S. and {Jevardat de Fombelle}, G. and {Jordan}, S. and {Krone-Martins}, A. and {Lanzafame}, A.~C. and {L{\"o}ffler}, W. and {Lorca}, A. and {Marchal}, O. and {Marrese}, P.~M. and {Moitinho}, A. and {Muinonen}, K. and {Nu{\~n}ez Campos}, M. and {Oreshina-Slezak}, I. and {Osborne}, P. and {Pancino}, E. and {Pauwels}, T. and {Recio-Blanco}, A. and {Riello}, M. and {Rimoldini}, L. and {Robin}, A.~C. and {Roegiers}, T. and {Sarro}, L.~M. and {Schultheis}, M. and {Smith}, M. and {Sozzetti}, A. and {Utrilla}, E. and {van Leeuwen}, M. and {Weingrill}, K. and {Abbas}, U. and {{\'A}brah{\'a}m}, P. and {Abreu Aramburu}, A. and {Ahmed}, S. and {Altavilla}, G. and {{\'A}lvarez}, M.~A. and {Anders}, F. and {Anderson}, R.~I. and {Anglada Varela}, E. and {Antoja}, T. and {Baig}, S. and {Baines}, D. and {Baker}, S.~G. and {Balaguer-N{\'u}{\~n}ez}, L. and {Balog}, Z. and {Barache}, C. and {Barros}, M. and {Barstow}, M.~A. and {Bartolom{\'e}}, S. and {Bashi}, D. and {Bassilana}, J.-L. and {Baudeau}, N. and {Becciani}, U. and {Bedin}, L.~R. and {Bellas-Velidis}, I. and {Bellazzini}, M. and {Beordo}, W. and {Bernet}, M. and {Bertolotto}, C. and {Bertone}, S. and {Bianchi}, L. and {Binnenfeld}, A. and {Blanco-Cuaresma}, S. and {Bland-Hawthorn}, J. and {Blazere}, A. and {Boch}, T. and {Bossini}, D. and {Bouquillon}, S. and {Bragaglia}, A. and {Braine}, J. and {Bratsolis}, E. and {Breedt}, E. and {Bressan}, A. and {Brouillet}, N. and {Brugaletta}, E. and {Bucciarelli}, B. and {Butkevich}, A.~G. and {Buzzi}, R. and {Camut}, A. and {Cancelliere}, R. and {Cantat-Gaudin}, T. and {Capilla Guilarte}, D. and {Carballo}, R. and {Carlucci}, T. and {Carnerero}, M.~I. and {Carretero}, J. and {Carton}, S. and {Casamiquela}, L. and {Casey}, A. and {Castellani}, M. and {Castro-Ginard}, A. and {Ceraj}, L. and {Cesare}, V. and {Charlot}, P. and {Chaudet}, C. and {Chemin}, L. and {Chiavassa}, A. and {Chornay}, N. and {Chosson}, D.},
        title = "{Discovery of a dormant 33 solar-mass black hole in pre-release Gaia astrometry}",
      journal = {\aap},
         year = 2024,
        month = jun,
       volume = {686},
          eid = {L2},
        pages = {L2},
          doi = {10.1051/0004-6361/202449763},
archivePrefix = {arXiv},
       eprint = {2404.10486},
 primaryClass = {astro-ph.GA},
       adsurl = {https://ui.adsabs.harvard.edu/abs/2024A&A...686L...2G}
}

@ARTICLE{Lindegren2018,
       author = {{Lindegren}, L. and {Hern{\'a}ndez}, J. and {Bombrun}, A. and
         {Klioner}, S. and {Bastian}, U. and {Ramos-Lerate}, M. and
         {de Torres}, A. and {Steidelm{\"u}ller}, H. and {Stephenson}, C. and
         {Hobbs}, D. and {Lammers}, U. and {Biermann}, M. and {Geyer}, R. and
         {Hilger}, T. and {Michalik}, D. and {Stampa}, U. and {McMillan}, P.~J. and
         {Casta{\~n}eda}, J. and {Clotet}, M. and {Comoretto}, G. and
         {Davidson}, M. and {Fabricius}, C. and {Gracia}, G. and
         {Hambly}, N.~C. and {Hutton}, A. and {Mora}, A. and {Portell}, J. and
         {van Leeuwen}, F. and {Abbas}, U. and {Abreu}, A. and {Altmann}, M. and
         {Andrei}, A. and {Anglada}, E. and {Balaguer-N{\'u}{\~n}ez}, L. and
         {Barache}, C. and {Becciani}, U. and {Bertone}, S. and {Bianchi}, L. and
         {Bouquillon}, S. and {Bourda}, G. and {Br{\"u}semeister}, T. and
         {Bucciarelli}, B. and {Busonero}, D. and {Buzzi}, R. and
         {Cancelliere}, R. and {Carlucci}, T. and {Charlot}, P. and {Cheek}, N. and
         {Crosta}, M. and {Crowley}, C. and {de Bruijne}, J. and
         {de Felice}, F. and {Drimmel}, R. and {Esquej}, P. and {Fienga}, A. and
         {Fraile}, E. and {Gai}, M. and {Garralda}, N. and
         {Gonz{\'a}lez-Vidal}, J.~J. and {Guerra}, R. and {Hauser}, M. and
         {Hofmann}, W. and {Holl}, B. and {Jordan}, S. and {Lattanzi}, M.~G. and
         {Lenhardt}, H. and {Liao}, S. and {Licata}, E. and {Lister}, T. and
         {L{\"o}ffler}, W. and {Marchant}, J. and {Martin-Fleitas}, J. -M. and
         {Messineo}, R. and {Mignard}, F. and {Morbidelli}, R. and {Poggio}, E. and
         {Riva}, A. and {Rowell}, N. and {Salguero}, E. and {Sarasso}, M. and
         {Sciacca}, E. and {Siddiqui}, H. and {Smart}, R.~L. and {Spagna}, A. and
         {Steele}, I. and {Taris}, F. and {Torra}, J. and {van Elteren}, A. and
         {van Reeven}, W. and {Vecchiato}, A.},
        title = "{Gaia Data Release 2. The astrometric solution}",
      journal = {\aap},
         year = 2018,
        month = aug,
       volume = {616},
          eid = {A2},
        pages = {A2},
          doi = {10.1051/0004-6361/201832727},
archivePrefix = {arXiv},
       eprint = {1804.09366},
 primaryClass = {astro-ph.IM},
       adsurl = {https://ui.adsabs.harvard.edu/abs/2018A&A...616A...2L}
}

@ARTICLE{Ertel2020,
       author = {{Ertel}, S. and {Defr{\`e}re}, D. and {Hinz}, P. and {Mennesson}, B. and {Kennedy}, G.~M. and {Danchi}, W.~C. and {Gelino}, C. and {Hill}, J.~M. and {Hoffmann}, W.~F. and {Mazoyer}, J. and {Rieke}, G. and {Shannon}, A. and {Stapelfeldt}, K. and {Spalding}, E. and {Stone}, J.~M. and {Vaz}, A. and {Weinberger}, A.~J. and {Willems}, P. and {Absil}, O. and {Arbo}, P. and {Bailey}, V.~P. and {Beichman}, C. and {Bryden}, G. and {Downey}, E.~C. and {Durney}, O. and {Esposito}, S. and {Gaspar}, A. and {Grenz}, P. and {Haniff}, C.~A. and {Leisenring}, J.~M. and {Marion}, L. and {McMahon}, T.~J. and {Millan-Gabet}, R. and {Montoya}, M. and {Morzinski}, K.~M. and {Perera}, S. and {Pinna}, E. and {Pott}, J. -U. and {Power}, J. and {Puglisi}, A. and {Roberge}, A. and {Serabyn}, E. and {Skemer}, A.~J. and {Su}, K.~Y.~L. and {Vaitheeswaran}, V. and {Wyatt}, M.~C.},
        title = "{The HOSTS Survey for Exozodiacal Dust: Observational Results from the Complete Survey}",
      journal = {"Astronomical Journal"},
         year = 2020,
        month = apr,
       volume = {159},
       number = {4},
          eid = {177},
        pages = {177},
          doi = {10.3847/1538-3881/ab7817},
archivePrefix = {arXiv},
       eprint = {2003.03499},
 primaryClass = {astro-ph.SR},
       adsurl = {https://ui.adsabs.harvard.edu/abs/2020AJ....159..177E}
}

@article{astropy,
doi = {10.3847/1538-4357/ac7c74},
url = {https://doi.org/10.3847/1538-4357/ac7c74},
year = {2022},
month = {aug},
publisher = {The American Astronomical Society},
volume = {935},
number = {2},
pages = {167},
author = {The Astropy Collaboration and Price-Whelan, Adrian M. and Lim, Pey Lian and Earl, Nicholas and Starkman, Nathaniel and Bradley, Larry and Shupe, David L. and Patil, Aarya A. and Corrales, Lia and Brasseur, C. E. and Nöthe, Maximilian and Donath, Axel and Tollerud, Erik and Morris, Brett M. and Ginsburg, Adam and Vaher, Eero and Weaver, Benjamin A. and Tocknell, James and Jamieson, William and van Kerkwijk, Marten H. and Robitaille, Thomas P. and Merry, Bruce and Bachetti, Matteo and Günther, H. Moritz and Paper Authors and Aldcroft, Thomas L. and Alvarado-Montes, Jaime A. and Archibald, Anne M. and Bódi, Attila and Bapat, Shreyas and Barentsen, Geert and Bazán, Juanjo and Biswas, Manish and Boquien, Médéric and Burke, D. J. and Cara, Daria and Cara, Mihai and Conroy, Kyle E and Conseil, Simon and Craig, Matthew W. and Cross, Robert M. and Cruz, Kelle L. and D’Eugenio, Francesco and Dencheva, Nadia and Devillepoix, Hadrien A. R. and Dietrich, Jörg P. and Eigenbrot, Arthur Davis and Erben, Thomas and Ferreira, Leonardo and Foreman-Mackey, Daniel and Fox, Ryan and Freij, Nabil and Garg, Suyog and Geda, Robel and Glattly, Lauren and Gondhalekar, Yash and Gordon, Karl D. and Grant, David and Greenfield, Perry and Groener, Austen M. and Guest, Steve and Gurovich, Sebastian and Handberg, Rasmus and Hart, Akeem and Hatfield-Dodds, Zac and Homeier, Derek and Hosseinzadeh, Griffin and Jenness, Tim and Jones, Craig K. and Joseph, Prajwel and Kalmbach, J. Bryce and Karamehmetoglu, Emir and Kałuszyński, Mikołaj and Kelley, Michael S. P. and Kern, Nicholas and Kerzendorf, Wolfgang E. and Koch, Eric W. and Kulumani, Shankar and Lee, Antony and Ly, Chun and Ma, Zhiyuan and MacBride, Conor and Maljaars, Jakob M. and Muna, Demitri and Murphy, N. A. and Norman, Henrik and O’Steen, Richard and Oman, Kyle A. and Pacifici, Camilla and Pascual, Sergio and Pascual-Granado, J. and Patil, Rohit R. and Perren, Gabriel I and Pickering, Timothy E. and Rastogi, Tanuj and Roulston, Benjamin R. and Ryan, Daniel F and Rykoff, Eli S. and Sabater, Jose and Sakurikar, Parikshit and Salgado, Jesús and Sanghi, Aniket and Saunders, Nicholas and Savchenko, Volodymyr and Schwardt, Ludwig and Seifert-Eckert, Michael and Shih, Albert Y. and Jain, Anany Shrey and Shukla, Gyanendra and Sick, Jonathan and Simpson, Chris and Singanamalla, Sudheesh and Singer, Leo P. and Singhal, Jaladh and Sinha, Manodeep and Sipőcz, Brigitta M. and Spitler, Lee R. and Stansby, David and Streicher, Ole and Šumak, Jani and Swinbank, John D. and Taranu, Dan S. and Tewary, Nikita and Tremblay, Grant R. and Val-Borro, Miguel de and Van Kooten, Samuel J. and Vasović, Zlatan and Verma, Shresth and de Miranda Cardoso, José Vinícius and Williams, Peter K. G. and Wilson, Tom J. and Winkel, Benjamin and Wood-Vasey, W. M. and Xue, Rui and Yoachim, Peter and Zhang, Chen and Zonca, Andrea and Astropy Project Contributors},
title = {The Astropy Project: Sustaining and Growing a Community-oriented Open-source Project and the Latest Major Release (v5.0) of the Core Package*},
journal = {The Astrophysical Journal}
}

@ARTICLE{Stark2019,
       author = {{Stark}, Christopher C. and {Belikov}, Rus and {Bolcar}, Matthew R. and {Cady}, Eric and {Crill}, Brendan P. and {Ertel}, Steve and {Groff}, Tyler and {Hildebrandt}, Sergi and {Krist}, John and {Lisman}, P. Douglas and {Mazoyer}, Johan and {Mennesson}, Bertrand and {Nemati}, Bijan and {Pueyo}, Laurent and {Rauscher}, Bernard J. and {Riggs}, A.~J. and {Ruane}, Garreth and {Shaklan}, Stuart B. and {Sirbu}, Dan and {Soummer}, Remi and {Laurent}, Kathryn St. and {Zimmerman}, Neil},
        title = "{ExoEarth yield landscape for future direct imaging space telescopes}",
      journal = {Journal of Astronomical Telescopes, Instruments, and Systems},
         year = 2019,
        month = apr,
       volume = {5},
          eid = {024009},
        pages = {024009},
          doi = {10.1117/1.JATIS.5.2.024009},
archivePrefix = {arXiv},
       eprint = {1904.11988},
 primaryClass = {astro-ph.EP},
       adsurl = {https://ui.adsabs.harvard.edu/abs/2019JATIS...5b4009S}
}

@techreport{Ygouf2025,
  author = {Ygouf, M. and Krist, J. and Konomos, A. and Zvonarek, K. and Zellem, R. and Zimmerman, N. and Bailey, V.},
  title = {Roman Coronagraph Instrument Post Processing Report - OS11 HLC Distribution},
  institution = {Jet Propulsion Laboratory}, year = {2025},
  url = {https://roman.ipac.caltech.edu/docs/sims/20250107_Roman_CGI_OS11_post_processing_report_URS.pdf}}

@misc{10.5281/zenodo.21203363,
  author       = {{Roman Coronagraph Community Participation Program}},
  title        = {The Roman Coronagraph Community Participation Program: results from the August 2025 white paper call},
  month        = jul,
  year         = 2026,
  publisher    = {Zenodo},
  doi          = {10.5281/zenodo.21203363},
  url          = {https://doi.org/10.5281/zenodo.21203363}
}

@misc{10.17909/g6k6-bx64,
  author = {{Sanghi}, Aniket},
  title = {PID 9431 Data},
  publisher = {STScI/MAST},
  year = {2025},
  doi = {10.17909/g6k6-bx64},
  url = {https://doi.org/10.17909/g6k6-bx64}
}
\bibliographystyle{aasjournalv7}

\end{document}